\documentclass[11pt,a4paper]{article}
\usepackage{jheppub}
\usepackage{geometry}
\usepackage{amsfonts}
\usepackage{amsmath}
\usepackage{amssymb}
\usepackage{xcolor}
\usepackage{graphicx}
\usepackage{afterpage}
\usepackage[normalem]{ulem}

\font\mybb=msbm10 at 10pt
\def\bb#1{\hbox{\mybb#1}}
\def\be{\begin{equation}}
\def\ee{\end{equation}}
\def\bea{\begin{eqnarray}}
\def\eea{\end{eqnarray}}

\begin{document}
\title{{Towards field theory of multiple D0-branes.} {\rm Hamiltonian mechanics and quantization of simplest 3D prototype of multiple D$0$-brane system}
}
\author{
Igor Bandos $^{\dagger\ddagger}$ and Unai D.M. Sarraga $^{\dagger}$
\\ \bigskip
$^{\dagger}$ Department of Physics and EHU Quantum Center, University of the Basque Country UPV/EHU,
\\
P.O. Box 644, 48080 Bilbao, Spain
 \\
 $^{\ddagger}$
IKERBASQUE, Basque Foundation for Science, Plaza Euskadi 5,
48009, Bilbao, Spain}

\date{23/04/2024--26/06/2024}

\bigskip
\abstract{
Recently we have constructed a completely supersymmetric nonlinear action possessing the properties expected from multiple D$0$-brane system. Its quantization should result in an interesting supersymmetric field theory in the (super)space with additional matrix coordinates which can provide an important insights in the study of String Theory. As a first stage toward this aim, in this paper  we construct the Hamiltonian mechanics and perform covariant quantization of the simplest three dimensional  counterpart of the ten dimensional multiple D$0$-brane model. We obtain a supersymmetric system of equations in a (super)spacetime enlarged by bosonic and fermionic matrix
coordinates which appears  as a result of such quantization and discuss some of its properties.}
\def\theequation{\arabic{section}.\arabic{equation}}
\maketitle

\thispagestyle{empty}

\section{Introduction}

The ``second (super)string revolution'' of the middle 90th brought us understanding of an important role of supersymmetric extended objects, super-$p$-branes, in String Theory (presently also known as M-theory). Here $p$ is the number of spatial dimensions of the worldvolume of the brane which is a surface in spacetime (actually in superspace), so that $p=1$ corresponds to string, $p=2$ to membrane and $p=0$ to particles. The bosonic degrees of freedom of simplest  $p$-branes are described (in a suitably chosen coordinate system)  by a $\text{D}-p-1$ coordinate functions which fix the position of the $\text{d}=p+1$ dimensional $p$-brane worldvolume in the $\text{D}$-dimensional spacetime. This is the case for ten dimensional ($\text{D}=10$) fundamental string (F$1$-brane) \cite{Green:1981yb,Green:1983wt} and  $\text{D}=11$ supermembrane \cite{Bergshoeff:1987cm} (also known as M2-brane). However, there are branes which carry additional fields on their worldvolume.

In particular, this is the case for  Dirichlet $p$-branes or D$p$-branes which are the supersymmetric extended objects where the fundamental string can have its endpoints \cite{Sagnotti:1987tw,Dai:1989ua,Horava:1989ga} which carry $\text{d}=p+1$ dimensional gauge fields on their worldvolumes. Super-D$p$-branes can be described by supersymmetric solutions of 10D supergravity equations \cite{Polchinski:1995mt},
by worldvolume actions \cite{Cederwall:1996pv,Aganagic:1996pe,Cederwall:1996ri,Aganagic:1996nn,Bergshoeff:1996tu,Bandos:1997rq} or by the superfield equations obtained in  \cite{Howe:1996mx} (before the actions were known!) in the frame of superembedding approach \cite{Bandos:1995zw,Sorokin:1999jx,Bandos:2009xy,Bandos:2023web}.

In type IIA superstring theory there exist supersymmetric D$p$-branes with $p=0,2,4,6$ and $8$, while the zoo  of  type IIB superstring theory is populated by super-D$p$-branes with $p=1,3,5,7$ and $9$. The $\text{D}$9-brane of type IIB theory is spacetime filling as far as superstring is consistent in spacetime of  dimension $\text{D}=10$, and this object was studies much earlier than its lower $p$ cousins.

Its counterpart in the bosonic string theory, which is consistent in 26-dimensional spacetime, $\text{D}=26$, is the D$25$-brane. A bosonic string ending on a D$25$-brane is actually an open string the endpoints of which are free; it was studied beginning the very first years  of String Theory \cite{Goddard:1973qh}. In fact, it was more interesting than closed string for considering string model as an effective description of QCD \cite{Nambu:1970unpub} with the aim to describe the confinement of quarks.

The free endpoints of open strings can be coupled minimally to gauge fields.  The consistency of such a coupling  was studied by Fradkin and  Tseytlin \cite{Fradkin:1985qd} who found that it requires the gauge field to obey equations which can be obtained from the Born-Infeld (BI) action \cite{Born:1934gh}. In the modern language, this implies that the BI action is an effective action of spacetime filling D$25$-brane of the bosonic string theory. The spacetime filling D$9$-brane of type IIB superstring theory is described by a (nontrivial) supersymmetric extension of $\text{D}=10$ BI action \cite{Cederwall:1996ri,Aganagic:1996nn,Bergshoeff:1996tu,Akulov:1998bq}.

The discovery of lower $p$, not spacetime filling D$p$-branes  was delayed till late 80th by believing that the open string with free endpoints and closed strings were the only possible boundary conditions which did not break the target space (i.e. $\text{D}$-dimensional spacetime) Lorentz invariance $\text{SO}(1,\text{D}-1)$. Indeed the string with endpoints located on $\text{d}=p+1$ dimensional planes in spacetime  breaks $\text{SO}(1,\text{D}-1)$ down to its $\text{SO}(1,p)\otimes \text{SO}(\text{D}-p-1)$ subgroup. The breakthrough in  \cite{Sagnotti:1987tw,Dai:1989ua,Horava:1989ga} used the brilliant result of Scherk and Schwarz  who discovered that closed string model actually describes a theory of gravity \cite{Scherk:1974ca}.  Since open string theory inevitably contains the closed string sector, it describes a theory of gravity too. In a theory of gravity one cannot imagine appearance of fixed planes so that the $\text{d}=p+1$ dimensional surfaces where the open  string can have endpoints should be a dynamical objects and these objects were called Dirichlet $p$-branes in  \cite{Dai:1989ua}. Then the breaking of the target space Lorentz symmetry $\text{SO}(1,\text{D}-1)$ down to  $\text{SO}(1,p)\otimes \text{SO}(\text{D}-p-1)$  is a spontaneous breaking characteristic for any $p$-brane \cite{Hughes:1986fa}.

The endpoints of an open string on the D$p$-brane can be coupled to $\text{d}=p+1$ dimensional gauge field, and the consistency of such coupling was found to require the gauge field to obey the equations which follow from  the Dirac-Born-Infeld (DBI) action \cite{Leigh:1989jq}. The actions of
super-D$p$-branes  \cite{Cederwall:1996pv,Aganagic:1996pe,Cederwall:1996ri,Aganagic:1996nn,Bergshoeff:1996tu} are given by quite nontrivial
supersymmetric generalization of DBI action which, in addition to a straightforward  supersymmetric generalization of the DBI term, contains the so-called Wess-Zumino term describing the coupling of D$p$-brane to (a superspace generalization of) so-called Ramond-Ramond gauge fields of the supergravity multiplet. The characteristic property of these super-D$p$-brane actions is their invariance under local fermionic $\kappa$-symmetry. It is important because its existence guarantees that the ground state of the dynamical system preserves a half of target space supersymmetry\footnote{Of course, this property is common for supersymmetric branes in their complete and Lorentz invariant formulations; in particular this is the case of Green-Schwarz superstring \cite{Green:1983wt} and 11D supermembrane \cite{Bergshoeff:1987cm}. } and thus is a stable 1/2 BPS state\footnote{BPS abbreviates for Bogomol'nyi-Prasad-Sommerfield, the authors of \cite{Bogomolny:1975de,Prasad:1975kr}. BPS state is a state which saturates the so-called Bogomol'nyi bound and is stable due to this saturation. $1/2$ BPS  implies that the saturation is related to the preservation of 1/2 of supersymmetry by the solution.}  \cite{Bergshoeff:1997kr} (see \cite{Bandos:2001jx} for further development).

The quest for the action of  multiple  D$p$-brane (mD$p$) system, which should provide an effective description of the maximally supersymmetric system of $N$ nearly coincident D$p$-branes and $N^2$ fundamental strings ending on these, was created by seminal paper  \cite{Witten:1995im} in which E. Witten argued that such a system exhibits the enhanced U$(N)$ gauge symmetry and, at very low energy, allows for the gauge fixed description by U$(N)$ Supersymmetric-Yang-Mills (SYM) action. Setting $N=1$, we arrive at the Abelian SYM or super-Maxwell action with maximal supersymmetry which can be obtained as a weak field approximation from  the gauge fixed version of the action for single super-D$p$-brane \cite{Cederwall:1996pv,Aganagic:1996pe,Cederwall:1996ri,Aganagic:1996nn,Bergshoeff:1996tu} which is given by the sum of Dirac-Born-Infeld (DBI) and Wess-Zumino (WZ) terms. Hence it was expected that the mD$p$ action contains some non-Abelian generalization of the DBI term or Born-Infeld (BI) action in the case of spacetime filling D$9$-brane.

This latter case deserves some comments. The set of worldvolume fields of single super-D$9$-brane consists of fermionic coordinate functions, a pair of $16$-component  fermionic spinors of the same chirality
 and an Abelian ten dimensional gauge field. Upon  gauge fixing of $\kappa$-symmetry by reducing twice the number of fermionic fields, the worldvolume field content of D$9$-brane literally coincides with 10d SYM multiplet, the Wess-Zumino term vanishes and its action becomes a maximally supersymmetric extension of the Born-Infeld functional  with nonlinearly-realized supersymmetry \`a la Volkov-Akulov \cite{Aganagic:1996nn}. The multiple D$9$-brane action should contain a non-Abelian generalization of this latter and even in the purely bosonic case such a generalization requires some additional specification.  In 1997 Tseytlin proposed to use  the symmetric trace prescription for this purposes \cite{Tseytlin:1997csa}.  However, the supersymmetric generalization of  Tseytlin's action, which could be considered as a complete action of
multiple D$9$-brane action, is still unknown.

Although the problem of constructing supersymmetric actions for multiple D$p$-branes is still unsolved in its complete form, many interesting results were obtained during these years
\cite{Tseytlin:1997csa,Emparan:1997rt,Taylor:1999gq,Taylor:1999pr,Myers:1999ps,Bergshoeff:2000ik,
Bergshoeff:2001dc,Sorokin:2001av,Janssen:2002vb,Drummond:2002kg,Panda:2003dj,Janssen:2002cf,Janssen:2003ri,
Lozano:2005kf,Howe:2005jz,Howe:2006rv,Howe:2007eb,Bandos:2009yp,Bandos:2009gk,Bandos:2010hc,McGuirk:2012sb,
Bandos:2012jz,Bandos:2013uoa,Bandos:2013swa,Choi:2017kxf,Choi:2018fqw,Bandos:2018ntt,Brennan:2019azg,Bandos:2021vrq,Bandos:2022uoz,Bandos:2022dpx}
(see \cite{Bandos:2022uoz} for a  description of these results in some details). In particular, an action which possesses  the  properties expected from mD$0$ system in flat type IIA superspace was found in our \cite{Bandos:2022uoz}.

More specifically, the model in \cite{Bandos:2022uoz} is formulated in terms of bosonic and fermionic coordinate functions and the worldline fields of 1d SU$(N)$ SYM multiplet with ${\cal N}=16$ supersymmetry and it is invariant, besides the target space supersymmetry, under local worldline supersymmetry. This local worldline supersymmetry acts on the coordinate functions as $\kappa$-symmetry of single D$0$-brane and on the  SYM fields as local version of the supersymmetry of SYM supermultiplet. Its existence guarantees that the ground state of the system is a supersymmetric 1/2 BPS state which can be identified with the BPS state of multiple D$0$-brane which can be described by supergravity solution. Upon gauge fixing of the worldline supersymmetry and reparametrization symmetry the field content reduces to that of 1d U$(N)$ SYM multiplet  with ${\cal N}=16$ supersymmetry, or with the dimensional reduction of  $\text{D}=10$ SYM for U$(N)$ gauge group, as expected for the mD$0$ system. This allows us to conclude that in \cite{Bandos:2022uoz} we have found a good candidate for mD$0$-brane action.

A single D$0$-brane is just a  massive  superparticle in type IIA superspace \cite{Bergshoeff:1996tu} and  its worldvolume is a one dimensional worldline. A one  dimensional (1d) gauge field  is known to be pure gauge, thus such  is the D$0$-brane worldline gauge field. Furthermore its field strength cannot be defined so that in D$0$-brane case the  DBI part of the  action is reduced  to just a standard superparticle kinetic term. Furthermore, the WZ term is constructed from fermionic coordinates of flat type IIA superspace and, hence,  the D$0$-brane  action is given by 10D generalization of the de Azc\'arraga-Lukierski action for the $\text{D}=4$ ${\cal N}=2$ massive superparticle \cite{deAzcarraga:1982dhu,deAzcarraga:1982njd}. Thus, there is no reason to expect a kind of non-Abelian DBI-like contribution to the mD$0$ action, and our candidate mD$0$ action(s) in \cite{Bandos:2022uoz} do not contain such.

The unexpected property of the action of \cite{Bandos:2022uoz} is the presence of an arbitrary positive definite function ${\cal M}({\cal H})$ of an SU$(N)$ invariant function of matrix fields ${\cal H}$ in which one can recognize the Hamiltonian of SYM model \footnote{In  \cite{Bandos:2022uoz}  and \cite{Bandos:2022dpx}, for shortness, we allowed ourselves to call it 'relative motion Hamiltonian', but in the context of present study such a name may be confusing and we avoid to use it. }.
Surprisingly, the action possesses local fermionic worldline supersymmetry, characteristic property expected from mD$0$ system, for arbitrary choice of  ${\cal M}({\cal H})$, so that we can speak about a family of candidate actions for the description of mD$0$.
Of this family, the simplest representative with  ${\cal M}({\cal H})=m$ for constant $m$ appearing also in the center of mass part of the action had been found before in \cite{Bandos:2018ntt}. However, the most promising candidate for the role of mD$0$ action has a particular nontrivial form of  ${\cal M}({\cal H})$ given in Eq. \eqref{cM=m+} below:  as we showed in \cite{Bandos:2022dpx} it  can be obtained by dimensional reduction from the 11D multiple M-wave (also called multiple M$0$ or mM$0$) action of \cite{Bandos:2012jz}.

The quantization of mD$0$ system should result in an interesting field theory in a (super)spacetime extended by a number of bosonic and fermionic matrix coordinates which can bring  us a new comprehension in String/M-theory. To argue in favour of such hope, let us recall the concept of $p$-brane democracy by Paul Townsend \cite{Townsend:1995gp} which states that any of the String/M-theory $p$-branes can be considered as fundamental, thus taking on the role usually attributed to fundamental strings, while all other $p'$-branes emerge as solitonic objects  in such  a theory  of  fundamental $p$-brane. From this point of view, the standard choice of string as  the fundamental object is preferable just because of convenience  to have equations which can be linearized in a suitable gauge and well defined perturbation theory. Furthermore, the interaction in this case is described by a topology of the worldsheet thus allowing one to use the free string action also in description of interacting strings. However the string field theory (see \cite{Witten:1985cc,Siegel:1988yz,Zwiebach:1992ie,Erler:2019vhl,Sen:2024nfd} and refs therein) is not easy at all, and is still under construction, at least in its complete form.

Among other possibilities, the choice of D$0$-branes as fundamental objects was discussed as very intriguing
opportunity, as it seemed to indicate  a possibility to come back from extended objects to particles as basic  building blocks for  constructing a fundamental theory. However, to this end one should consider rather a theory of many  D-particles interacting among themselves, which is to say
the dynamical system which we call multiple D$0$-brane or  mD$0$. This is what suggests us  to expect that the field theory of mD$0$-brane system can give us new insights in String/M-theory.

The study of mD$0$ system using the low energy effective action  suggested by Witten, i.e. by using just the action of 10D U$(N)$ SYM dimensionally reduced to $\text{d}=1$  \cite{Danielsson:1996uw}, was one of the roots of the M(atrix)-theory conjecture
\cite{Banks:1996vh} which also used the 1d ${\cal N}=16$ U$(N)$ SYM action, presently also known as the action of BFSS matrix model\footnote{BFSS is for Banks, Fischler,  Shenker and Susskind, the authors of \cite{Banks:1996vh}.}.
The ground state wavefunction of the quantum mechanics obtained by quantization of BFSS action was studied
in \cite{Halpern:1997fv,Graf:1998bm,Frohlich:1999zf,Hoppe:2000tj,Hasler:2002wt,Lin:2014wka,Hoppe:2023vph}; related studies can be found also in \cite{Komatsu:2024vnb}.

Our action \cite{Bandos:2022uoz,Bandos:2022dpx} can be considered as 10D Lorentz invariant and completely gauge invariant (supersymmetric and $\kappa$-symmetric) generalization of this BFSS action. Hence its quantization should provide us with the complete supersymmetric and manifestly Lorentz invariant mD$0$ field theory.

The aim of this paper is to start the program of constructing the mD$0$-brane field theory. As this problem has happened to be quite complicated, we begin with quantization of a 3D counterpart of mD$0$ system (which we will roughly call 3D mD$0$) described by 3D version of 10D mD$0$ action found in our \cite{Bandos:2021vrq}. Moreover, we will restrict ourselves by the simplest case  described by the action from \cite{Bandos:2021vrq} with constant ${\cal M}({\cal H})=m$ (i.e. 3D counterpart of the action from \cite{Bandos:2018ntt}).
This problem (which also appeared to be quite complex) has provided us with a good toy model to approach 10D mD$0$ system: it has indicated  where the problems can appear in 10D mD$0$ quantization and has suggested the method to resolve these.

Furthermore, the superfield theory obtained in this work may be  interesting on its own as it provides a generalization
(supersymmetric and 3D Lorentz covariant) of a simplest version of the problem of the quantization of
Yang-Mills theory in Schr\"odinger representation (see recent \cite{Nair:2023hsy} for the comprehensive  review of $\text{D}=3$ case and for references). To consider this just for the case of 1d dimensional reduction of  SYM
(of 3D ${\cal N}=2$ SYM in our case) can be relevant e.g. for the  description of effective Hamiltonian of zero modes in this theory \cite{Smilga:2002mra}.

The paper is organized according the Table of Contents. The brief description of the work can be found in the concluding Sec. \ref{section:Conclusion}.

\setcounter{equation}{0}

\section{Simplest 3D counterpart of our  10D mD$0$ action }

\subsection{Physical and auxiliary fields of  a 3D counterpart of mD$0$ system  }

The dynamical variables of 3D mD0 system consisting of $N$ D0-branes and strings connecting the different D$0$'s or ending on the same D0-brane, can be split into the center of mass variables and the matrix fields describing the relative motion of the mD0 constituents. This latter set contains the fields of $\text{D}=3$ SU($N$) Yang-Mills model dimensionally reduced to $\text{d}=1$ and thus includes traceless $N \times N$  bosonic matrix fields
${\bb Z}=|| {\bb Z}_i{}^j(\tau)  || $ and $\bar{{\bb Z}}= ||\bar{{\bb Z}}_i{}^j(\tau) ||$  related by Hermitian conjugation,
\be\label{Z=bZ*}
{{\bb Z}} = (\bar{{\bb Z}})^\dagger \qquad \Leftrightarrow \qquad {{\bb Z}}_i{}^j = (  \bar{{\bb Z}}_j{}^i)^* \; , \qquad {\rm tr}\, {{\bb Z}} = {{\bb Z}}_i{}^i=0\; ,  \qquad i,j=1,...,N\; ,
\ee
two Hermitian conjugate  traceless $N \times N$ fermionic matrix fields $\mathbf{\Psi}=||{\mathbf{\Psi}}_i{}^j(\tau) ||$ and $\bar{\mathbf{\Psi}}=|| \bar{\mathbf{\Psi}}_i{}^j (\tau) ||$
\be\label{Psi=bPsi*}
{\mathbf{\Psi}}=\bar{\mathbf{\Psi}}{}^\dagger \qquad \Leftrightarrow \qquad {\mathbf{\Psi}}_i{}^j =( \bar{\mathbf{\Psi}}_j{}^i )^*\; , \qquad {\rm tr}\,  {\mathbf{\Psi}} \equiv  {\mathbf{\Psi}}_i{}^i=0
\ee
and anti-Hermitian 1-form
\be
{\bb A}=\text{d}\tau {\bb A}_\tau= \text{d}\tau || {\bb A}_\tau{}_i{}^j (\tau)|| =- {\bb A}^\dagger \qquad \Leftrightarrow \qquad   {\bb A}_\tau{}_i{}^j=- ( {\bb A}_\tau{}_j{}^i)^*\; .
\ee
All these fields depend  on the proper time $\tau$ which parametrizes the center of mass worldline ${\cal W}^1$ of our 3D mD0 system.

The set of the center of mass variables is the same as one used for the description of single D$0$-brane in the spinor moving frame formulation. This contains real three-vector bosonic and two complex conjugate  fermionic spinor coordinate functions
\be
z^M(\tau) = (x^\mu(\tau), \theta^\alpha (\tau), \bar{\theta}{}^\alpha (\tau)), \qquad {}  \qquad  \bar{\theta}{}^\alpha (\tau)=(\theta^\alpha (\tau))^*\; , \qquad  {}  \qquad   \mu=0,1,2\; , \qquad \alpha=1,2 \; , \qquad
\ee
which define parametrically the worldline ${\cal W}^1$ as a surface in flat D=3 ${\cal N}=2$ superspace $\Sigma^{(3|4)}$ with coordinates
$z^M = (x^\mu, \theta^\alpha , \bar{\theta}{}^\alpha)$
\bea
{\cal W}^1 \; \subset \; \Sigma^{(3|4)}\; : \qquad z^M=z^M(\tau) \; ,
\eea
and the spinor moving frame variables which we will describe below.

We have denoted the coordinate functions by the same symbol as coordinates believing that this cannot produce a confusion since only the coordinate functions will be used in the study of classical mechanics while only coordinates will be used then in the description of field theory obtained upon quantization.

\subsection{3D version of the de Azc\'arraga-Lukierski action for single D0-brane }

When single D0-brane is considered, its action can be written in terms of the above described coordinate functions only. We present it here as a reference point which also allows us to fix our notation.

The standard action for single D0-brane \cite{Bergshoeff:1996tu}
 is given by 10D counterpart of the $\text{D}=4$ action by the de Azc\'arraga and Lukierski \cite{deAzcarraga:1982dhu,deAzcarraga:1982njd}. Here we will discuss its 3D counterpart given by
\be\label{SD0=A+L}
S_{\text{A+L}}=- m\int \text{d}\tau \left(\sqrt{E_\tau^a E_{\tau}^b\eta_{ab}}+\dot{\theta}^\alpha \bar{\theta}_\alpha- \theta^\alpha \dot{\bar{\theta}}_\alpha \right)\,
\ee
where dot denotes derivative with respect to proper time $\tau$ and
\be\label{Eta=}
E_\tau^a =\dot{x}{}^a - i \dot{\theta}\gamma^a \bar{\theta} + i {\theta}\gamma^a \dot{\bar{\theta}}\; , \qquad
\ee
is the (coefficient for $\text{d}\tau$ of the) pull-back of the Volkov-Akulov 1-form,
\be\label{Ea=}
E^a=  \text{d}{x}{}^a - i \text{d}{\theta}\gamma^a \bar{\theta} + i {\theta}\gamma^a \text{d}{\bar{\theta}}=: E^a(z)\; , \qquad E^a(z(\tau))=
\text{d}\tau E_\tau^a \; , \qquad a=0,1,2\; . \qquad
\ee
In it
\be \dot{\theta}\gamma^a \bar{\theta}= \dot{\theta}{}^\alpha \gamma^a_{\alpha\beta} \bar{\theta}{}^\beta ,
\ee
with symmetric 2$\times$2 matrices
\be
 \gamma^a_{\alpha\beta}
 = -i \gamma^a_\alpha{}^\sigma\epsilon_{\sigma\beta}\; , \qquad \tilde{\gamma}{}^{a\alpha\beta}
= i \epsilon^{\alpha\sigma}\gamma^a{}_\sigma{}^{\beta} \, \qquad
\ee
constructed from unit antisymmetric spin tensor
\be \label{eps=} \epsilon^{\alpha\beta}= i\sigma_2= - \epsilon_{\alpha\beta}
 = \left(\begin{matrix}0& 1 \cr -1 & 0\end{matrix}\right)\; . \ee
and 3D Dirac matrices. These  are imaginary in our mostly minus metric notation and obey
\be
\gamma^a\gamma^b = \eta^{ab}+ i\epsilon^{abc}\gamma_c\; , \qquad \eta^{ab}=\text{diag}(+1,-1,-1)\; .
\ee
In all other cases the spinor indices are raised and lowered by the epsilon symbol \eqref{eps=}, e.g.
\be
\theta_\alpha = \epsilon_{\alpha\beta} \theta^\beta \; , \qquad \theta^\alpha = \epsilon^{\alpha\beta} \theta_\beta \; . \qquad
\ee

The characteristic property of the action \eqref{SD0=A+L} is its invariance under local fermionic $\kappa$-symmetry
\bea
&& \delta_\kappa x^\mu =  i \delta_\kappa{\theta}\gamma^a \bar{\theta} - i {\theta}\gamma^a \delta_\kappa{\bar{\theta}}\; , \qquad \nonumber\\ \nonumber \\
&& \delta_\kappa{\theta}{}^\alpha=\kappa_\beta \left(\epsilon -i\tilde{\gamma}_a E_\tau^a /\sqrt{E_\tau^cE_{\tau c}}\right)^{\beta\alpha} \; , \qquad \delta_\kappa\bar{\theta}{}^\alpha=\bar{\kappa}_\beta \left(\epsilon +i\tilde{\gamma}_a E_\tau^a /\sqrt{E_\tau^cE_{\tau c}}\right)^{\beta\alpha} \; . \qquad \\ \nonumber
\eea
This property is important because it implies that the ground state of the system preserves 1/2 of the spacetime supersymmetry \cite{Bergshoeff:1997kr} and hence is a BPS state, a 3D counterpart of the D$0$-brane  BPS state of String Theory.

Thus the moving frame and spinor moving frame variables are not obligatory to describe a single D$0$-brane: in this case spinor moving frame formulation, which can be found  in \cite{Bandos:2021vrq}, is classically  equivalent to the one provided by the above described de Azc\'arraga-Lukierski action. However, the action for 3D mD0 system, which was found in \cite{Bandos:2021vrq}, is known only in its form involving the spinor moving frame variables. We describe these in the next (sub)section.

\subsection{Spinor moving frame variables (Lorentz harmonics)}

In \cite{Bandos:2021vrq} we have used the real spinor moving frame variables $(v_{\alpha}^1, v_{\alpha}^2)$  constrained by
$ v^{\alpha 2}v_{\alpha}^1=1$. \footnote{This constraint implies that $2\times 2$ matrix
$(v_{\alpha}^1, v_{\alpha}^2)$ belongs to  double covering SL$(2,{\mathbb R})$ of the 3D Lorentz group
SO$(1,2)$. Hence the second name of Lorentz harmonics; see below, in particular  footnote \ref{HarmFoot}, for more details and references. } However, we later found it much more convenient  to describe the spinor frame variables using complex spinors
\be\label{w=} w_\alpha= \frac 1 {\sqrt{2}} (v_{\alpha}^1-iv_{\alpha}^2)\qquad \text{  and its c.c.} \qquad
\bar{w}_\alpha= \frac 1 { \sqrt{2}} (v_{\alpha}^1+iv_{\alpha}^2)\ee
which obey
\be \label{bww=i}
\bar{w}{}^\alpha w_\alpha=i\,
\ee so that

 \be\delta_\alpha{}^\beta = \, i \bar{w}{}_\alpha w^\beta- i w_\alpha  \bar{w}{}^\beta\; , \qquad \epsilon_{\alpha\beta} = -i \bar{w}{}_\alpha w_\beta+ i w_\alpha  \bar{w}{}_\beta\; .
\ee
Taking these constraints into account, we find that the derivatives of these complex spinors (providing a 3D version of the 4D Newman-Penrose diad \cite{Newman:1961qr}) are expressed by
\be\label{Dw=}
\text{D}w_\alpha :=\text{d}w_\alpha + ia w_\alpha =  if \bar{w}_\alpha \; , \qquad \text{D} \bar{w}_\alpha :=\text{d} \bar{w}_\alpha - ia  \bar{w}_\alpha =- i \bar{f} {w}_\alpha  \; , \qquad
\ee
\noindent in terms of Cartan forms \footnote{These are related with real Cartan forms $f^{pq}=v^{\alpha p}\text{d}v_{\alpha}^q$  used in  \cite{Bandos:2021vrq} by
\be
f=\frac 1 2 (f^{11}-f^{22}) -if^{12} \; , \qquad \bar{f}=\frac 1 2 (f^{11}-f^{22}) +if^{12}  \; , \qquad a=\frac 12 (f^{11}+f^{22})=\frac 12 f^{qq} \; . \nonumber \qquad
\ee}

\be\label{f=wdw}
f= w^\alpha \text{d}w_\alpha\; , \qquad \bar{f} = \bar{w}^\alpha \text{d}\bar{w}_\alpha\; , \qquad a = w^\alpha \text{d}\bar{w}_\alpha= \bar{w}^\alpha \text{d}w_\alpha\; . \qquad
\ee

\noindent
The first two of these  provide a  covariant basis of the space co-tangent to the coset  SU$(1,1)/\text{U}(1)= \text{SL}(2,{\bb R})/\text{SO}(2)$, while the third, $a$, transforms as a connection under U$(1)$ symmetry of our construction. This allows us to introduce the U$(1)$-covariant derivatives $\text{D}$ in \eqref{Dw=}.
It is easy to check that these Cartan forms obey the following structure equations (Maurer-Cartan equations)

\be
\text{D}f=\text{d}f+2i f\wedge a =0\; , \qquad \text{D}\bar{f}= \text{d}\bar{f}-2i \bar{f}\wedge a =0\; , \qquad \text{d}a=-if\wedge \bar{f} \; . \qquad
\ee

The set of moving frame vectors (a 3D version of the complex 4D light-like Newman-Penrose tetrade \cite{Newman:1961qr}) is composed from the complex spinor $w$ and its c.c.   as
\be\label{u=wgw}
u_a^{(0)}=w\gamma_a\bar{w}\; , \qquad u_a=w\gamma_aw\; , \qquad \bar{u}_a=\bar{w}\gamma_a\bar{w}\; . \qquad
\ee
Due to the properties of $\text{D}=3$ gamma matrices, these obey

\bea\label{u0u0=1}
u_a^{(0)}u^{a(0)}=1\; , \qquad u_a^{(0)}u^{a}=0\; , \qquad u_a^{(0)}\bar{u}{}^{a}=0 \; , \qquad  \\ \nonumber \\
\label{uu=0} u_au^{a}=0\; , \qquad \bar{u}_{a}\bar{u}{}^{a}=0 \; , \qquad u_a\bar{u}{}^{a}=-2 \;
\eea
and

\be\label{ug=ww}
u_a^{(0)}\gamma^a_{\alpha\beta}=2 w_{(\alpha}\bar{w}_{\beta )}\; , \qquad u_a\gamma^a_{\alpha\beta}=2 w_{\alpha}{w}_{\beta}\; , \qquad \bar{u}_a\gamma^a_{\alpha\beta}=2 \bar{w}_{\alpha}\bar{w}_{\beta}\; . \qquad
\ee
Eqs. \eqref{u0u0=1} and  \eqref{uu=0} imply that the real 3$\times$3  matrix
\be\label{uinO} u_a^{(b)}=\left(u_a^{(0)}, \frac 1 2 (u_a+\bar{u}_a), \frac 1 {2i} (u_a-\bar{u}_a)\right)\;\in \;  \text{O}(1,2) \; , \qquad \ee  belongs to O$(1,2)$ group. Furthermore, the way of constructing these vectors from bilinear of spinors implies that \footnote{To see this we can take the constant ``vacuum'' value of the spinors $v_\alpha^q=\delta_\alpha^q$ in \eqref{w=} and find that with our representation for 3D gamma-matrices (see p.5 of \cite{Bandos:2021vrq}) the vectors are  $u_a^{(0)}=\delta_a^0$, $u_a=-\delta_a^0+i\delta_a^1$, $\bar{u}_a=-\delta_a^0-i\delta_a^1$ and obey \eqref{euuu=-2i}.}
\be\label{euuu=-2i}
\epsilon^{abc}u_a^{(0)}u_b\bar{u}_c =-2i\qquad \Leftrightarrow \qquad u_{[b}\bar{u}_{c]} =-i\epsilon_{abc}u^{a(0)}\; , \qquad
\ee
which makes this matrix to be SO$(1,2)$ valued,
\be\label{uinSO} u_a^{(b)}=\left(u_a^{(0)}, \frac 1 2 (u_a+\bar{u}_a), \frac 1 {2i} (u_a-\bar{u}_a)\right)\;\in \;  \text{SO}(1,2) \; , \qquad \ee
which is tantamount to saying   that this set of vectors describes an oriented Lorentz  frame.

Then \eqref{ug=ww}, or equivalently
\eqref{u=wgw}, describes the relation of moving  frame with spinor moving frame, i.e. to show as the former is constructed from the latter.

The derivatives of the moving frame vectors are expressed through the Cartan forms by
\be\label{du0}
\text{d}u^{(0)}_a=if\bar{u}_a - i\bar{f}u_a \; , \qquad \text{D}u_a= \text{d}{u}_a+ 2ia {u}_a=2ifu^{(0)}_a  \; , \qquad \text{D}\bar{u}_a=\text{d}\bar{u}_a- 2ia \bar{u}_a=-2i\bar{f}u^{(0)}_a   \; , \qquad
\ee
so that \eqref{f=wdw} are equivalent to
\be\label{f=udu}
f=\frac i 2 {u}{}^a \text{d}u^{(0)}_a \; , \qquad \bar{f}=-\frac i 2 \bar{u}{}^a \text{d}u^{(0)}_a  \; , \qquad a=-\frac i 4 \bar{u}{}^a \text{d}u_a  =\frac i 4 u^a \text{d}\bar{u}_a  \; . \qquad
\ee

When functions of the spinor frame variables $f(\bar{w}, w)$ are considered, due to the constraint \eqref{bww=i}, their differential (exterior derivative) can be decomposed in the basis of Cartan forms \eqref{f=wdw},

\be
\text{d}f= \text{d}\bar{w}_\alpha \frac {\partial f}  {\partial \bar{w}} + \text{d}{w}_\alpha \frac {\partial f}  {\partial  w}= if \bar{{\bb D}} - i\bar{f} {\bb D} + ia {\bb D}^{(0)}\; ,
\ee

\noindent where at the second stage we have used Eqs.  \eqref{Dw=} and introduced the covariant derivatives
\bea\label{bbD:=}
{\bb D} = w_\alpha \frac {\partial} {\partial \bar{w}_\alpha}  \; , \qquad  \bar{{\bb D}}= \bar{w}_\alpha\frac {\partial} {\partial {w}_\alpha}  \; , \qquad {\bb D}^{(0)}= \bar{w}_\alpha \frac {\partial} {\partial \bar{w}_\alpha} - {w}_\alpha \frac {\partial} {\partial {w}_\alpha}  \; . \qquad \\
\nonumber
\eea
These have the  characteristic property that their action on the l.h.s. of the constraint \eqref{bww=i} vanish
\be {\bb D} ( \bar{w}{}^{\alpha} w_\alpha ) =0  \; , \qquad  \bar{{\bb D}} (\bar{w}{}^{\alpha} w_\alpha )=0 \; , \qquad {\bb D}^{(0)} (\bar{w}{}^{\alpha} w_\alpha )=0  \; \qquad \ee

 \noindent
 and obey  $\mathfrak{su}(1,1)\simeq \mathfrak{su}(2,{\bb R})$ algebra

 \be
 {}[{\bb D} \, , \, \bar{{\bb D}}]= {\bb D}^{(0)} \; , \qquad  [ {\bb D}^{(0)}\, , \, {\bb D}]=- 2{\bb D}   \; , \qquad  [{\bb D} \, , \, \bar{{\bb D}}]= 2 \bar{{\bb D}}\; . \qquad
 \ee

Using spinor moving frame and moving frame vectors we can split (the pull-back of) the supervielbein forms of $\text{D}=3$ ${\cal N}=2$ superspace
\be\label{Ea==}
E^a=  \delta^a_\mu \text{d}{x}{}^\mu - i \text{d}{\theta}\gamma^a \bar{\theta} + i {\theta}\gamma^a \text{d}{\bar{\theta}}\; , \qquad E^\alpha = \text{d}\theta^\alpha \; , \qquad \bar{E}{}^\alpha =\text{d}\bar{\theta}{}^\alpha \qquad
\ee
into the set of 3 bosonic and 4 fermionic Lorentz invariant one forms

\bea\label{E0=}
{\rm E}^{(0)}=E^au_a^{(0)}\; , \qquad {\rm E}=E^au_a\; , \qquad \bar{{\rm E}}=E{}^a\bar{u}_a\; , \qquad
\\ \nonumber  \\ \label{Ew=}
\text{E}^w= \text{d}\theta^\alpha w_\alpha \; , \qquad \text{E}^{\bar{w}}= \text{d}\theta^\alpha {\bar{w}}_\alpha \; , \qquad
\\ \nonumber  \\  \label{bEw=}
\bar{\text{E}}{}^w= \text{d}\bar{\theta}{}^\alpha w_\alpha \; , \qquad \bar{\text{E}}{}^{\bar{w}}= \text{d}\bar{\theta}{}^\alpha {\bar{w}}_\alpha\; . \qquad
\eea

\subsection{Lagrangian of the simplest 3D counterpart of the mD$0$ system}
The Lagrangian of the  3D counterpart of 10D mD$0$ action proposed in \cite{Bandos:2022uoz} contains two constants of dimension of mass, $m$ and $\mu$, as well as  arbitrary positive definite  function  \be\label{cM=cMcH}
{\cal M}={\cal M}({\cal H})\quad
\ee
of the following bosonic composite  of the matrix fields
\begin{eqnarray}
\label{cH=} && {\cal H}=    {\rm tr}\left( {\bb P} \bar{\bb P} +  [{\bb Z},  \bar{\bb Z}]^2 -
{i\over 2} {\bb Z}{\mathbf \Psi}{\mathbf \Psi} + {i\over 2} \bar{\bb Z} \bar{\mathbf \Psi}  \bar{\mathbf \Psi} \right) \;
.\qquad
\end{eqnarray}
This can be recognized as Hamiltonian of 3D SU($N$) SYM model dimensionally reduced to $\text{d}=1$. Introducing for convenience also a pair of  conjugate fermionic currents
\begin{equation}
{\nu} := \text{tr}(\mathbf \Psi \mathbb{P} + \bar{\mathbf \Psi}[\mathbb{Z}, \bar{\mathbb{Z}}])~, \qquad {\bar{\nu}} := \text{tr}(\bar{\mathbf \Psi} \bar{\mathbb{P}} + \mathbf \Psi[\mathbb{Z}, \bar{\mathbb{Z}}])~,\qquad
\end{equation}
which can be recognized as the SYM supercurrents, we can write the Lagrangian from \cite{Bandos:2022uoz} as follows

\begin{eqnarray}\label{eq:3DmD0_L}
L[m,\mu, \mathcal{M}({\cal H})]&:=& m {\rm E}_\tau^{0}+ m  (\dot{\theta}^\alpha \bar{\theta}_\alpha- \theta^\alpha \dot{\bar{\theta}}_\alpha ) + {1\over \mu^6}
 {\rm tr}\left(\bar{\bb P}{\rm D}_\tau {\bb Z} + {\bb P}{\rm D}_\tau \bar{\bb Z} - {i\over 8} {\rm D}_\tau{ {\mathbf \Psi}}\,  \bar{\mathbf \Psi} + {i\over 8} {\mathbf \Psi} {\rm D}_\tau \bar{\mathbf \Psi} \right) +\nonumber \\
&
+& {1\over \mu^6} {\rm E}_\tau^{0}\frac 2 {{\cal M}}  {{\cal H}}  + {1\over \mu^6}  \frac i {\sqrt{{\cal M}}}\, \dot{\theta}{}^\alpha\, w_\alpha\,
 \bar{\nu} +  {1\over \mu^6}   \frac i {\sqrt{{\cal M}}}\dot{\bar{\theta}}{}^{\alpha}\bar{w}_{\alpha}\, \nu  +{1\over \mu^6}
 \frac {\dot{{\cal M}}} {{\cal M}}
 {\rm tr}\left(\bar{\bb P} {\bb Z}+ {\bb P} \bar{\bb Z}\right) \, .  \qquad \; \\ \nonumber
\end{eqnarray}
Here ${\rm D}_\tau $ are  covariant derivatives which contain, besides the $\mathfrak{su}(N)$ valued connection ${\bb A}=\text{d}\tau {\bb A}_\tau$,  also composite U(1) connection given by the  Cartan form $a=\text{d}\tau a_\tau$ defined in  \eqref{f=wdw} (see also \eqref{f=udu}),
\bea\label{DtZ:=a}
\text{D}_\tau {\bb Z}= \dot{{\bb Z}}+2ia_\tau  {\bb Z} +[{\bb A}_\tau ,{\bb Z}] \; , \qquad \text{D}_\tau {\mathbf \Psi}= \dot{{\mathbf \Psi}}-i a_\tau  {\mathbf \Psi} +[{\bb A}_\tau ,{\mathbf \Psi}]\; , \qquad \nonumber  \\ \nonumber \\ \text{D}_\tau \bar{{\bb Z}}= \dot{\bar{{\bb Z}}}-2ia_\tau  \bar{{\bb Z}} +[{\bb A}_\tau ,\bar{{\bb Z}}] \; , \qquad \text{D}_\tau \bar{{\mathbf \Psi}}= \dot{ \bar{{\mathbf \Psi}}}+i a_\tau   \bar{{\mathbf \Psi}} +[{\bb A}_\tau , \bar{{\mathbf \Psi}}]\; .  \qquad \\ \nonumber
\eea

Two special members of this family of Lagrangians have this function \eqref{cM=cMcH} being  constant
\be\label{cM=m}
{\cal M} = m={\text{const}}\; ,
\ee
and of the following special form
\be\label{cM=m+}
{\cal M} = \frac m 2 +  \sqrt{\frac {m^2} {4}+\frac {\cal H} {\mu^6}}\; .
\ee
The latter was obtained in \cite{Bandos:2021vrq} by dimensional reduction of the 4D counterpart of the multiple M0-brane (mM0) action \cite{Bandos:2012jz} while the former is the simplest representative of the family which can be recognized as the 3D counterpart of the 10D action  discussed previously in \cite{Bandos:2018ntt}. This simplest  model with constant ${\cal M}=m$ and Lagrangian

\begin{eqnarray}\label{eq:3DmD0_L_mConstw}
L:=L[m,\mu, m] &=& m {\rm E}_\tau^{0}+ m  (\dot{\theta}^\alpha \bar{\theta}_\alpha- \theta^\alpha \dot{\bar{\theta}}_\alpha ) +  {1\over \mu^6}
 {\rm tr}\left(\bar{\bb P}{\rm D}_\tau {\bb Z} + {\bb P}{\rm D}_\tau \bar{\bb Z} - {i\over 8} {\rm D}_\tau{\mathbf \Psi}\,  \bar{\mathbf \Psi} + {i\over 8} {\mathbf \Psi} {\rm D}_\tau \bar{\mathbf \Psi}  \right) +\nonumber \\ \nonumber \\
&
+& {1\over \mu^6} {\rm E}_\tau^{0}\frac 2 {m}  {{\cal H}} + {1\over \mu^6}  \frac i {\sqrt{m}}\, \dot{\theta}{}^\alpha\, w_\alpha\,
 \bar{\nu} +  {1\over \mu^6}   \frac i {\sqrt{m}}\dot{\bar{\theta}}{}^{\alpha}\bar{w}_{\alpha}\,
\nu \;  \qquad \\ \nonumber
\end{eqnarray}
will be the subject of this paper. We will elaborate it in  Hamiltonian formalism and then quantize it, thus arriving at the equation of the
simplest 3D counterpart of the mD$0$-brane field theory.


\setcounter{equation}{0}

\section{Hamiltonian formalism of the simplest 3D prototype of mD$0$ brane model}

\subsection{Canonical momenta and Poisson brackets}

Generically, the canonical momenta $p_A$ are defined as derivatives of the Lagrangian $L(q^A,\dot{q}{}^A)$ with respect to velocities $\dot{q}^A=\frac{\text{d}}{\text{d}\tau}q^A$ in direction of the coordinates $q^A$,
\be\label{pA:=} p_A=  \dfrac{\partial {L}}{\partial \dot{q}^A}~ \ee
and  are canonically conjugate to this coordinate. This statement implies the diagonal structure of the Poisson brackets algebra
\bea\label{PB}
&& {}[p_A, q^B\}_{\text{PB}}= -(-)^{\varepsilon (A)\varepsilon (B) }[q^B, p_A\}_{\text{PB}}=-\delta_A{}^B \; , \qquad \qquad \nonumber \\ \nonumber \\ && {}[q^A, q^B\}_{\text{PB}}= -(-)^{\varepsilon (A)\varepsilon (B) } [q^B, q^A\}_{\text{PB}}= 0 \; , \qquad
[p_A, p_B\}_{\text{PB}}= -(-)^{\varepsilon (A)\varepsilon (B) }[p_A, p_B\}_{\text{PB}}=  0\; ,  \\ \nonumber
\eea
where $\varepsilon (A)$ is Grassmann (fermionic) parity of $p_A$ and $q^A$ which is equal to $0$ for bosons and to $1$ for fermions~\footnote{The Poisson brackets are defined essentially by relations \eqref{PB} and Leibniz rules $[fg,h\}_{\text{PB}}=f [g,h\}_{\text{PB}}+ (-)^{\varepsilon (f)\varepsilon (h) } [f,h\}_{\text{PB}}g$, where $f$, $g$ and $h$ are arbitrary functions of coordinates $q^A$  and momenta $p_A$. One can also deduce an explicit expression $[f, g\}_{\text{PB}}= (-)^{\varepsilon (A)\varepsilon (f) } \left( \frac {\partial f}  {\partial q^A } \, \frac {\partial g}  {\partial p_A } - (-)^{\varepsilon (A)}  \frac {\partial f}  {\partial p_A } \, \frac {\partial g}  {\partial q^A }\right)$.}.

Then the canonical Hamiltonian $H_0(p_A, q^A)$ is defined by Legendre transform of the Lagrangian
\footnote{We simplify the definition of the Legendre transform a bit,  as this should not cause problems in our context.  See e.g.  sec. 3 of \cite{Bandos:2021rqy} and references therein for a more exact definition and examples of dynamical system where its use is essential.}
\be\label{H0=dqp-L}
H_0 (p_A,q^A)=\dot{q}{}^Ap_A - L(q^A,\dot{q}{}^A)\;
\ee
and the evolution of the dynamical system is described by the Hamiltonian equations
\be\label{evol=H0}
\dot{f}(q,p)= [{f}(q,p)\, ,\, H_0]_{\text{PB}}\; \qquad \Leftrightarrow \qquad  \begin{cases} \dot{q}^A=  \dfrac {\partial H_0}  {\partial  p_A} \; , \cr \cr \dot{p}_A= - \dfrac {\partial H_0}  {\partial  q^A}\; .
 \end{cases}
\ee
In our case we firstly denote   the momenta conjugate to the bosonic and fermionic coordinate functions by
\begin{equation}
\begin{array}{ccccc}\label{Pa:=}
p_a = \dfrac{\partial {L}}{\partial \dot{x}^a}~,&~& \bar{\Pi}_\alpha = \dfrac{\partial {L}}{\partial \dot{\theta}^\alpha}~,&~& {\Pi}_\alpha = \dfrac{\partial {L}}{\partial \dot{\bar{\theta}}^\alpha}
\end{array}
\end{equation}
so that their nonvanishing Poisson brackets are
\begin{eqnarray}
&& [p_a, x^b]_{\text{PB}} =-[x^b, p_a ]_{\text{PB}} = -\delta_a^b~,\qquad  \nonumber \\ \nonumber \\
&&
\{\bar{\Pi}_\alpha, \theta^\beta\}_{\text{PB}} =\{\theta^\beta, \bar{\Pi}_\alpha\}_{\text{PB}} = -\delta_\alpha^\beta~, \qquad  \nonumber \\ \\ \nonumber  && \{{\Pi}_\alpha, \bar{\theta}^\beta\}_{\text{PB}} =\{ \bar{\theta}^\beta, {\Pi}_\alpha\}_{\text{PB}} = -\delta_\alpha^\beta~.
\end{eqnarray}

\subsection{Covariant momenta and Poisson/Dirac brackets in the spinor moving frame sector}

Similarly, we can introduce the canonical momenta conjugate to spinor frame variables
\be
\bar{P}_\alpha =\frac {\partial {\cal L}} {\partial {\dot{w}}^\alpha} \; , \qquad P_\alpha =\frac {\partial {\cal L}} {\partial {\dot{\bar{w}}}{}^\alpha}
\ee
which have  nonvanishing Poisson brackets
\be
{}[\bar{P}{}^\alpha, w_\beta]_{\text{PB}}={- \delta^\alpha{}_\beta{}}\; ,  \qquad {}[P^\alpha, \bar{w}_\beta]_{\text{PB}}={- \delta^\alpha{}_\beta{}}\; .   \qquad
\ee
If working with these canonical momenta, we have to consider the condition \eqref{bww=i} as a second class constraint (in terminology of Dirac \cite{Dirac:1963}, see below for more details and discussion), and find that this is conjugate to the constraint\footnote{Here and below $\approx$ implies a weak equality in the sense of \cite{Dirac:1963}, i.e. the relations between coordinates and momenta which can be used only after all Poisson brackets are calculated.}
\be\label{wP+cc=0}
w_\alpha\bar{P}{}^\alpha+ \bar{w}_\alpha{P}{}^\alpha \approx 0 \; ,
\ee
 which will also appear in our model. Then we could resolve these by passing from Poisson brackets to corresponding Dirac brackets (see \cite{Bandos:2018qqo} and refs. therein for more discussion) after which  both
\eqref{bww=i} and \eqref{wP+cc=0} can be considered as satisfied in the strong sense.

Actually this stage can be overcame by introducing from the very beginning the so-called covariant momenta

\be\label{frak-d=}
{\mathfrak d}= w_\alpha P^\alpha  \; , \qquad  \bar{{\mathfrak d}}= \bar{w}_\alpha\bar{P}{}^\alpha \; , \qquad {\mathfrak d}^{(0)} =\bar{w}_\alpha P^\alpha - w_\alpha \bar{P}{}^\alpha  \; , \qquad
\ee
which have the following Poisson brackets with  spinor frame variables

\bea\label{fdw=PB}
&  {} [{\mathfrak d}, w_\alpha ]_{\text{PB}}= 0\; ,  \qquad &[{\mathfrak d}, \bar{w}_\alpha ]_{\text{PB}}=  - {w}_\alpha  \; , \qquad \nonumber \\ \nonumber \\
 & {} [\bar{{\mathfrak d}}, w_\alpha ]_{\text{PB}}= - \bar{w}_\alpha \; ,  \qquad &[\bar{{\mathfrak d}}, \bar{w}_\alpha ]_{\text{PB}}=  0\; , \qquad \nonumber \\ \nonumber  \\ & {} [{\mathfrak d}^{(0)}, w_\alpha ]_{\text{PB}}=  w_\alpha \; ,   \qquad &[{\mathfrak d}^{(0)}, \bar{w}_\alpha ]_{\text{PB}}=  - \bar{w}_\alpha \; . \qquad \\ \nonumber
\eea
It is easy to check that these covariant momenta   have vanishing Poisson brackets with  the constraints \eqref{bww=i}.
Hence, if using only the covariant momenta, we can consider the constraint \eqref{bww=i} to be satisfied in the strong sense.

Notice that \eqref{fdw=PB} imply the following brackets of the covariant momenta with moving frame vectors

\bea
  {} [{\mathfrak d},u_a^{(0)} ]_{\text{PB}}=-u_a \; ,  \qquad &[{\mathfrak d}, u_a ]_{\text{PB}}= 0  \; , \qquad & [{\mathfrak d}, \bar{u}_a ]_{\text{PB}}=-2u_a^{(0)}   \; , \qquad   \nonumber \\ \nonumber  \\ {} [\bar{{\mathfrak d}},u_a^{(0)} ]_{\text{PB}}=-\bar{u}_a  \; ,  \qquad &[\bar{{\mathfrak d}}, u_a ]_{\text{PB}}=-2 u_a^{(0)}  \; , \qquad & [\bar{{\mathfrak d}}, \bar{u}_a ]_{\text{PB}}=0   \; , \qquad
    \nonumber \\ \nonumber  \\  {} [{\mathfrak d}^{(0)}, u_a^{(0)} ]_{\text{PB}}=0\; , \;  \qquad & [{\mathfrak d}^{(0)}, u_a ]_{\text{PB}}=2u_a\; ,  \qquad  & [{\mathfrak d}^{(0)}, \bar{u}_a ]_{\text{PB}}=-2 \bar{u}_a\; .  \qquad \\ \nonumber
\eea
On the Poisson brackets,  the  covariant momenta represent the algebra $\mathfrak{so}(1,2)\simeq \mathfrak{su}(1,1)$

\bea
 & {} [{\mathfrak d}^{(0)},{\mathfrak d} ]_{\text{PB}}=2{\mathfrak d} \; , \;  \qquad  [{\mathfrak d}^{(0)}, \bar{{\mathfrak d}} ]_{\text{PB}}=-2\bar{{\mathfrak d}}\; , & \qquad \nonumber \\ \nonumber \\
 & {} [{\mathfrak d},\bar{{\mathfrak d}} ]_{\text{PB}}={\mathfrak d}^{(0)} \; .  & \qquad \\ \nonumber
\eea

To find the expression for canonical Hamiltonian \eqref{H0=dqp-L} in terms of Lagrangian and covariant momenta we start from the standard
$\text{d}\tau H_0= \text{d}w_\alpha \bar{P}^\alpha + \text{d}\bar{w}_\alpha {P}^\alpha +...-L$, substitute there the expression for the derivatives of $w_\alpha $ and  $\bar{w}_\alpha$ in  \eqref{Dw=} valid when constraint \eqref{bww=i} is satisfied, and use \eqref{frak-d=} arriving at

\be\label{H0=frak-d}
\text{d}\tau H_0=  ia {\mathfrak d}^{(0)}+ if\bar{{\mathfrak d}}-i\bar{f} {\mathfrak d}+...-\text{d}\tau L\; \qquad \Leftrightarrow \qquad H_0=  ia_\tau  {\mathfrak d}^{(0)}+ if_\tau \bar{{\mathfrak d}}-i\bar{f}_\tau  {\mathfrak d}+...-\text{d}\tau L\; . \qquad
\ee
This relation makes transparent that the covariant momenta can be identified with derivatives of the Lagrangian with respect to Cartan forms
\be\label{fd=dL-df}
 {\mathfrak d}^{(0)}=- i \frac {\partial L}{\partial a_\tau} \; , \qquad {\mathfrak d}= i \frac {\partial L}{\partial \bar{f}_\tau} \; , \qquad   \bar{{\mathfrak d}}=-i \frac {\partial L}{\partial {f}_\tau}\; . \qquad
\ee

\subsection{Simplifying  Poisson/Dirac brackets in the sector of matrix fields }

The canonical momentum of the bosonic matrix fields are
\bea
 \dfrac{\partial \mathcal{L}}{\partial \dot{\mathbb{Z}}_j^i}\approx\dfrac{1}{\mu^6} \bar{\mathbb{P}}^j_i ~, \qquad \dfrac{\partial \mathcal{L}}{\partial \dot{\bar{\mathbb{P}}}{}^j_i} \approx 0 \; , \qquad \nonumber \\  \nonumber \\  \dfrac{\partial \mathcal{L}}{\partial \dot{\bar{\mathbb{Z}}}_j^i}\approx \dfrac{1}{\mu^6} \mathbb{P}^j_i ~,
\qquad \dfrac{\partial \mathcal{L}}{\partial \dot{{\mathbb{P}}}{}^j_i} \approx 0 \; . \qquad  \eea
These relations have the form of resolved pairs of second class constraints allowing to exclude the variables $\mathbb{P}^j_i$, $ \bar{\mathbb{P}}^j_i$ and their momenta from the consideration.

We however prefer an equivalent but more convenient way consisting in removing the momenta conjugate to $ \mathbb{P}^j_i$ and $ \bar{\mathbb{P}}^j_i$ and replacing the momenta conjugate to $\mathbb{Z}= \mathbb{Z}_i^j$ and to  $\bar{\mathbb{Z}}= \bar{\mathbb{Z}}_i^j$
by $ \bar{\mathbb{P}}^j_i$ and  $ \mathbb{P}^j_i$. Then the nonvanishing Poisson brackets for  new phase space  bosonic matrix variables (actually these are Dirac brackets) are  \footnote{Notice that the last contributions on the r.h.s.'s of Eqs. \eqref{bbPbbZ=PB}, $- \dfrac{1}{N} \delta^j_i \delta^l_k$, yield their  consistency with tracelessness of  matrices, \eqref{Z=bZ*}.}

\begin{equation}\label{bbPbbZ=PB}
\begin{array}{ccc}
[\bar{\mathbb{P}}^j_i, \mathbb{Z}_k{}^l ]_{\text{PB}} = -\mu^6 \left( \delta^l_i \delta_k^j - \dfrac{1}{N} \delta^j_i \delta^l_k\right)~, &~\qquad & [\mathbb{P}^j_i, \bar{\mathbb{Z}}_k{}^l]_{\text{PB}} = -\mu^6 \left( \delta^l_i \delta_k^j - \dfrac{1}{N} \delta^j_i \delta^l_k\right)~.
\end{array}
\end{equation}

Similar shortcut approach is usually applied when constructing the Hamiltonian mechanics of fermionic fields with canonical kinetic term in which case the conjugate of fermionic field is identified with its momentum. In our case the fermionic matrix fields have such a kinetic term.

The canonical momentum for the fermionic matrix field $\mathbf{\Psi} = \mathbf{\Psi}_i^j$ and its Hermitian conjugate $\bar{\mathbf{\Psi}} = \bar{\mathbf{\Psi}}_i^j$ are given by
\begin{equation} \label{PPsi=bPsi}
\begin{array}{ccc}
\Pi_{\bar{\mathbf \Psi}} = \dfrac{\partial \mathcal{L}}{\partial \dot{\bar{\mathbf \Psi}}} \approx  - \dfrac{i}{8 \mu^6} \mathbf \Psi~,&\qquad  &\bar{\Pi}_{\mathbf \Psi} = \dfrac{\partial \mathcal{L}}{\partial \dot{\mathbf \Psi}} \approx - \dfrac{i}{8 \mu^6} \bar{\mathbf \Psi}~.
\end{array}
\end{equation}

\noindent The canonical Poisson brackets  with nonvanishing r.h.s. are

\begin{equation}
\left\lbrace \mathbf\Pi_i{}^j, \bar{\mathbf \Psi}_k{}^l \right\rbrace_{\text{PB}} =\left\lbrace \mathbf \Psi_i{}^j, \bar{\mathbf\Pi}_k{}^l \right\rbrace_{\text{PB}} =  - ( \delta^l_i \delta_k^j - \dfrac{1}{N} \delta^j_i \delta^l_k )~
\end{equation}
where for shortness we write $\mathbf \Pi_i{}^j:=(\Pi_{\bar{\mathbf \Psi}})_i{}^j$ and $\bar{\mathbf\Pi}_i{}^j:=(\bar{\Pi}_{\bar{\mathbf \Psi}})_i{}^j$.  Using these we find that \eqref{PPsi=bPsi} are second class constraints. As they are not written in an explicitly resolved form, to treat them as equations which are satisfied in the strong sense, we have to replace Poisson brackets by Dirac brackets
\begin{equation}
\begin{split}
[...,...\rbrace_{\text{D}} = [...,... \rbrace_{\text{PB}} &- 4i\mu^6 [...,(\Pi_{\bar{\mathbf \Psi}} + \frac{i}{8\mu^6} \mathbf \Psi)_i^j \rbrace_{\text{PB}}[(\bar{\Pi}_{{\mathbf \Psi}} + \frac{i}{8\mu^6} \bar{\mathbf \Psi})_j^i,... \rbrace_{\text{PB}}-\\ \\
&- 4i\mu^6 [...,(\bar{\Pi}_{{\mathbf \Psi}}  \frac{i}{8\mu^6} \bar{\mathbf \Psi})_i^j \rbrace_{\text{PB}}[(\Pi_{\bar{\mathbf \Psi}} + \frac{i}{8\mu^6} \mathbf \Psi)_j^i,... \rbrace_{\text{PB}}~.
\end{split}
\end{equation}
Then the complex fermionic matrix field and its Hermitean conjugate become also canonically conjugate with this Dirac brackets,
\begin{equation}\label{PsibPsi=DB}
\left\lbrace \mathbf \Psi_i{}^j, \bar{\mathbf \Psi}_k{}^l \right\rbrace_{\text{D}} = -4i\mu^6 ( \delta^l_i \delta_k^j - \dfrac{1}{N} \delta^j_i \delta^l_k )~, \qquad \left\lbrace \mathbf \Psi_i{}^j, {\mathbf \Psi}_k{}^l \right\rbrace_{\text{D}} = 0 \; , \qquad \left\lbrace \bar{\mathbf \Psi}_i{}^j, \bar{\mathbf \Psi}_k{}^l \right\rbrace_{\text{D}} = 0\; . \qquad
\end{equation}
Below, for simplicity, and also taking into account the custom to impose relations like \eqref{PsibPsi=DB} as canonical, we will refer to these Dirac brackets as Poisson brackets and write \eqref{PsibPsi=DB} as
\begin{equation}\label{PsibPsi=PB}
\left\lbrace \mathbf \Psi_i{}^j, \bar{\mathbf \Psi}_k{}^l \right\rbrace_{\text{PB}} = -4i\mu^6 ( \delta^l_i \delta_k^j - \dfrac{1}{N} \delta^j_i \delta^l_k )~, \qquad \left\lbrace \mathbf \Psi_i{}^j, {\mathbf \Psi}_k{}^l \right\rbrace_{\text{PB}} = 0 \; , \qquad \left\lbrace \bar{\mathbf \Psi}_i{}^j, \bar{\mathbf \Psi}_k{}^l \right\rbrace_{\text{PB}} = 0\; . \qquad
\end{equation}
This is also convenient because we have already used implicitly the passage to Dirac brackets in the spinor moving frame sector of our model.

\subsection{Primary constraints}

With our Lagrangian \eqref{eq:3DmD0_L_mConstw},
computing the canonical momenta  \eqref{pA:=}  and covariant momenta \eqref{fd=dL-df}  of the  fields from center of mass sector  and of the gauge field,  we find the following set of  \textit{primary constraints} of the system
\begin{equation}
\Phi_a := p_a - mu_a^{(0)} - \dfrac{2}{\mu^6}\dfrac{\mathcal{H}}{m} u_a^{(0)} \approx 0~,
\label{eq:Phi_a}
\end{equation}

\begin{equation}
d_\alpha :=  \bar{\Pi}_\alpha + i p_a (\gamma^a \bar{\theta})_\alpha - m \bar{\theta}_\alpha - \dfrac{i}{\mu^6 \sqrt{ m}} w_\alpha\, \bar{\nu} \approx 0~, \qquad \bar{\nu}=\text{tr}(\bar{\mathbf \Psi} \bar{\mathbb{P}} + \mathbf \Psi [\mathbb{Z}, \bar{\mathbb{Z}}])\; ,
\label{eq:Phif}
\end{equation}

\begin{equation}
\bar{d}_\alpha := {\Pi}_\alpha + i p_a (\gamma^a \theta)_\alpha + m \theta_\alpha - \dfrac{i}{\mu^6 \sqrt{m}} \bar{w}_\alpha\, \nu \approx 0~,
\qquad \nu=\text{tr}(\mathbf \Psi \mathbb{P} + \bar{\mathbf \Psi} [\mathbb{Z}, \bar{\mathbb{Z}}]) \; , \label{eq:barPhif}
\end{equation}

\bea\label{fd=0}
&& {\mathfrak d}\approx 0 \; , \qquad  \\ \nonumber \\  \label{bfd=0} && \bar{{\mathfrak d}} \approx 0 \; , \qquad \\ \nonumber \\  \label{U=0}&& U:= {\mathfrak d}^{(0)}-  \frac 2{\mu^6} \, {\cal B}\approx 0 \; , \qquad  {\cal B}:= {\rm tr}\left(\bar{\mathbb{P}}\mathbb{Z}-\mathbb{P}\bar{\mathbb{Z}}+\frac i 8 \mathbf{\Psi}\bar{\mathbf{\Psi}}\right) \;  \qquad
\eea
and
\begin{equation}
\mathbb{P}_{\mathbb{A}}:= \frac {\partial L}{\partial \dot{\mathbb{A}}_\tau} \approx 0~.
\label{eq:PA}
\end{equation}

Although the latter relation is the only primary constraints imposed just on the matrix  variables, the matrix fields also contribute to the constraints that appear from the definition of momenta of the center of mass variables through ${\cal H}$ of \eqref{cH=} in \eqref{eq:Phi_a}, through

\begin{equation}
 \bar{\nu}=\text{tr}(\bar{\mathbf \Psi} \bar{\mathbb{P}} + \mathbf \Psi [\mathbb{Z}, \bar{\mathbb{Z}}])\; \qquad {\text{and}} \qquad \nu=\text{tr}(\mathbf \Psi \mathbb{P} + \bar{\mathbf \Psi} [\mathbb{Z}, \bar{\mathbb{Z}}]) \;
\label{nu=}
\end{equation}
in \eqref{eq:Phif}  and \eqref{eq:barPhif}, as well as through
\be\label{cB:=} {\cal B}:= {\rm tr}\left(\bar{\mathbb{P}}\mathbb{Z}-\mathbb{P}\bar{\mathbb{Z}}+\frac i 8 \mathbf{\Psi}\bar{\mathbf{\Psi}}\right) \;  \qquad
\ee
in \eqref{U=0}.

\subsection{Canonical Hamiltonian and secondary constraints}

Canonical Hamiltonian of our system is defined by the Legendre transform of the Lagrangian, which we write already in terms of SO$(1,2)/\text{SO}(2)$ Cartan forms and covariant momenta (see \eqref{H0=frak-d}) as

\begin{equation}\label{H0:=}
\begin{split}
\text{d}\tau H_0 &= \text{d}x^{a}p_{a}  + \text{d}\theta^\alpha  \bar{\Pi}_\alpha +\text{d} \bar{\theta}^\alpha {\Pi}_\alpha  + ia \tilde{\mathfrak{d}}^{(0)} + i f \bar{{\mathfrak{d}}} - i \bar{f} {\mathfrak{d}} +\\
&~\\
&+ \frac{1}{\mu^6} \text{tr}(\text{d}\mathbb{Z}\bar{\mathbb{P}}) + \frac{1}{\mu^6} \text{tr}(\text{d}\bar{\mathbb{Z}}\mathbb{P}) -	\frac{i}{8 \mu^6} \text{tr}(\text{d}\mathbf{\Psi} \bar{\mathbf{\Psi}}) -\frac{i}{8 \mu^6} \text{tr}(\text{d}\bar{\mathbf{\Psi}} \mathbf{\Psi}) + \text{tr}(\text{d}\mathbb{A} \mathbb{P}_{\mathbb{A}}) - \mathcal{L}_{\text{mD}0} ~. \\ {}
\end{split}
\end{equation}

\noindent For our Lagrangian  \eqref{eq:3DmD0_L_mConstw},  after  extracting the primary constraints,  \eqref{H0:=} reduces to
\begin{eqnarray}\label{H0=trAG}
H_0 = E_\tau^a\Phi_a+\dot{\theta}{}^\alpha d_\alpha +\dot{\bar{\theta}}{}^\alpha \bar{d}_\alpha
+ ia_\tau U + i f_\tau \bar{\tilde{\mathfrak{d}}} - i \bar{f}_\tau \tilde{\mathfrak{d}} +
 \text{tr}(\text{d}\mathbb{A} \mathbb{P}_{\mathbb{A}}) + {\rm tr} ({\bb A}_\tau {\bb G})  \approx  {\rm tr}  ({\bb A}_\tau {\bb G})~,
\end{eqnarray}
where $\mathbb{G}$ is $N\times N$ anti-Hermitian traceless matrix
\begin{equation}\label{bbG:=}
\mathbb{G}:= \dfrac{1}{\mu^6} \left( [\bar{\mathbb{Z}}, \mathbb{P}] + [\mathbb{Z}, \bar{\mathbb{P}}] - \frac{i}{4} \{\mathbf \Psi, \bar{\mathbf \Psi} \} \right) \; .
\end{equation}

The last expression valid in the weak sense can be used to check whether the preservation of the primary constraints in the evolution, as defined by Hamiltonian equations \eqref{evol=H0}, produce secondary constraints.
In our case this happens: the requirement of preservation of the primary constraint \eqref{eq:PA},  $\dot{\mathbb{P}}_{\mathbb{A}} = [\mathbb{P}_{\mathbb{A}}, H_0] \approx 0$, leads to the \textit{secondary constraint}

\begin{equation}
\mathbb{G}= \dfrac{1}{\mu^6} \left( [\bar{\mathbb{Z}}, \mathbb{P}] + [\mathbb{Z}, \bar{\mathbb{P}}] - \frac{i}{4} \{\mathbf \Psi, \bar{\mathbf \Psi} \} \right) \approx 0\; .
\label{eq:Gauss}
\end{equation}

This reveals the role of the 1d gauge field ${\mathbb A}_\tau$ as a Lagrange multiplier for the constraint \eqref{eq:Gauss}  which we will call the {\it Gauss law} keeping in mind its counterpart in higher dimensional supersymmetric gauge theories. Furthermore, as it can be shown that the constraint \eqref{eq:PA} is of the first class and generates a gauge symmetry consisting in an arbitrary shift of ${\mathbb A}_\tau$, we will allow ourself
to streamline the presentation by just skipping the conjugate variables  ${\mathbb A}_\tau$ and
${\mathbb P}_{\mathbb A}$ from the consideration below \footnote{More precisely, one can understand this reduction of the phase space as fixing the gauge
${\mathbb A}_\tau =0$ by the symmetry generated by the constraint ${\mathbb P}_{\mathbb A}$ and then  treating the constraint  \eqref{eq:PA} as strong equality. }.

Then, the Gauss law \eqref{eq:Gauss} results in a vanishing of the canonical Hamiltonian in the weak sense
  \begin{equation}
  \label{H0=0}
  H_0\approx 0\; .
\end{equation}

\subsection{First class constraints and gauge symmetries}

Eq. \eqref{H0=0} implies that the total Hamiltonian, as introduced by Dirac in \cite{Dirac:1963}, for our system  is equal to a linear combination of the constraints,
\begin{eqnarray}\label{H=LMs}
H &=& b^a \Phi_a + {\kappa}^\alpha d_\alpha + \bar{ {\kappa}}^\alpha \bar{d}_\alpha + ik\bar{{\mathfrak d}} - i\bar{k} {\mathfrak d} + ik^{(0)} \left({\mathfrak d}{}^{(0)} -   \frac 2{\mu^6} \, {\cal B}\right) + \text{tr}(\mathbb{Y}\mathbb{G})~.
\end{eqnarray}
The coefficients in this expression, the functions of proper time called Lagrange multipliers \footnote{Notice that a reincarnation of the  pure gauge 1d gauge field $ {\bb A}_\tau $, which we removed from consideration using the gauge symmetry generated by \eqref{eq:PA}, in the Lagrange multiplier ${\bb Y}$ for the Gauss constraints \eqref{eq:Gauss}.},
are restricted by the conditions of  the preservation of all the constraints under the evolution
\cite{Dirac:1963},
\be\label{evol=H}
\frac{\text{d}}{\text{d}\tau} (\text{constraints})= [\text{constraints}\,,\, H]_{\text{PB}}\approx 0 \; \qquad
\ee
 ({\it cf.} \eqref{evol=H0}) \footnote{Generically this procedure might result in appearance of further secondary constraints, but this does not happen in our case, where it produces a set of equations on the Lagrange multipliers.}.  The solution of Eqs. \eqref{evol=H}   expresses the original Lagrange multipliers in terms of fewer fields serving as parameters  for the so-called first class constraints, which generate gauge symmetries of the dynamical system under consideration on the Poisson brackets.

To specify the  system of equations, it is useful first to obtain the algebra on the constraints on the Poisson brackets. Let us begin from the Gauss law constraint  \eqref{bbG:=} which is a(n anti-Hermitian traceless) matrix, ${\bb G}= ({\bb G}_i{}^j)$, so that the brackets between different matrix elements of this matrix can be, and some  are indeed, nonvanishing. Actually these provide a representation of $\mathfrak{su}(N)$ algebra.
To write it in a compact and comprehensive form, it is convenient  to introduce
``reference'' traceless matrices $\mathbb{Y}$ and $\mathbb{Y}^\prime$ and write the brackets of
the traces of  ${\bb G}$ with these matrices. In such a way we find the  $\mathfrak{su}(N)$ algebra in the form
\begin{equation} \label{GG=G}
[\text{tr}(\mathbb{Y}\mathbb{G}), \text{tr}(\mathbb{Y}^\prime \mathbb{G})]_{\text{PB}} = \text{tr}\left([\mathbb{Y}, \mathbb{Y}^\prime ]\mathbb{G}\right)~.
\end{equation}

Furthermore, ${\bb G}$ has vanishing Poisson brackets with all other constraint, which just reflects their
SU$(N)$ invariance (see Appendix \ref{App=PB} for technical details). Indeed ${\bb G}$  is thus the first class constraint which generates the SU$(N)$ gauge symmetry of the model.

The algebra of the other constraints, given by Eqs. \eqref{eq:Phi_a}-\eqref{eq:PA}, is  described by the following nonvanishing Poisson brackets (see Appendix \ref{App=PB}):

\bea
[\mathfrak{d},\mathfrak{d}^{(0)} - \frac{2}{\mu^6} \mathcal{B}]_{\text{PB}}= -2 \mathfrak{d}~, \qquad [\mathfrak{d},\Phi_a]{}_{\text{PB}}= \left(m + \frac{2}{\mu^6}\frac{\mathcal{H}}{m} \right) u_a~, \qquad
\\ \nonumber \\ {}[\mathfrak{d},\bar{d}_\alpha]_{\text{PB}} = \frac{i}{\mu^6 \sqrt{m}} w_\alpha \nu~, \qquad
\\ \nonumber \\ {}  [\mathfrak{d},\bar{\mathfrak{d}}]_{\text{PB}} = \mathfrak{d}^{(0)}~, \qquad  \\ \nonumber \\
{}[\bar{\mathfrak{d}},\mathfrak{d}^{(0)} - \frac{2}{\mu^6} \mathcal{B}]_{\text{PB}} = 2 \bar{\mathfrak{d}}~, \qquad [\bar{\mathfrak{d}},\Phi_a]_{\text{PB}} = \left(m + \frac{2}{\mu^6}\frac{\mathcal{H}}{m} \right) \bar{u}_a~,  \qquad \\ \nonumber \\ {} [\bar{\mathfrak{d}},d_\alpha]_{\text{PB}} = \frac{i}{\mu^6 \sqrt{m}} \bar{w}_\alpha \bar{\nu}~, \qquad   \\ \nonumber \\
{}[\Phi_a ,d_\alpha]_{\rm{PB}}= \; \frac {2i}{m\sqrt{m}}\, u^{(0)}_a w_\alpha \,  {\rm tr}({\mathbf \Psi}{\bb G}) \; , \qquad  \\ \nonumber \\ {}[\Phi_a  ,\bar{d}_\alpha]_{\rm{PB}}= -\frac {2i}{m\sqrt{m}}\, u^{(0)}_a \bar{w}_\alpha\, {\rm tr}(\bar{{\mathbf \Psi}}{\bb G}) \, , \qquad
\eea
and
\bea
{}\{d_\alpha ,d_\beta \}_{\rm{PB}}&=& -\frac {8i}{m}\, w_\alpha w_\beta\,  {\rm tr}(\bar{{\bb Z}}{\bb G}) \; , \qquad  \\ \nonumber \\ {}\{\bar{d}_\alpha ,\bar{d}_\beta \}_{\rm{PB}}&=& \; \; \frac {8i}{m}\, \bar{w}_\alpha\bar{w}_\beta\, {\rm tr}({\bb Z}{\bb G}) \, , \qquad
\\
\nonumber
\\
{}\{d_\alpha ,\bar{d}_\beta \}_{\rm{PB}}&=&-2iP_a\gamma^a_{\alpha\beta}+2m\epsilon_{\alpha\beta}+\frac {4i}{\mu^6}\, w_\alpha \bar{w}_\beta \, \frac {{\cal H}}{m} \qquad \nonumber  \\  \nonumber \\
{}&=&-2i\Phi_a\gamma^a_{\alpha\beta} -4i  w_\beta\bar{w}_\alpha  \,  \left(m +\frac {1}{\mu^6}\,\frac {{\cal H}}{m}\right)\; .
\\ \nonumber \eea

Using this algebra,  we find that the preservation of the constraints in the evolution, described by Eq. \eqref{evol=H} with Hamiltonian \eqref{H=LMs},  requires the Lagrange multipliers to obey a system of equations which is solved by
\bea k=0\; , \qquad \bar{k}=0\; , \qquad && \\ \nonumber \\
b^au_a=0\; , \qquad b^a\bar{u}_a=0\;  \qquad && \Rightarrow \qquad b^a= b u^{(0)a}\; , \\ \nonumber \\
\bar{\kappa}^\beta w_\beta=0 \qquad && \Rightarrow \qquad  \bar{\kappa}^\beta=\bar{\kappa}w^\beta \; , \qquad \\ \nonumber \\
\kappa^\beta \bar{w}_\beta=0 \qquad && \Rightarrow \qquad \kappa^\beta =\kappa  \bar{w}^\beta\; ,  \nonumber \\
\eea
so that the total Hamiltonian reads
\begin{eqnarray}\label{H=1st}
H &=& b \, u^{a(0)} \Phi_a + {\kappa}  \, \bar{w}^\alpha d_\alpha + \bar{ {\kappa}}\, w^\alpha \bar{d}_\alpha+ ik^{(0)} \left({\mathfrak d}{}^{(0)} -   \frac 2{\mu^6} \, {\cal B}\right) + \text{tr}(\mathbb{Y}\mathbb{G})~.
\end{eqnarray}
The arbitrary Lagrange multipliers $b$, $\kappa$, $\bar{ {\kappa}}$, $k^{(0)}$ and $\mathbb{Y}_j{}^i$ obeying  $\mathbb{Y}_i{}^i=0$ reflect the  gauge symmetries of our model: reparametrization, local worldline supersymmetry ($\kappa$-symmetry), U$(1)$ and SU$(N)$ symmetries. In \eqref{H=1st} they are multiplied by the  constraints generating these symmetries,
\begin{equation}\label{Phi0=0}
\Phi^{(0)} := p_a u^{a(0)} - \left( m + \frac {2}{\mu^6}\,\frac {{\cal H}}{m}\right)\approx 0~,
\end{equation}
\begin{equation}\label{bwd=0}
\bar{w}^\alpha d_\alpha := \bar{w}^\alpha (  \bar{\Pi}_\alpha + i P_a (\gamma^a \bar{\theta})_\alpha -m \bar{\theta}_\alpha) + \frac{1}{\mu^6}  \frac{1}{\sqrt{m}}\bar{{\nu}}  \approx 0~,
\end{equation}
\begin{equation}\label{wbd=0}
w^\alpha \bar{d}_\alpha := w^\alpha ({\Pi}_\alpha + i P_a (\gamma^a \theta)_\alpha + m\theta_\alpha ) - \frac{1}{\mu^6}  \frac{1}{\sqrt{m}} {\nu} \approx 0~,
\end{equation}
\begin{equation}\label{U0=0}
U^{(0)}:= \mathfrak{d}^{(0)} -  \frac 2 {\mu^6} {\cal B} \approx 0~, \qquad
\end{equation}
with ${\cal B}$ defined in  \eqref{cB:=}, and \eqref{eq:Gauss}.

\subsection{Second class constraints }

The remaining constraints are the second class ones. They can be collected in conjugate  pairs

\begin{eqnarray}
\mathfrak{d} \approx 0~, & \qquad  & \bar{\mathfrak{d}} \approx 0~, \label{eqs:frak_d} \\ \nonumber \\
\Phi := u^a \Phi_a = u^a p_a \approx 0~, & \qquad  &\bar{\Phi} := \bar{u}^a \Phi_a = \bar{u}^a p_a \approx 0~, \label{eqs:Phi} \\ \nonumber \\
 w^{\alpha} d_{\alpha} = w^\alpha ( \bar{\Pi}_{\alpha} + i p_a(\gamma^a \bar{\theta})_\alpha -m \bar{\theta}_\alpha) \approx 0~, &  \qquad & \bar{w}^{\alpha} \bar{d}_{\alpha} = \bar{w}^\alpha({\Pi}_\alpha + ip_a(\gamma^a \theta )_\alpha + m\theta_\alpha) \approx 0~, \label{eqs:wd} \\ \nonumber
\end{eqnarray}
as can be seen from their Poisson brackets, particularly

\begin{eqnarray}
  \label{fdbP=}  [\mathfrak{d}, \bar{\Phi}]_{\text{PB}} &=& - 2 \Phi^{(0)} - 2 \left( m + \frac{2}{\mu^6}\frac{\mathcal{H}}{m} \right) \approx - 2 \left( m + \frac{2}{\mu^6}\frac{\mathcal{H}}{m} \right)~, \\ \nonumber \\  \label{bfdP=}
 [\bar{\mathfrak{d}}, \Phi]_{\text{PB}} & = & - 2 \Phi^{(0)} - 2 \left( m + \frac{2}{\mu^6}\frac{\mathcal{H}}{m} \right) \approx - 2 \left( m + \frac{2}{\mu^6}\frac{\mathcal{H}}{m} \right)~,\\  \nonumber \\   \label{wdbwbd=}
\{w^\alpha d_\alpha, \bar{w}^\alpha \bar{d}_\alpha\}_{\text{PB}} & = & - 2i \Phi^{(0)} - 4i \left(m + \frac{1}{\mu^6} \frac{\mathcal{H}}{m} \right) \approx - 4i \left(m + \frac{1}{\mu^6} \frac{\mathcal{H}}{m} \right)~. \\ \nonumber
\end{eqnarray}

\subsection{The algebra of first and second class constraints}
Notice  that there are other nonvanishing brackets of the second class constraints:
\begin{eqnarray} \label{fdfbd=}
{}[ \mathfrak{d}, \bar{\mathfrak{d}} ]_{\text{PB}} = \mathfrak{d}^{(0)} & = & \left( \mathfrak{d}^{(0)} - \frac{2}{\mu^6} \mathcal{B} \right) +\frac{2}{\mu^6} \mathcal{B}  \approx \frac{2}{\mu^6}\mathcal{B}~, \qquad \\ \nonumber \\ \label{fdbwbd=}
{}[\mathfrak{d}, \bar{w}^\alpha \bar{d}_\alpha]_{\text{PB}} &=& -w^\alpha \bar{d}_\alpha - \frac{1}{\mu^6 \sqrt{m}}\nu \approx - \frac{1}{\mu^6 \sqrt{m}}\nu~,\qquad \\  \nonumber \\   \label{bfdwd=}
 [\bar{\mathfrak{d}},  w^\alpha d_\alpha]_{\text{PB}} & =&-\bar{w}^\alpha {d}_\alpha+ \frac{1}{\mu^6 \sqrt{m}}\bar{\nu} \approx \;  \frac{1}{\mu^6 \sqrt{m}}\bar{\nu}~.\qquad~
\end{eqnarray}
The algebra of the I and II class constraints is resumed  in the  Table~\ref{table:mD0-centr} where we used the notation
\be\label{M(H)}
M({\cal H}):= m+\frac {1}{\mu^6}\frac {{\cal H}}{m} \; .
\ee
\begin{table}[h!]
\resizebox{\textwidth}{!}{\begin{tabular}{c||cccc||cccccc}
 $[...,... \}_{\text{PB}}$
&   ${U}{}^{(0)}$ & $\Phi^{(0)}$& $\bar{w}d$ & $w\bar{d}$ & $\mathfrak{d}$ & $\bar{\mathfrak{d}}$ & $\Phi$ &
$\bar{\Phi}$ &  $wd$ & $\bar{w}\bar{d}$ \\
 \hline \hline
 \\
$U^{(0)}$& 0 & 0 & $-\bar{w}d$ & $w\bar{d}$ &  2$\mathfrak{d}$ & -2$\bar{\mathfrak{d}}$ & 2$\Phi$ &
$-2\bar{\Phi}$ &  $wd$ & $-\bar{w}\bar{d}$
\\
$\Phi^{(0)}$ & 0 & 0 & $-\frac {2{\rm tr} (\Psi {\mathbb G})}{m\sqrt{m}}$ & $-\frac {2{\rm tr} (\bar{\Psi} {\mathbb G})}{m\sqrt{m}}$ & $\Phi$ & $\bar{\Phi}$ & 0 & 0 & 0 & 0 \\
$\bar{w}d$  & $\bar{w}d$  & $\frac {2{\rm tr} (\Psi {\mathbb G})}{m\sqrt{m}}$ & $\frac {8i{\rm tr} (\bar{Z}{\mathbb G})}{m}$ &  $-2i\Phi^{(0)}$ & $wd$ & 0 & 0 & 0 & 0 & $-2i\bar{\Phi}$ \\
$w\bar{d}$  & $-w\bar{d}$  & $\frac {2{\rm tr} (\bar{\Psi} {\mathbb G})}{m\sqrt{m}}$ &  $-2i\Phi^{(0)}$  & $-\frac {8i{\rm tr} ({Z}{\mathbb G})}{m}$  & 0 & $\bar{w}\bar{d}$  & 0 & 0 & $-2i\Phi$ & 0 \\
 \hline \hline
 \\
${\mathfrak{d}}$ &  $-2{\mathfrak{d}}$ & $-\Phi $ & $-wd$ & 0 & 0 & \fbox{$\begin{matrix}U^{(0)}+\cr +\frac {2{\cal B}}{\mu^6}\end{matrix}$ }& 0 & \fbox{$\begin{matrix}-2\Phi^{(0)}-\cr -2M(2{\cal H})\end{matrix}$} & 0 & \fbox{$\begin{matrix}-w\bar{d}-\cr -\frac {\nu}{\mu^6\sqrt{m}}\end{matrix}$}\\
$\bar{\mathfrak{d}}$  &  $2\bar{\mathfrak{d}}$ & $-\bar{\Phi}$ & 0 &$-\bar{w}\bar{d}$ & \fbox{$\begin{matrix}-U^{(0)}-\cr -\frac {2{\cal B}}{\mu^6}\end{matrix}$ }& 0 & \fbox{$\begin{matrix}-2\Phi^{(0)}-\cr -2M(2{\cal H})\end{matrix}$} & 0 & \fbox{$\begin{matrix}-\bar{w}d+\cr +\frac {\bar{\nu}}{\mu^6\sqrt{m}}\end{matrix}$}  & 0 \\
 $\Phi$ & $-2\Phi$&  0 & 0 & 0 & 0 & \fbox{$\begin{matrix}2\Phi^{(0)}+\cr +2M(2{\cal H})\end{matrix}$}  & 0 & 0  & 0 & 0 \\
 $\bar{\Phi}$ &  $2\bar{\Phi}$  & 0 &0 & 0  & \fbox{$\begin{matrix}2\Phi^{(0)}+\cr +2M(2{\cal H})\end{matrix}$} & 0 & 0 & 0 & 0 & 0 \\
  $wd$ & -$wd$ & 0 & 0 & $-2i\Phi $ & 0 & \fbox{$\begin{matrix}\bar{w}d-\cr -\frac {\bar{\nu}}{\mu^6\sqrt{m}}\end{matrix}$}  & 0 & 0 & 0 & \fbox{$\begin{matrix}-2i\Phi^{(0)}-\cr -4iM({\cal H})\end{matrix}$}\\
   $ \bar{w}\bar{d}$ &  $ \bar{w}\bar{d}$  & 0&  $-2i\bar{\Phi} $ & 0 &  \fbox{$\begin{matrix}w\bar{d}+\cr +\frac {{\nu}}{\mu^6\sqrt{m}}\end{matrix}$} & 0  & 0 & 0 & \fbox{$\begin{matrix}-2i\Phi^{(0)}-\cr -4iM({\cal H})\end{matrix}$} & 0\\
 \hline \hline
\end{tabular}}
\caption{Algebra of the first and second class constraints on the Poisson brackets  of the simplest 3D mD0 system; $M({\cal H}):= m+\frac {1}{\mu^6}\frac {{\cal H}}{m}$ is defined in \eqref{M(H)}.
}
\label{table:mD0-centr}
\end{table}

\noindent We did not include in this Table~\ref{table:mD0-centr} the first class constraints ${\bb G}_i{}^j$ \eqref{bbG:=}, as these nonvanishing brackets with themselves, \eqref{GG=G}.

(This should not be confused with ${\cal M}({\cal H})$ from \cite{Bandos:2021vrq} which was discussed in the first sections of this paper). The appearance of the positively definite additive contributions $M({\cal H})$ or $M(2{\cal H})$ in the right hand side of the Poisson brackets clearly indicates  the pairs of fermionic and bosonic second class constraints.

The above algebra of constraints is complicated enough to make difficult to proceed with the BRST quantization along the scheme of ``abelization'' of the second class constraints   \cite{Egorian:1993sc,Batalin:1989dm}.
The technical problem comes, in particular, from the non-canonical form of the bosonic second class constraints algebra in which, besides \eqref{fdbP=},  \eqref{bfdP=} and \eqref{wdbwbd=}, also the brackets \eqref{fdfbd=}, \eqref{fdbwbd=} and  \eqref{bfdwd=}  do not vanish in the weak sense.

Furthermore, when  trying to apply the Gupta-Bleuler quantization scheme in its canonical forms, with this algebra of constraints we arrived (in supercoordinate representation) at a system of equations which has only trivial solution: we discuss this in the simplest case of vanishing all the matrix and fermionic fields in Appendix \ref{GB-failed}. This is the only example known to the authors where the Gupta-Bleuler simplified method of quantization of dynamical systems fails.

Hence the only option which remains, which is consistent and allows to overcome difficulties, is to use the Dirac brackets procedure in the bosonic sector of our dynamical system.
Furthermore, this procedure can be significantly simplified, actually reduced to the  explicit resolution of the bosonic second class constraints, after we  pass to the so-called analytical basis in the center of mass sector of the configuration superspace of our system. We will make this change of coordinate basis in the next section.

\setcounter{equation}{0}

\section{Hamiltonian mechanics in the analytical basis }

\subsection{Constraints in the analytical basis}

Let us begin by noticing that the center of mass sector of the configurational space of our dynamical system can be identified with the so-called Lorentz harmonic superspace, an extended enlarged superspace with coordinates  \footnote{\label{HarmFoot} Here ``extended'' refers to ${\cal N}=2$ supersymmetry and ``enlarged'' reflects the presence of additional bosonic spinor (spinor moving frame or Lorentz harmonic) coordinates.
The terminology goes back to
 \cite{Bandos:1990ji} which developed the line of seminal papers on Harmonic superspaces  \cite{Galperin:1984av,Galperin:1984bu,Sokatchev:1985tc,Sokatchev:1987nk}.
 See Appendix \ref{App=D0} for more discussion. }

\be\label{c-basis}
\Sigma^{(3+3|4)} = \{ (x^a,\theta^\alpha , \bar{\theta}^\alpha , w_\alpha , \bar{w}_\alpha )\} =: \{ (Z^M , w_\alpha , \bar{w}_\alpha )\} \; , \qquad \bar{w}^\alpha{w}_\alpha =i\; .
\ee
The complete configuration superspace includes also $2(N^2-1)$ bosonic and $(N^2-1)$  fermionic matrix coordinates, and its basis can be chosen as

 \be\label{c-basisALL}
\Sigma^{(4+2N^2|3+N^2)}  =  \{ (x^a,\theta^\alpha , \bar{\theta}^\alpha , w_\alpha , \bar{w}_\alpha
; {\bb Z}, \bar{{\bb Z}}, \mathbf{\Psi}  )\} =: \{ (Z^M , w_\alpha , \bar{w}_\alpha; {\bb Z}, \bar{{\bb Z}}, \mathbf{\Psi}  )\} \; . \qquad \; .
\ee
This implies that $ \bar{\mathbf{\Psi}}$ is considered to be the momentum conjugate to ${\mathbf{\Psi}}$.

Clearly, there is a possibility to choose another  coordinate basis. In our case it is convenient {\it to make a change of coordinates in the center of mass sector} \eqref{c-basis} of the whole configurational superspace choosing  in it  the  analytical coordinate  basis

\be
 \Sigma^{(3+3|4)} = \{ ({\rm x}{}^{(0)},{\rm x}_A,\bar{{\rm x}}_A\; ,  \theta^w , {\theta}^{\bar{w}} , \bar{\theta}^{{w}} ,\bar{\theta}^{\bar{w}}  ; w_\alpha , \bar{w}_\alpha  )\} =:\{ ({\rm Z}{}^{(M)}_{An}, \; w_\alpha , \bar{w}_\alpha\; , {\bb Z}, \bar{{\bb Z}}, \mathbf{\Psi},  \bar{\mathbf{\Psi}}   )\} \; \qquad
\ee
with
\bea\label{x0=}
& {\rm x}{}^{(0)}:=x^au_a^{(0)}\; , \qquad & \\ \nonumber \\ \label{xA=}
& {\rm x}_A : = {\rm x}\, -2i \theta^w \bar{\theta}^{{w}}\;  , & \qquad  {\rm x}:= x^au_a\; , \qquad \\ \nonumber \\ \label{bxA=} & \bar{{\rm x}}_A:= \bar{{\rm x}} \, +2i{\theta}^{\bar{w}} \bar{\theta}^{\bar{w}} \; ,  & \qquad \bar{{\rm x}}:= x^a \bar{u}_a\; , \\
\nonumber
\\ \label{thw=}
& \theta^w := \theta^\alpha w_\alpha~, \qquad & \theta^{\bar{w}}:= \theta^\alpha \bar{w}_\alpha~, \qquad \\
\nonumber \\ \label{bthw=}
& \bar{\theta}^w := \bar{\theta}^\alpha w_\alpha~, \qquad & \bar{\theta}^{\bar{w}}:= \bar{\theta}^\alpha \bar{w}_\alpha~ \qquad
\eea
  (see Appendix \ref{App=D0}, particularly Eqs. \eqref{xA=D0}-\eqref{bthw=D0}, {\it cf.}  \cite{Sokatchev:1985tc,Sokatchev:1987nk,Bandos:1990ji}).

In this analytical basis our Lagrangian form, ${\cal L}=\text{d}\tau L$ with $L$ from \eqref{eq:3DmD0_L_mConstw}, reads

\begin{eqnarray}\label{cL=mD0=An}
\mathcal{L}&=&   \text{d}{\rm x}^{(0)} \left(m + \frac{2}{\mu^6} \frac{\mathcal{H}}{m} \right)  -  ifm\; \left[\bar{{\rm x}}_A \, \left(1+ \frac 2 {\mu^6} \frac {\cal H}{m^2}\right) -  \frac {4i} {\mu^6} \frac {\cal H}{m^2} \,{\theta}^{\bar{w}} \bar{\theta}^{\bar{w}} + \frac i {\mu^6 } \frac 1{m\sqrt{m}} {\theta}^{\bar{w}}\bar{\nu}\right] + \nonumber \\   \nonumber
\\ && + i \bar{f}m\;
\left[ {\rm x}_A\, \left(1+ \frac 2 {\mu^6} \frac {\cal H}{m^2}\right)+ \frac {4i} {\mu^6} \frac {\cal H}{m^2} \,\theta^w \bar{\theta}^{{w}}+ \frac i {\mu^6 } \frac 1{m\sqrt{m}} \bar{\theta}^w\nu \;\right]  - \nonumber \\   \nonumber
\\ && -2i \, \left( \text{d}\theta^{\bar{w}} -ia\theta^{\bar{w}}  \right) \, \bar{\theta}^w\,  \left(m + \frac{1}{\mu^6} \frac{\mathcal{H}}{m} \right)  -2i \, \left(\text{d}\bar{\theta}^w+ia\bar{\theta}^w  \right)\, \theta^{\bar{w}} \, \left(m + \frac{1}{\mu^6} \frac{\mathcal{H}}{m} \right)
- \nonumber \\   \nonumber
\\ && -2i \, \left( \text{d}\theta^{{w}} +ia\theta^{{w}}  \right) \, \left( \frac{1}{\mu^6} \frac{\mathcal{H}}{m}\, \bar{\theta}^{\bar{w}} -\frac{1}{\mu^6}  \frac{1}{2\sqrt{m}}\bar{\nu} \right)
-2i \, \left(\text{d}\bar{\theta}^{\bar{w}}-ia\bar{\theta}^{\bar{w}} \right)\, \left( \frac{1}{\mu^6} \frac{\mathcal{H}}{m}\, {\theta}^{w} -\frac{1}{\mu^6}  \frac{1}{2\sqrt{m}}\nu \right)+
 \nonumber \\  \nonumber \\
&& + {1\over \mu^6}{\rm tr}\left(\bar{\bb P}{\rm D}_\tau {\bb Z} + {\bb P}{\rm D}_\tau \bar{\bb Z} -{i\over 8} {\rm D}_\tau{\mathbf \Psi}\,  \bar{\mathbf \Psi} + {i\over 8} {\mathbf \Psi} {\rm D}_\tau \bar{\mathbf \Psi}  \right)~. \nonumber \\
\end{eqnarray}
It is convenient to introduce the notation
\be\label{tcH:=}
\tilde{{\cal H}}= \frac{1}{\mu^6} \frac{\mathcal{H}}{m}\; , \qquad
\tilde{\nu}= \frac{1}{\mu^6}  \frac{1}{2\sqrt{m}}\nu = (\bar{\tilde{\nu}})^*  \qquad
\ee

\noindent and to write the above Lagrangian in a bit more compact form
\begin{eqnarray}\label{cL=mD0=An=}
\mathcal{L}&=&  \text{d}{\rm x}^{(0)}\, (m+2\tilde{\mathcal{H}})  -  i f\; \left[\bar{{\rm x}}_A \, (m+2\tilde{\mathcal{H}}) -  {4i} \tilde{\cal H} \,{\theta}^{\bar{w}} \bar{\theta}^{\bar{w}} + 2i {\theta}^{\bar{w}}\bar{\tilde{\nu}}\right] + \nonumber \\   \nonumber \\ &&
 + i \bar{f}\;
\left[ {{\rm x}}_A \, (m+2\tilde{\mathcal{H}}) +  {4i} \tilde{\cal H} \,{\theta}^{{w}} \bar{\theta}^{{w}} + 2i \bar{\theta}^w\tilde{\nu} \;\right]  - \nonumber \\   \nonumber \\ && -2i \, \left( \text{d}\theta^{\bar{w}} -ia\theta^{\bar{w}}  \right) \, \bar{\theta}^w \, (m+\tilde{\mathcal{H}}) -2i \, \left(\text{d}\bar{\theta}^w+ia\bar{\theta}^w  \right)\, \theta^{\bar{w}}  \, (m+\tilde{\mathcal{H}})
- \nonumber \\   \nonumber \\ &&- 2i \, \left( \text{d}\theta^{{w}} +ia\theta^{{w}}  \right) \, \left( \tilde{\mathcal{H}}\, \bar{\theta}^{\bar{w}} -\bar{\tilde{\nu}} \right) -2i \, \left(\text{d}\bar{\theta}^{\bar{w}}-ia\bar{\theta}^{\bar{w}} \right)\, \left(\tilde{\mathcal{H}}\, {\theta}^{w} -\tilde{\nu} \right)+
 \nonumber \\  \nonumber \\
&& + {1\over \mu^6}{\rm tr}\left(\bar{\bb P}{\rm D}_\tau {\bb Z} + {\bb P}{\rm D}_\tau \bar{\bb Z}-{i\over 8} {\rm D}_\tau{\mathbf \Psi}\,  \bar{\mathbf \Psi} + {i\over 8} {\mathbf \Psi} {\rm D}_\tau \bar{\mathbf \Psi}  \right)~. \nonumber \\
\end{eqnarray}
Then the canonical  Hamiltonian is defined by
\begin{equation}\label{H0=analyt}
\begin{split}
\text{d}\tau H_0 &= \text{d}{\rm x}^{(0)}p^{(0)} + \text{d}{\rm x}_A \bar{p} + \text{d}\bar{\rm x}_A p + \text{d}\theta^w \Pi^\theta_w + \text{d}\theta^{\bar{w}} \Pi^\theta_{\bar{w}} + \text{d} \bar{\theta}^w \bar{\Pi}^{\bar{\theta}}_w + \text{d} \bar{\theta}^{\bar{w}} \bar{\Pi}^{\bar{\theta}}_{\bar{w}} + ia \tilde{\mathfrak{d}}^{(0)} + i f \bar{\tilde{\mathfrak{d}}} - i \bar{f} \tilde{\mathfrak{d}} +\\
&~\\
&+ \frac{1}{\mu^6} \text{tr}(\text{d}\mathbb{Z}\bar{\mathbb{P}})+ \frac{1}{\mu^6} \text{tr}(\text{d}\bar{\mathbb{Z}}\mathbb{P}) -	\frac{i}{8 \mu^6} \text{tr}(\text{d}\mathbf{\Psi} \bar{\mathbf{\Psi}}) -\frac{i}{8 \mu^6} \text{tr}(\text{d}\bar{\mathbf{\Psi}} \mathbf{\Psi}) + \text{tr}(\text{d}\mathbb{A} \mathbb{P}_{\mathbb{A}}) - \mathcal{L}_{\text{mD}0}~.
\end{split}
\end{equation}

In \eqref{H0=analyt}  the momenta for the bosonic coordinate functions are related to their central basis counterparts by

\begin{equation}\label{up=-2p}
\begin{array}{ccccc}
p^{(0)}:= u^{a(0)}p_a~,&~&p := -\dfrac{1}{2}u^a p_a~,&~&\bar{p} := -\dfrac{1}{2}\bar{u}^a p_a
\end{array}
\end{equation}
and have  the nonvanishing Poisson brackets

\begin{equation}
\begin{array}{ccccc}
\left[p^{(0)}, {\rm x}^{(0)} \right]_{\text{PB}} = -1~,&~&\left[p, \bar{{\rm x}}_A \right]_{\text{PB}} = -1~,&~&\left[\bar{p}, {\rm x}_A \right]_{\text{PB}} = -1~.
\end{array}
\end{equation}

\noindent The momenta conjugate to the fermionic coordinate functions of the analytical basis obeying

\begin{equation}
\begin{array}{ccccccc}
{}\{{\Pi}^\theta_w , \theta^w\}_{\text{PB}} = -1~,&~& \{ {\Pi}^\theta_{\bar{w}} , \theta^{\bar{w}}\}_{\text{PB}} = -1~, &~& \{ \bar{\Pi}^{\bar{\theta}}_w , \bar{\theta}^w\}_{\text{PB}} = -1~, &~& \{ \bar{\Pi}^{\bar{\theta}}_{\bar{w}} , \bar{\theta}^{\bar{w}}\}_{\text{PB}} = -1~
\end{array}
\end{equation}

\noindent are related to their central basis cousins by a bit more complicated relation

\begin{equation}\label{Pi-PiAn}
\begin{array}{ccc}
\Pi^\theta_{w}:= -i\bar{w}^\alpha \Pi_\alpha+2i \bar{\theta}^w\bar{p}~,&~& \Pi^\theta_{\bar{w}}:= iw^\alpha \Pi_\alpha - 2i \bar{\theta}^{\bar{w}}p~,\\
~\\
\bar{\Pi}^{\bar{\theta}}_{w}:= -i\bar{w}^\alpha \bar{\Pi}_\alpha -2i \theta^w \bar{p} ~,&~& \bar{\Pi}^{\bar{\theta}}_{\bar{w}}:= iw^\alpha \bar{\Pi}_\alpha+ 2i \theta^{\bar{w}}p~.
\end{array}
\end{equation}

The covariant momenta  of the analytical basis, distinguished by tilde from their central basis counterparts, obey

\begin{equation}
\begin{array}{lll}
{}[\tilde{\mathfrak{d}}^{(0)} , {\rm Z}_{An}^{(M)}]_{\text{PB}} =0 \; , \qquad {}[\tilde{\mathfrak{d}}, {\rm Z}_{An}^{(M)}]_{\text{PB}} =0 \; , \qquad  {}[\bar{\tilde{\mathfrak{d}}}, {\rm Z}_{An}^{(M)}]_{\text{PB}} =0 \; , \qquad
\end{array}
\end{equation}
while
\begin{equation}
\begin{array}{lll}
{}[{\mathfrak{d}}^{(0)} , Z^{M}]_{\text{PB}} =0 \; , \qquad {}[{\mathfrak{d}},  Z^{M}]_{\text{PB}} =0 \; , \qquad  {}[\bar{{\mathfrak{d}}}, Z^{M}]_{\text{PB}} =0 \; . \qquad
\end{array}
\end{equation}

Eq. \eqref{H0=analyt} also encodes the definitions of the canonical momenta as derivatives of the Lagrangian with respect to velocities, \eqref{pA:=},  as well as of the covariant momenta as derivatives of the Lagrangian with respect to Cartan forms,  \eqref{fd=dL-df}. Calculating these with our Lagrangian \eqref{cL=mD0=An=} we find the primary constraints ({\it cf.} \eqref{eq:Phi_a}-\eqref{eq:PA})
\begin{eqnarray}\label{p0=An}
\Phi^{(0)} := p^{(0)} - (m + 2\tilde{\mathcal{H}}) \approx 0~,   \qquad
 \\ \nonumber \\  \label{pA=0} - \frac 1 2\  \Phi := p \approx 0~,    \qquad \\ \nonumber \\  \label{bpA=0} - \frac 1 2\bar{\Phi} :=  \bar{p} \approx 0~,    \qquad \\ \nonumber \\ \label{fd=An}
 \tilde{{\mathfrak d}}+ {{\rm x}}_A \, (m+2\tilde{{\mathcal{H}}}) +  {4i} \tilde{{\cal H}} \,{\theta}^{{w}} \bar{\theta}^{{w}} + 2i \bar{\theta}^w\tilde{\nu}  \approx 0 \; , \qquad  \\ \nonumber
 \\  \label{bfd=An}  \bar{\tilde{{\mathfrak d}}} + \bar{{\rm x}}_A \, (m+2\tilde{\mathcal{H}}) -  {4i} \tilde{\cal H} \,{\theta}^{\bar{w}} \bar{\theta}^{\bar{w}} + 2i {\theta}^{\bar{w}}\bar{\tilde{\nu}}\approx 0 \; , \qquad \\ \nonumber
 \\  \label{U=An} \tilde{U}:= \tilde{{\mathfrak d}}^{(0)}-4i \theta^{\bar{w}}  \bar{\theta}^w  \, (m+\tilde{\mathcal{H}})+ 4i \theta^{{w}}  \bar{\theta}^{\bar{w}}\,\tilde{\mathcal{H}}\,   -2i \theta^{{w}} \bar{\tilde{\nu}} +2i \bar{\theta}^{\bar{w}} \tilde{{\nu}}  -   2\tilde{{\cal B}}\approx 0 \; , \qquad \\ \nonumber \\   \label{tB=} \text{where} \qquad \tilde{{\cal B}} := \frac 1 {\mu^6} {\cal B}:= \frac 1 {\mu^6}{\rm tr}\left(\bar{{\mathbb P}}{\mathbb Z}-{\mathbb P}\bar{{\mathbb Z}}+\frac i 8 {\mathbf \Psi}\bar{{\mathbf \Psi}}\right) ,   \qquad
\end{eqnarray}

\begin{eqnarray}\label{dw=}
 d_w:= \Pi^\theta_w - 2i \left( \bar{\tilde{\nu}}- \bar{\theta}^{\bar{w}} \tilde{{\mathcal{H}}} \right) \approx 0~, \qquad  \\ \nonumber \\  \label{dbw=}  \qquad d_{\bar{w}}:= \Pi^\theta_{\bar{w}}+ 2i\left(m+    \tilde{{\mathcal{H}}} \right)\bar{\theta}^w \approx 0~, \qquad  \\ \nonumber  \\  \label{bdw=}  \bar{d}_w := \bar{\Pi}^{\bar{\theta}}_w + 2i\left(m+    \tilde{{\mathcal{H}}} \right) \theta^{\bar{w}}\approx 0~, \qquad   \\ \nonumber
 \\ \label{bdbw=} \bar{d}_{\bar{w}} := \bar{\Pi}^{\bar{\theta}}_{\bar{w}}  - 2i \left( \tilde{{\nu}}- {\theta}^{{w}} \tilde{{\mathcal{H}}} \right) \approx 0~,  \qquad
\end{eqnarray}

\noindent and \eqref{eq:PA},
\begin{equation}
\mathbb{P}_{\mathbb{A}}:= \frac {\partial L}{\partial \dot{\mathbb{A}}_\tau} \approx 0~.
\label{eq:PA=An}
\end{equation}

\noindent
Exactly as in the central basis calculations, the requirement of preservation of this latter  leads to the secondary constraint \eqref{eq:Gauss},

\begin{equation}
\mathbb{G}= \dfrac{1}{\mu^6} \left( [\bar{\mathbb{Z}}, \mathbb{P}] + [\mathbb{Z}, \bar{\mathbb{P}}] - \frac{i}{4} \{\mathbf \Psi, \bar{\mathbf \Psi} \} \right) \approx 0
\label{eq:Gauss=An}
\end{equation}
and, as a result, the canonical Hamiltonian vanishes in the weak sense

  \begin{equation}
  \label{H0=0=An}
  H_0\approx 0\; .
\end{equation}
This means that the total Hamiltonian is given by linear combination of the constraints or, being more precise, of the first class constraints.

The set of these first class constraints
includes  Eqs.  \eqref{p0=An}, \eqref{dw=}, \eqref{bdbw=},

\begin{eqnarray}
\label{Phi0=0I}
\Phi^{(0)}\approx 0\; , \qquad d_w\approx 0\; , \qquad \bar{d}_{\bar{w}}\approx 0\; , \qquad
\end{eqnarray}

\noindent the following sum of \eqref{U=An} with the linear combination of the  second class constraints \eqref{pA=0},  \eqref{bpA=0}, \eqref{dbw=} and \eqref{bdw=},
\begin{eqnarray}
\label{ttU=}
\tilde{\tilde{U}}^{(0)}& := &\tilde{U}^{(0)} - 2 \text{x}_A \bar{p} + 2 \bar{\text{x}}_A p - \bar{\theta}^w \bar{d}_w + \theta^{\bar{w}}d_{\bar{w}} = \nonumber \\ \nonumber \\
&\; = & \tilde{{\mathfrak d}}^{(0)} - 2 \text{x}_A \bar{p} + 2 \bar{\text{x}}_A p - \bar{\theta}^w \bar{\Pi}^{\bar{\theta}}_w + \theta^{\bar{w}} \Pi^{\theta}_{\bar{w}} +4i \theta^{w}  \bar{\theta}^{\bar{w}} \tilde{\mathcal{H}} - 2i \theta^w \bar{\tilde{\nu}} + 2i \bar{\theta}^{\bar{w}} \tilde{\nu} - 2 \tilde{{\cal B}} \approx 0~ \\ \nonumber
\end{eqnarray}
as well as  \eqref{eq:PA=An} (just expressing the pure gauge nature of the 1d gauge field) and the Gauss constraint \eqref{eq:Gauss=An}.
The remaining constraints  are of the second class.

Below we will not need the complete knowledge of the algebra of the constraints on the Poisson brackets. Let us just notice that,
instead of calculating it directly, we can use their relation  with their central basis counterparts and knowledge of the algebra of these latter.
In the case of first class constraint \eqref{Phi0=0} and second class constraints \eqref{eqs:Phi}, this relation is  just a coincidence,

\begin{equation}
\Phi^{(0)} := p_a u^{(0)} - (m + 2 \tilde{\cal H}) = p^{(0)} -( m+2 \tilde{{\cal H}})~,
\end{equation}
\begin{equation}
\Phi := \Phi_a u^a = - 2p~,
\end{equation}
\begin{equation}
\bar{\Phi} := \Phi_a \bar{u}^a = -2 \bar{p}~.
\end{equation}

The same is true for  the Gauss constraints \eqref{eq:Gauss=An} and for \eqref{eq:PA=An} while for  the remaining constraints the relations are nontrivial:
\begin{eqnarray}
U^{(0)}:= \mathfrak{d}^{(0)} - 2 \tilde{\mathcal{B}} = \tilde{\mathfrak{d}}^{(0)} - 2 \text{x}_A \bar{p} + 2 \bar{\text{x}}_Ap - \theta^w \Pi^\theta_w - \bar{\theta}^w \bar{\Pi}^{\bar{\theta}}_{w} +  \theta^{\bar{w}} \Pi^\theta_{\bar{w}} + \bar{\theta}^{\bar{w}} \bar{\Pi}^{\bar{\theta}}_{\bar{w}} - 2 \tilde{\mathcal{B}} = \nonumber \\ \nonumber \\
=\tilde{\tilde{U}}{}^{(0)}- \theta^w d_w + \bar{\theta}^{\bar{w}} \bar{d}_{\bar{w}}
~,
\end{eqnarray}

\begin{equation}
\bar{w}^\alpha d_\alpha := \bar{w}^\alpha (  \bar{\Pi}_\alpha + i p_a (\gamma^a \bar{\theta})_\alpha - m \bar{\theta}_\alpha) + 2\bar{\tilde{\nu}}  =
i d_w- \Phi^{(0)} \bar{\theta}^{\bar{w}}~,
\end{equation}
\begin{equation}
w^\alpha \bar{d}_\alpha := w^\alpha ({\Pi}_\alpha + i p_a (\gamma^a \theta)_\alpha + m\theta_\alpha ) - 2\tilde{\nu}  =- i \bar{d}_{\bar{w}}  + \Phi^{(0)} \theta^w~,
\end{equation}
\begin{equation}
w^\alpha d_\alpha := w^\alpha(  \bar{\Pi}_\alpha + i p_a (\gamma^a \bar{\theta})_\alpha - m \bar{\theta}_\alpha) =  -i d_{\bar{w}} + 4 p \bar{\theta}^{\bar{w}}  + \Phi^{(0)} \bar{\theta}^w~,
\end{equation}
\begin{equation}
\bar{w}^\alpha \bar{d}_\alpha := \bar{w}^\alpha({\Pi}_\alpha + i p_a (\gamma^a \theta)_\alpha + m\theta_\alpha )  =i \bar{d}_w - 4p \theta^w - \Phi^{(0)} \theta^{\bar{w}}~,
\end{equation}
\begin{equation}
\mathfrak{d} = \tilde{\mathfrak{d}} + 2(\text{x}^{(0)} + i \theta^{\bar{w}}\bar{\theta}^w + i \theta^w \bar{\theta}^{\bar{w}})p + (\text{x}_A + 2i \theta^w \bar{\theta}^w)p^{(0)} + \bar{\theta}^w \bar{\Pi}^{\bar{\theta}}_{\bar{w}} + \theta^w \Pi^{\theta}_{\bar{w}}~,
\end{equation}
\begin{equation}
\bar{\mathfrak{d}} = \bar{\tilde{\mathfrak{d}}} + 2(\text{x}^{(0)} - i \theta^{\bar{w}}\bar{\theta}^w - i \theta^w \bar{\theta}^{\bar{w}})\bar{p} + (\bar{\text{x}}_A - 2i \theta^{\bar{w}} \bar{\theta}^{\bar{w}})p^{(0)} + \theta^{\bar{w}} \Pi^\theta_w + \bar{\theta}^{\bar{w}} \bar{\Pi}^{\bar{\theta}}_{{w}}~.
\end{equation}

\noindent Using  these expressions, the algebra of the constraints  in the analytical basis can be  read off Table \ref{table:mD0-centr} and  Eqs. \eqref{fdbP=}-\eqref{bfdwd=}. Below we will need  a part of this algebra which is  resumed (essentially) in Table \ref{table:mD0=An} below.

\subsection{Resolving bosonic second class constraints}

Actually, as we have already stated, we do not need to know the complete  algebra of the constraints in detail. Instead we can observe that the bosonic second class constraints in the analytical basis, Eqs.~\eqref{bpA=0}, \eqref{fd=An} and  \eqref{pA=0}, \eqref{bfd=An}, are explicitly resolved (in terminology of Dirac \cite{Dirac:1963}). This implies that
in these pairs of constraints one element,  \eqref{fd=An} and \eqref{bfd=An}, can be rewritten as expression for coordinate ${\rm x}_A$ and $\bar{\rm x}_A$, respectively,  in terms of other variables, while the other, conjugate element,  \eqref{bpA=0} and \eqref{pA=0}, respectively,  implies vanishing of the momenta conjugate to this dependent coordinate. In such cases, the procedure of changing the Poisson brackets by Dirac brackets can be replaced by just setting momenta to zero in the strong sense

\be
\bar{p}=0\; , \qquad p=0\; ,
\ee

\noindent and substituting ${\rm x}_A$ and $\bar{\rm x}_A$ in all places when these appear by their expressions obtained by solving  \eqref{fd=An} and \eqref{bfd=An},

\bea  \label{xA=An}
 {{\rm x}}_A = -\frac{ \tilde{{\mathfrak d}}+  {4i} \tilde{{\cal H}} \,{\theta}^{{w}} \bar{\theta}^{{w}} +2i \bar{\theta}^w\tilde{\nu}}{m+2\tilde{{\mathcal{H}}}}   \; , \qquad   \bar{{\rm x}}_A = -\frac{\bar{\tilde{{\mathfrak d}}} - {4i} \tilde{\cal H} \,{\theta}^{\bar{w}} \bar{\theta}^{\bar{w}} + 2i {\theta}^{\bar{w}}\bar{\tilde{\nu}}} {m+2\tilde{{\mathcal{H}}}}   \; . \qquad
\eea

Fortunately in our case ${\rm x}_A$ and $\bar{{\rm x}}_A$ do not contribute to other constraints so that the above prescription is tantamount to the reduction of configuration space by omitting the directions corresponding to these coordinates. The same effect can be reached by omitting  \eqref{fd=An} and \eqref{bfd=An} from the set of constraints and thus converting \eqref{bpA=0} and \eqref{pA=0} into the first class constraints. Then  ${{\rm x}}_A$ and $\bar{{\rm x}}_A$ can be gauged away using the symmetries generated by \eqref{bpA=0} and \eqref{pA=0}, thus arriving at the same result as described above.

As far as the fermionic second class constraints are concerned, since the canonically conjugate constraints in this case are also complex conjugate, we can quantize these using the Gupta-Bleuler method. At the classical level this can be reflected by omitting one of the conjugate constraint, say $d_{\bar{w}}$, thus converting its conjugate,  $\bar{d}_w$ into an effectively first class constraint.

\begin{table}[h]
\resizebox{\textwidth}{!}{\begin{tabular}{c||c|c|c|c|c}
 $[...,... \}_{\text{PB}}$
& $\Phi^{(0)}$&  $\tilde{\tilde{U}}^{(0)}$  & $d_w$ & $\bar{d}_{\bar{w}}$ &
 $ \bar{d}_w$ \\
 \hline \hline
 ~\\
$\Phi^{(0)}$& 0 & $ \boxed{\begin{matrix} 2i (\theta^w \text{tr}(\tilde{\mathbf{\Psi}} \tilde{{\mathbb{G}}}) +\cr  +\bar{\theta}^{\bar{w}} \text{tr}(\bar{\tilde{\mathbf{\Psi}}}\tilde{{\mathbb{G}}}))\end{matrix}}$ & $2i\text{tr}(\tilde{\mathbf{\Psi}} \tilde{{\mathbb{G}}})$ & $-2i \text{tr}(\bar{\tilde{\mathbf{\Psi}}} \tilde{{\mathbb{G}}})$   & 0 \\
~\\
$\tilde{\tilde{U}}^{(0)}$ & $\boxed{\begin{matrix}-2i (\theta^w \text{tr}(\tilde{\mathbf{\Psi}} \tilde{{\mathbb{G}}}) +\cr +\bar{\theta}^{\bar{w}} \text{tr}(\bar{\tilde{\mathbf{\Psi}}} \tilde{{\mathbb{G}}}))\end{matrix}}$ & 0 &  $\boxed{\begin{matrix}2\theta^w \left(3 \bar{\theta}^{\bar{w}}\text{tr}(\tilde{\mathbf{\Psi}} \tilde{{\mathbb{G}}}) \right. -\cr -\left. 4i \text{tr}(\bar{\mathbb{Z}} \tilde{{\mathbb{G}}}) \right)\end{matrix}}$ & $\boxed{\begin{matrix}2 \bar{\theta}^{\bar{w}}\left(3\theta^w \text{tr}(\bar {\tilde{\mathbf{\Psi}}} \tilde{{\mathbb{G}}}) \right. -\cr -\left.
4i \text{tr}(\mathbb{Z} \tilde{{\mathbb{G}}}) \right)\end{matrix}}$ &  $\boxed{\begin{matrix} -\bar{d}_w -\cr -2 \theta^{\bar{w}} \bar{\theta}^{\bar{w}} \text{tr}(\bar{\tilde{\mathbf{\Psi}}} \tilde{{\mathbb{G}}}) + \cr +2 \theta^w\theta^{\bar{w}}  \text{tr}(\tilde{{\mathbf{\Psi}}} \tilde{{\mathbb{G}}}) \end{matrix}}$  \\
~\\
 $d_w$ &  $-2i \text{tr}(\tilde{\mathbf{\Psi}} \tilde{{\mathbb{G}}})$ & $\boxed{\begin{matrix}-2\theta^w \left(3 \bar{\theta}^{\bar{w}}\text{tr}(\tilde{\mathbf{\Psi}} \tilde{{\mathbb{G}}}) \right. -\cr -\left. 4i \text{tr}(\bar{\mathbb{Z}} \tilde{{\mathbb{G}}}) \right)\end{matrix}}$ & $\boxed{\begin{matrix} 4 \bar{\theta}^{\bar{w}} \text{tr}(\tilde{\mathbf{\Psi}}\tilde{{\mathbb{G}}}) -\cr - 8i \text{tr}(\bar{\mathbb{Z}}\tilde{{\mathbb{G}}})\end{matrix}}$ & $\boxed{\begin{matrix}2 (\theta^w \text{tr}(\tilde{\mathbf{\Psi}} \tilde{{\mathbb{G}}}) +\cr  +\bar{\theta}^{\bar{w}} \text{tr}(\bar{\tilde{\mathbf{\Psi}}} \tilde{{\mathbb{G}}}))\end{matrix}}$ &  $2 \theta^{\bar{w}}  \text{tr}(\tilde{\mathbf{\Psi}} \tilde{{\mathbb{G}}})$ \\
 ~\\
 $\bar{d}_{\bar{w}}$ & $2i \text{tr}(\bar{\tilde{\mathbf{\Psi}}} \tilde{{\mathbb{G}}})$ & $\boxed{\begin{matrix}-2\bar{\theta}^{\bar{w}}\left(3\theta^w \text{tr}(\bar {\tilde{\mathbf{\Psi}}} \tilde{{\mathbb{G}}}) \right. -\cr -\left. 4i \text{tr}(\mathbb{Z} \tilde{{\mathbb{G}}}) \right)\end{matrix}}$  & $\boxed{\begin{matrix}2 (\theta^w \text{tr}(\tilde{\mathbf{\Psi}} \tilde{{\mathbb{G}}}) +\cr  +\bar{\theta}^{\bar{w}} \text{tr}(\bar{\tilde{\mathbf{\Psi}}} \tilde{{\mathbb{G}}}))\end{matrix}}$ & $\boxed{\begin{matrix}-4  \theta^{w} \text{tr}(\bar{\tilde{\mathbf{\Psi}}}\tilde{{\mathbb{G}}})+ \cr +8i \text{tr}({\mathbb{Z}}\tilde{{\mathbb{G}}})\end{matrix}}$ &  $-2 \theta^{\bar{w}}  \text{tr}(\bar{\tilde{\mathbf{\Psi}}} \tilde{{\mathbb{G}}})$  \\
 ~\\
 $ \bar{d}_w$ &  0 &   $\boxed{\begin{matrix} \bar{d}_w +\cr +2 \theta^{\bar{w}} \bar{\theta}^{\bar{w}} \text{tr}(\bar{\tilde{\mathbf{\Psi}}} \tilde{{\mathbb{G}}}) - \cr -2 \theta^w\theta^{\bar{w}}  \text{tr}(\tilde{{\mathbf{\Psi}}} \tilde{{\mathbb{G}}}) \end{matrix}}$ & $2 \theta^{\bar{w}}  \text{tr}(\tilde{\mathbf{\Psi}} \tilde{{\mathbb{G}}})$ & $-2 \theta^{\bar{w}}  \text{tr}(\bar{\tilde{\mathbf{\Psi}}} \tilde{{\mathbb{G}}})$ & 0  \\
 ~\\
 \hline \hline
\end{tabular}}
\caption{Closed algebra of the ``effective first class constraints'' of the mD0 system. For compactness, we have omitted the lines and columns  corresponding to the Gauss constraint \eqref{eq:Gauss=An} (which has only one nonvanishing element on their crossing) and to the first class  constraint  \eqref{eq:PA=An} (which are line and column of zeros),  and  have also renormalized the Gauss constraint and the fermionic matrix fields as follows: $\tilde{{\bb G}}:= \frac 1 m {\bb G}\,$, $ \tilde{{\bf \Psi}} := \frac 1 {\sqrt{m}} \; {\bf \Psi}\,$,  $ \bar{\tilde{\mathbf{\Psi}}} := \frac 1 {\sqrt{m}} \; \bar{\mathbf{\Psi}}$. }
\label{table:mD0=An}
\end{table}

The essential part of the algebra of the effective first class constraints thus obtained is collected in Table~\ref{table:mD0=An}, where we omitted the  lines and columns with the constraints \eqref{eq:PA=An}, which have vanishing brackets with all others, and with the Gauss constraint matrix \eqref{eq:Gauss=An} which has nonvanishing brackets only with itself, \eqref{GG=G}. To make the r.h.s's of the brackets shorter, we use in Table~\ref{table:mD0=An}  renormalized expressions for Gauss constraint and   for the fermionic matrix fields:

\be\label{tG=G/m}
\tilde{{\bb G}}:= \frac 1 m {\bb G}\, \qquad  \tilde{{\bf \Psi}} := \frac 1 {\sqrt{m}} \; {\bf \Psi}\, \qquad  \bar{\tilde{\mathbf{\Psi}}} := \frac 1 {\sqrt{m}} \; \bar{\mathbf{\Psi}}\; . \qquad
\ee

Notice that all the terms in the  r.h.s. of the brackets resumed in Table
\ref{table:mD0=An} are proportional to the Gauss law constraints and hence vanish when this is taken into account.

\bigskip

\setcounter{equation}{0}

\section{Quantization  of multiple D$0$-brane system and its field theory equations}
Quantization implies the replacement of the phase space variables by operators the commutators and anticommutators of which can be determined through  the Poisson brackets of their classical prototypes by the Dirac prescription
\begin{equation}\label{Dirac=quant}
[...,...\}_{\text{PB}}\quad \mapsto \quad \frac 1 i\, [...,...\}~.
\end{equation}
\noindent
In the supercoordinate representation the bosonic and fermionic configuration space variables maintain their $\text{(quasi-)}$classical description, i.e. they are numerical ($c$-number) or Grassmann algebra ($a$-number) valued, respectively, while the momenta are represented by differential operators such that the Dirac prescription \eqref{Dirac=quant} holds.

In particular, in our case the bosonic and fermionic center of mass coordinate functions and their momenta are replaced by corresponding coordinates and differential operator as
\begin{equation}
\begin{array}{rcccl}
\hat{\text{x}}^{(0)} = \text{x}^{(0)}~,&~& \hat{\text{x}}_A = \text{x}_{A}~,&~&\hat{\bar{\text{x}}}_A = \bar{\text{x}}_A~,\\
~\\
\hat{p}^{(0)} = -i \partial_{\text{x}^{(0)}} ~,&~& \hat{\bar{p}}_A = -i \partial_{\text{x}_A}~,&~&\hat{p}_A = -i \partial_{\bar{\text{x}}_A}~,
\end{array}
\label{eq:bosOperators}
\end{equation}

\noindent and

\begin{equation}
\begin{array}{rcrclcl}
\hat{\theta}^w = \theta^w~,&~&\hat{\theta}^{\bar{w}} = \theta^{\bar{w}}~,&~& \hat{\bar{\theta}}^w = \bar{\theta}^w~,&~&\hat{\bar{\theta}}^{\bar{w}} = \bar{\theta}^{\bar{w}}~,\\
~\\
\hat{\Pi}^{\theta}_w = -i \partial_{\theta^w} ~,&~& \hat{\Pi}^{\theta}_{\bar{w}}= -i \partial_{\theta^{\bar{w}}}~,&~&\hat{\bar{\Pi}}^{\bar{\theta}}_w = -i \partial_{\bar{\theta}^w}~,&~&\hat{\bar{\Pi}}^{\bar{w}}_{\bar{\theta}} = -i \partial_{\bar{\theta}^{\bar{w}}}~,
\end{array}
\label{eq:ferOperators}
\end{equation}

\noindent
where we have used the notation
\begin{equation}
\begin{array}{ccrcrcr}
~&~&\partial_{\text{x}^{(0)}} = \dfrac{\partial}{\partial \text{x}^{(0)}}~,&~&\partial_{\text{x}_A} = \dfrac{\partial}{\partial \text{x}_A}~,&~&\partial_{\bar{\text{x}}_A} = \dfrac{\partial}{\partial \bar{\text{x}}_A}~,\\
~\\
\partial_{\theta^w} = \dfrac{\partial}{\partial \theta^w}~,&~&\partial_{\theta^{\bar{w}}} = \dfrac{\partial}{\partial \theta^{\bar{w}}}~,&~&\partial_{\bar{\theta}^w} = \dfrac{\partial}{\partial \bar{\theta}^w}~,&~&\partial_{\bar{\theta}^{\bar{w}}} = \dfrac{\partial}{\partial \bar{\theta}^{\bar{w}}}~.
\end{array}
\end{equation}

For the complex spinors frame variables the covariant momenta are expressed in terms of covariant derivatives

\be
\mathbb{D}^{(0)} =  \bar{w}_\alpha \dfrac{\partial}{\partial \bar{w}_\alpha} - {w}_\alpha \dfrac{\partial}{\partial {w}_\alpha}~,\qquad \mathbb{D} =w_\alpha \dfrac{\partial}{\partial \bar{w}_\alpha}~,\qquad \bar{\mathbb{D}} =  \bar{w}_\alpha\dfrac{\partial}{\partial {w}_\alpha}~,
\ee

\noindent so that
\begin{equation}
\begin{array}{lcrcr}
~&~&\hat{w}_\alpha = w_\alpha~,&~&\hat{\bar{w}}_\alpha = \bar{w}_\alpha~,\\
~\\
\hat{\mathfrak{d}}{}^{(0)}=-i\mathbb{D}^{(0)} ~,&~&\hat{\mathfrak{d}}=-i \mathbb{D} ~,&~&\hat{\bar{\mathfrak{d}}}= -i\bar{\mathbb{D}} ~.
\end{array}
\end{equation}

For the bosonic  matrix fields the quantization in coordinate representation is standard

\begin{equation}\label{PZ=ddZ}
\begin{array}{rccl}
\hat{\mathbb{Z}}_i{}^j = \mathbb{Z}_i{}^j ~,&~& \hat{\bar{\mathbb{Z}}}_i{}^j  = \bar{\mathbb{Z}}_i{}^j ~,\\
~\\
\hat{\bar{\mathbb{P}}}_i{}^j  = -i \mu^6 \dfrac{\partial}{\partial \mathbb{Z}_j{}^i}~,&~&\hat{\mathbb{P}}_i{}^j  = -i \mu^6 \dfrac{\partial}{\partial \bar{\mathbb{Z}}_j{}^i}~,
\end{array}
\end{equation}

\noindent
while for the fermionic matrix variables obeying

\begin{equation}\label{hPsihbPsi=}
\left\lbrace \hat{\mathbf \Psi}_i{}^j, \hat{\bar{\mathbf \Psi}}_k{}^l \right\rbrace = \; 4\mu^6 ( \delta^l_i \delta_k^j - \dfrac{1}{N} \delta^j_i \delta^l_k )~, \qquad \left\lbrace \hat{\mathbf \Psi}^j_i, \hat{{\mathbf \Psi}}_k^l \right\rbrace = 0 \; , \qquad \left\lbrace \hat{\bar{\mathbf \Psi}}^j_i, \hat{\bar{\mathbf \Psi}}_k^l \right\rbrace = 0\;  \qquad
\end{equation}
we should use the counterpart of the so-called holomorphic representation for Heisenberg algebra in which the annihilation operator is represented by derivative with respect to (classical counterpart of the) annihilation operator,

\begin{equation}\label{bPsi=ddPsi} \hat{\mathbf{\Psi}}^j_i = \mathbf{\Psi}^j_i~,\qquad \hat{\bar{\mathbf{\Psi}}}^j_i = \; 4 \mu^6 \dfrac{\partial}{\partial \mathbf{\Psi}^i_j}~.
\end{equation}

\noindent
Then the quantum counterpart of the SYM  supercurrents and Hamiltonian are represented by the following differential operators

\begin{equation}\label{hnu=}
\hat{\tilde{\nu}}=\frac{1}{\mu^6}  \frac{1}{2\sqrt{m}} \hat{\nu}=  \frac{1}{2\sqrt{m}} \text{tr}\left(-i \mathbf{\Psi} \dfrac{\partial}{\partial \overline{\mathbb{Z}}} +4 [\mathbb{Z}, \bar{\mathbb{Z}}]  \dfrac{\partial}{\partial \mathbf{\Psi}}\right)~, \qquad \end{equation}

 \begin{equation} \label{hbnu=}
\hat{\bar{\tilde{\nu}}}=\frac{1}{\mu^6}  \frac{1}{2\sqrt{m}} \hat{\bar{\nu}}= \frac{1}{2\sqrt{m}} \text{tr}\left(-4i  {\mu^6} \dfrac{\partial}{\partial \mathbb{Z}}  \dfrac{\partial}{\partial \mathbf{\Psi}}+  \frac{1}{\mu^6} \mathbf{\Psi} [\mathbb{Z}, \bar{\mathbb{Z}}]\right)~,  \qquad
\end{equation}

\begin{equation}\label{hcH=}
\hat{\tilde{{\cal H}}}= \frac{1}{\mu^6} \frac{\hat{\mathcal{H}}}{m}= \frac{1}{m} {\rm tr}\left( -\mu^6 \dfrac{\partial}{\partial \mathbb{Z}}\dfrac{\partial}{\partial \bar{\mathbb{Z}}} +8i\mu^6\bar{\bb Z} \dfrac{\partial}{\partial \mathbf{\Psi}}\dfrac{\partial}{\partial \mathbf{\Psi}}  +  \frac{1}{\mu^6} [{\bb Z},  \bar{\bb Z}]^2 -
{i\over 2}  \frac{1}{\mu^6} {\bb Z}{\mathbf \Psi}{\mathbf \Psi} \right) ~,
\end{equation}

\noindent and the quantum operator generating the U(1) transformations of matrix variables  is

\begin{equation}\label{hcB=}
\hat{\tilde{\mathcal{B}}}= \frac{1}{\mu^6} \hat{\mathcal{B}}= i \text{tr} \left(
\bar{\mathbb{Z}}\dfrac{\partial}{\partial \bar{\mathbb{Z}}}- \mathbb{Z} \dfrac{\partial}{\partial \mathbb{Z}}+\frac{1}{2} \mathbf{\Psi} \dfrac{\partial}{\partial \mathbf{\Psi}}\right)~.
\end{equation}

\bigskip

The state vector is represented by a function dependent on the configuration space coordinates \footnote{To be precise, we should also say
that $ \Xi  $ depends on the 1d gauge field ${\bb A}_\tau$, but then we should subject it to the quantum constraint \eqref{eq:PA} which reads
$\dfrac{\partial}{\partial \dot{\mathbb{A}}_\tau}\Xi =0 $ and implies just independence on  ${\bb A}_\tau$. We allowed ourselves to ``straighten'' the presentation by omitting this stage.
},

 \begin{equation}\label{Xi=}
 \Xi = \Xi\left({\rm x}^{(0)}, \bar{w}, w; \theta^w , \theta^{\bar{w}} , \bar{\theta}{}^w , \bar{\theta}{}^{\bar{w}}; {\mathbb{Z}}, \bar{\mathbb{Z}};
 \mathbf{\Psi} \right)~.
\end{equation}

The physical states are represented by the state vectors obeying equations obtained as quantum version of the effective first class constraints which read
\begin{equation}\label{id-xx0}
\left[ -i \partial_{\text{x}^{(0)}} - \left(m + 2 \hat{\tilde{\mathcal{H}}} \right) \right]\Xi = 0~,
\end{equation}
\begin{equation}\label{D0Xi-=0}
i\hat{\tilde{\tilde{U}}}^{(0)}\Xi= \left( \mathbb{D}^{(0)} - \bar{\theta}^w \partial_{\bar{\theta}^w} + \theta^{\bar{w}} \partial_{\theta^{\bar{w}}} - 4 \theta^{w} \bar{\theta}^{\bar{w}} \hat{\tilde{\mathcal{H}}} + 2 \theta^w \hat{\bar{\tilde{\nu}}} - 2 \bar{\theta}^{\bar{w}} \hat{\tilde{\nu}} - 2i \hat{\tilde{\mathcal{B}}} - q \right)\Xi = 0~,
\end{equation}
\begin{equation}\label{d-dthw}
\left[ \partial_{\theta^w}+ 2 \left(\hat{\bar{\tilde{\nu}}} - \bar{\theta}^{\bar{w}} \hat{\tilde{\mathcal{H}}} \right) \right]\Xi = 0~,
\end{equation}
\begin{equation}\label{d-dbthbw}
\left[ \partial_{\bar{\theta}^{\bar{w}}} + 2 \left(\hat{\tilde{\nu}} - \theta^w \hat{\tilde{\mathcal{H}}} \right) \right]\Xi = 0~,
\end{equation}
\begin{equation}\label{d-dbthw}
\left[\partial_{\bar{\theta}^w} -2 \left( m + \hat{\tilde{\mathcal{H}}} \right) \theta^{\bar{w}} \right]\Xi = 0~,
\end{equation}

\noindent where $\hat{\tilde{\nu}}$, $\hat{\bar{\tilde{\nu}}}$ and $ \hat{\tilde{\mathcal{H}}} $ are given in
\eqref{hnu=}, \eqref{hbnu=} and \eqref{hcH=}, and the quantum SU$(N)$ Gauss law constraint

\begin{equation}\label{hGXi=0}
i \hat{\mathbb{G}}^j_i \Xi =  \left(\bar{\mathbb{Z}}^k_i \dfrac{\partial}{\partial \bar{\mathbb{Z}}^k_j} - \bar{\mathbb{Z}}^j_k \dfrac{\partial}{\partial  \bar{\mathbb{Z}}_k^i}  +  \mathbb{Z}^k_i \dfrac{\partial}{\partial \mathbb{Z}^k_j} - \mathbb{Z}^j_k \dfrac{\partial}{\partial  \mathbb{Z}_k^i} + \mathbf{\Psi}^k_i \frac{\partial}{\partial 	\mathbf{\Psi}_j^k} - \mathbf{\Psi}^j_k \dfrac{\partial}{\partial \mathbf{\Psi}_k^i}  \right)\Xi = 0~,
\end{equation}

Notice that when calculating the quantum generator of SU$(N)$ transformations ${\mathbb{G}}$ in \eqref{hGXi=0} we have to use the $\text{`qp' ordering}$ for the commutators and anticommutators  of matrices,

\begin{equation}
\begin{array}{lcr}
[{\bar{\mathbb{Z}}}, {\mathbb{P}}]{}_i{}^j \mapsto  - i \mu^6 \left(\bar{\mathbb{Z}}_i{}^k \dfrac{\partial}{\partial \bar{\mathbb{Z}}_j{}^k} - \bar{\mathbb{Z}}_k{}^j \dfrac{\partial}{\partial  \bar{\mathbb{Z}}_k^i} \right)~,&~& [{\mathbb{Z}}, {\bar{\mathbb{P}}}]{}_i{}^j \mapsto  - i \mu^6 \left(\mathbb{Z}_i{}^k \dfrac{\partial}{\partial \mathbb{Z}_j{}^k} - \mathbb{Z}_k{}^j\dfrac{\partial}{\partial  \mathbb{Z}_k{}^i} \right)~,\\
~\\
~&~&  \{{\mathbf{\Psi}},{\bar{\mathbf{\Psi}}}  \}{}_i{}^j \mapsto 4 \mu^6 \left( \mathbf{\Psi}_i^{\,k} \dfrac{\partial}{\partial 	\mathbf{\Psi}_j^{\,k}} - \mathbf{\Psi}_k^{\, j}  \dfrac{\partial}{\partial \mathbf{\Psi}_k^{\, i}} \right)~,
\end{array}
\end{equation}
as otherwise the expression for   ${\mathbb{G}}$ will not be traceless as it should be. Thus, Eqs. \eqref{id-xx0}, \eqref{D0Xi-=0}, \eqref{d-dthw}, \eqref{d-dbthbw}, \eqref{d-dbthw}, \eqref{hGXi=0} define the (super)field theory of simplest 3D counterpart of multiple D$0$-brane system in the superspace with coordinate presented in \eqref{Xi=}.

It is convenient to allow the state vector to depend also on additional complex bosonic coordinate ${\rm x}_A$ and its c.c. $\bar{\rm x}_A$,

 \begin{equation}\label{Xi==}
 \Xi = \Xi\left({\rm x}^{(0)}, {\rm x}_A, \bar{\rm x}_A, \bar{w}, w; \theta^w , \theta^{\bar{w}} , \bar{\theta}{}^w , \bar{\theta}{}^{\bar{w}}; {\mathbb{Z}}, \bar{\mathbb{Z}};
 \mathbf{\Psi} \right)~,
\end{equation}

\noindent
but impose on it the additional constraints removing this dependence:

\begin{equation}\label{d-dxAXi=0}
 \dfrac{\partial}{\partial {\rm x}_A}\Xi =0  \; ,  \qquad \dfrac{\partial}{\partial \bar{\rm x}_A}\Xi =0   \; .  \qquad
\end{equation}

\noindent Being equivalent to original, this way of describing state vector makes manifest that the state vector can be considered as dependent on the central basis coordinates,

\begin{equation}\label{Xi=centr}
 \Xi = \tilde{\Xi}\left(x^a; \theta^\alpha, \bar{\theta}{}^\alpha; \bar{w}, w;  {\mathbb{Z}}, \bar{\mathbb{Z}};
 \mathbf{\Psi} \right)~,
\end{equation}

\noindent and this is useful to understand better the obtained  (super)field theory equations, and write this in more conventional form at least in some particular cases, like  $N=1$ case discussed in the next subsection.
In particular, in this simple case the  equations can be rewritten in terms of derivatives and covariant derivatives of the central basis,

\bea\label{ddxa=an}
\partial_a= u_a^{(0)} \partial_{{\rm x}^{(0)} } + u_a \partial_{{\rm x}_A} + \bar{u}_a \partial_{\bar{{\rm x}}_A}\; , \qquad  \\ \nonumber
\\ \label{Dal=an}
D_\alpha = \partial_\alpha + i(\gamma^a\bar{\theta})_\alpha \partial_a =
w_\alpha \left(\partial_{\theta^w } +i\bar{\theta}{}^{\bar{w}}  \partial_{{\rm x}^{(0)} }  \right) + \bar{w}_\alpha \left(\partial_{\theta^{\bar{w}} } +4i\bar{\theta}{}^{\bar{w}} \partial_{\bar{\rm x}_A} +i\bar{\theta}{}^{{w}}\partial_{{\rm x}^{(0)}}  \right)\; ,  \\ \nonumber \\ \label{bDal=an}
\bar{D}_\alpha = \bar{\partial}_\alpha + i(\gamma^a{\theta})_\alpha \partial_a=
w_\alpha \left(\bar{\partial}_{\bar{\theta}{}^w} +4i{\theta}{}^{w} \partial_{{\rm x}_A} +i{\theta}{}^{\bar{w}}\partial_{{\rm x}^{(0)}} \right) +  \bar{w}_\alpha \left(\bar{\partial}_{\bar{\theta}}{}^{\bar{w}} +i{\theta}{}^{w}  \partial_{{\rm x}^{(0)}}  \right)\; .
\\ \nonumber
\eea
and the spinor frame variables can be integrated out thus arriving to a more conventional description of field theory in terms of usual (central basis) spacetime coordinates.

\bigskip

But before considering $N=1$ case, let us notice the possibility, in general case of arbitrary $N>1$,  to change the variables in the relative motion sector by  decomposing the matrix fields
\begin{equation}\label{Z=ZITI}
\begin{array}{lcccr}
\hat{\mathbb{Z}}_i^j = \sqrt{2} \hat{{Z}}^I T_{Ii}^{~j}~,&~& \hat{\bar{\mathbb{Z}}}_i^j = \sqrt{2} \hat{\bar{{Z}}}^I T_{Ii}^{~j}~,&~&
\hat{\mathbf{\Psi}}_i^j = \sqrt{2} \hat{{\Psi}}^I T_{Ii}^{~j}~,\\
~\\
\hat{\bar{\mathbb{P}}}_i^j = \sqrt{2} \hat{\bar{{P}}}^I T_{Ii}^{~j}~,~&~& \hat{\mathbb{P}}_i^j = \sqrt{2} \hat{{P}}^I T_{Ii}^{~j}~,&~& \hat{\bar{\mathbf{\Psi}}}_i^j =  \sqrt{2} \hat{\bar{{\Psi}}}^I T_{Ii}^{~j}~,
\end{array}
\end{equation}
on  SU$(N)$ generators $T_{Ii}^{~j}$ which obey
\bea
{}[T_I,  T_J]=if_{IJK}T_K \; , \qquad f_{IJK}=f_{[IJK]} \; , \qquad f_{IJK}f_{I'JK}=N\delta_{II'}\; , \qquad I,J,K=1,\ldots N^2-1\; , \qquad \\ \nonumber \\
{\rm tr} (T_I T_J)= \frac 1 2 \delta_{IJ}\; , \qquad T_I T_J = \frac 1 2 \left(\frac 1 N \delta_{IJ} {\mathbb{I}}+\left(d_{IJK}+if_{IJK}\right)T_K \right)\; , \qquad d_{IJK}=d_{(IJK)}\; , \qquad
\\ \nonumber \\
 T_{I}{}_i{}^j  T_{I}{}_k{}^l =\frac 1 2 \left(\delta_i{}^l  \delta{}_k{}^j- \frac 1 N\delta{}_i{}^j  \delta{}_k{}^l\right)\qquad \Rightarrow \qquad (T_{I}T_{I})_i{}^j= C_F\delta{}_i{}^j \; , \qquad C_F=\frac {N^2-1}{N}\;  \qquad
\eea
and some other relations which can be found in  e.g. \cite{Haber:2019sgz}. In the simplest case of SU(2) ($N=2$),

\be
\text{SU}(2):\qquad T_I{}_i{}^j= \frac 1 2 \sigma_I{}_i{}^j\; , \qquad  f_{IJK}=\epsilon_{IJK}\; , \qquad  d_{IJK}=0\; , \qquad I,J,K=1,2,3.
\ee

The coefficients in \eqref{Z=ZITI} are chosen in such a way that the nontrivial commutation relations of the
basic operators have (almost) the standard form

\begin{equation}
\begin{array}{rcccl}
[\hat{{P}}^I, \hat{\bar{Z}}^J] = - i\mu^6 \delta^{IJ}~,&~&[\hat{\bar{P}}^I, \hat{Z}^J] = - i\mu^6 \delta^{IJ}~,&~&\{\hat{{\Psi}}{}^I, \hat{\bar{{\Psi}}}{}^J\} = \; 4\mu^6 \delta^{IJ}~,
\end{array}
\end{equation}
so that in the coordinate representation the above operators are realized by

\begin{equation}
\begin{array}{lcrcclcl}
\hat{{Z}}^I ={Z}^I~,&~& \hat{\bar{{P}}}^I = -i \mu^6 \dfrac{\partial}{\partial {Z}^I}~, &~&~& \hat{\bar{{Z}}}^I = \bar{{Z}}^I~,&~&\hat{{P}}^I = -i \mu^6 \dfrac{\partial}{\partial \bar{{Z}}^I}~, \qquad\\
\\
~&~& \hat{{\Psi}}^I = {\Psi}^I~,&~&~&\hat{\bar{{\Psi}}}^I = \;4 \mu^6 \dfrac{\partial}{\partial {\Psi}^I}~,&~&~
\end{array}
\end{equation}

\noindent where ${Z}^I= ({Z}^1, {Z}^2,\ldots , {Z}^{N^2-1})$ are complex bosonic ($c$-number) coordinates and
$ {\Psi}^I= (\Psi^1, \Psi^2,\ldots ,\Psi^{N^2-1})$ are fermionic ($a$-number) coordinates,

\be
 {\Psi}^I {\Psi}^J=- {\Psi}^J {\Psi}^I\; .
\ee

In terms of these relative motion coordinates  we have the following expressions for the quantum generator of SU($N$) transformations \eqref{hGXi=0}
\begin{equation}\label{hGXi==0}
\hat{\mathbb{G}}^j_i  =: 2\hat{{G}}^IT^{~j}_{Ii}= 2 f_{IJK}\left(\bar{{Z}}^I \dfrac{\partial}{\partial \bar{{Z}}^J}  +  {Z}^I \dfrac{\partial}{\partial {Z}^J}  +  {\Psi}^I \frac{\partial}{\partial 	{\Psi}^J}  \right)T_{Ki}{}^{j}~,
\end{equation}

\noindent
for the part of the U(1) generator acting on the matrix field  \eqref{hcB=}

\begin{equation}\label{hcB==}
\hat{\tilde{\mathcal{B}}}= \frac{1}{\mu^6} \hat{\mathcal{B}}= i  \left(
\bar{{Z}}^I\dfrac{\partial}{\partial \bar{{Z}}^I}- {Z}^I \dfrac{\partial}{\partial {Z}^I}+ \frac 1 2 {\Psi}^I \dfrac{\partial}{\partial {\Psi}^I}\right)~,
\end{equation}

\noindent for the SYM supercharges \eqref{hnu=}, \eqref{hbnu=}

\begin{equation}\label{tnu=}
\hat{\tilde{\nu}}=\frac{1}{\mu^6}  \frac{1}{2\sqrt{m}} \hat{\nu}=  - \frac{i}{2\sqrt{m}}\left( {\Psi}^I \dfrac{\partial}{\partial \bar{{Z}}^I} - 4\sqrt{2} f_{IJK}  {Z}^I \bar{{Z}}^J  \dfrac{\partial}{\partial {\Psi}^K}\right)~, \qquad \end{equation}

\begin{equation}\label{tbnu=}
\hat{\bar{\tilde{\nu}}}=\frac{1}{\mu^6}  \frac{1}{2\sqrt{m}} \hat{\bar{\nu}}= -\frac{i}{2\sqrt{m}}
\left(4  {\mu^6} \dfrac{\partial}{\partial {Z}^I}  \dfrac{\partial}{\partial {\Psi}^I}-  \frac{\sqrt{2}}{\mu^6} f_{IJK}  {Z}^I \bar{{Z}}^J {\Psi}^K\right)~,  \qquad
\end{equation}

\noindent and for the SYM Hamiltonian \eqref{hcH=}

\begin{equation}\label{htcH=}
\hat{\tilde{{\cal H}}}= \frac{1}{m} \,\left( -\mu^6 \dfrac{\partial}{\partial {Z}^I}\dfrac{\partial}{\partial \bar{{Z}}^I}
 - 4\sqrt{2} \mu^6 \bar{Z}^I f_{IJK} \dfrac{\partial}{\partial {\Psi}^J}\dfrac{\partial}{\partial {\Psi}^K}  -\dfrac{2}{\mu^6} f_{IJM} f_{KLM} {Z}^I \bar{{Z}}^J {Z}^K \bar{\mathbb{Z}}^L + {\sqrt{2} \over 4\mu^6} f_{IJK}  {Z}^I{\Psi}^J {\Psi}^K \right) ~.
\end{equation}

These expressions can be used in the equations \eqref{id-xx0}-\eqref{d-dbthw} imposed on the state vector superfield
(``wavefunction'')

 \begin{equation}\label{Xi===}
 \Xi = \Xi\left({\rm x}^{(0)}, {\rm x}_A, \bar{{\rm x}}_A, \bar{w}, w; \theta^w , \theta^{\bar{w}} , \bar{\theta}{}^w , \bar{\theta}{}^{\bar{w}}; {{Z}}{}^I, \bar{{Z}}{}^I;
{\Psi}^I \right)~,
\end{equation}

\noindent which obeys also \eqref{d-dxAXi=0}. This is manifestly a superfield on superspace with
$5+N^2$ bosonic and $3+N^2$ fermionic directions ($9$ bosonic and $7$ fermionic directions in the simplest case
$N=2$).

\section{Some properties of 3D mD$0$ field theory equations}
\setcounter{equation}{0}

\subsection{$\mathbf{\textit{N}}$~=~1. Field theory of single D$0$-brane}

If we set the number $N$ of D0-branes to unity, $N=1$,  the matrix fields disappear so that the state vector becomes superfield in a center of mass superspace, which is usual superspace enlarged by spinor frame variables (Lorentz harmonic superspace in terminology of \cite{Bandos:1990ji})

\begin{equation}\label{Xi=cen1N}
 \Xi_0:= \Xi \vert_{N=1}= \tilde{\Xi}_0\left(x^a; \theta^\alpha, \bar{\theta}{}^\alpha ; \bar{w}, w\right)= \Xi_0\left({\rm x}^{(0)}, {\rm x}_A, \bar{{\rm x}}_A, \bar{w}, w; \theta^w , \theta^{\bar{w}} , \bar{\theta}{}^w , \bar{\theta}{}^{\bar{w}}\right)
\end{equation}

\noindent
and the above system of equations for this vector reduces to

\begin{equation}\label{d-dthw1N}
\partial_{\theta^w}\Xi_0 = 0~,
\qquad
\partial_{\bar{\theta}^{\bar{w}}}\Xi_0 = 0~,
\end{equation}
\begin{equation}\label{d-dbthw1N}
\left[\partial_{\bar{\theta}^w} -2  m  \theta^{\bar{w}} \right]\Xi_0 = 0~,
\end{equation}

\begin{equation}\label{d-dxA=0}
\partial_{ {\rm x}_A}\Xi_0 =0  \; ,  \qquad \partial_{ \bar{\rm x}_A}\Xi_0 =0  \; ,  \qquad
\end{equation}
\begin{equation}\label{id-xx01N}
\left( -i \partial_{\text{x}^{(0)}} - m  \right)\Xi_0 = 0~,
\end{equation}

\begin{equation}\label{D0Xi-=01N}
\left( \mathbb{D}^{(0)} - \bar{\theta}^w \partial_{\bar{\theta}^w} + \theta^{\bar{w}} \partial_{\theta^{\bar{w}}} - q \right)\Xi_0 = 0~.
\end{equation}

This system of equations describes the free (super)field theory of single 3D counterpart of D$0$ brane, i.e. of a massive ${\cal N}=2$ superparticle in $\text{D}=3$. In Appendix \ref{App=D0} we show how to obtain those straightforwardly by quantization of massive superparticle in its spinor moving frame formulation. While the quantization of such a simple system in the standard formulation described by the de Azc\'arraga-Lukierski action \eqref{SD0=A+L} is much simpler, in our case to follow this more complicated quantization makes sense as it describes the $N=1$ limit of our simplest counterpart of the  multiple D$0$ system which is currently known only in its spinor moving frame formulation.

 Eqs. \eqref{d-dxA=0} and  \eqref{d-dthw1N} result in independence of   the state vector superfield  $\Xi$ on ${\rm x}_A, \bar{{\rm x}}_A; \theta^w, \bar{\theta}^{\bar{w}}$,

 \begin{equation}\label{Xi0=}
 \Xi_0=  \Xi_0\left({\rm x}^{(0)},  \bar{w}, w;  \theta^{\bar{w}} , \bar{\theta}{}^w \right)\; ,
\end{equation}

\noindent Eqs. \eqref{d-dbthw1N} and
\eqref{id-xx01N} are easily solved by
\be
\Xi_0 = e^{im{\rm x}^{(0)} - 2m {\theta}^{\bar{w}}\bar{\theta}^{{w}}} \, \chi ( {\theta}^{\bar{w}}, \bar{w}_\alpha,  w_\alpha) \; \equiv  e^{im{\rm x}^{(0)} } \, (1- 2 m{\theta}^{\bar{w}}\bar{\theta}^{{w}})\, \chi ( {\theta}^{\bar{w}}, \bar{w}_\alpha,  w_\alpha) \;  \qquad
\ee

\noindent  and then \eqref{D0Xi-=01N} implies

\be\label{D0chi=}
\left({\bb D}^{(0)} + {\theta}^{\bar{w}}\partial_{{\theta}^{\bar{w}}}-q\right) \chi ( {\theta}^{\bar{w}}, \bar{w}_\alpha,  w_\alpha)
=0 \; .  \qquad
\ee
Decomposing $\chi$ in the powers of (fermionic and hence nilpotent) ${\theta}^{\bar{w}}$,

\be
\chi ( {\theta}^{\bar{w}}, \bar{w}_\alpha,  w_\alpha)= \phi (\bar{w}_\alpha,  w_\alpha) + i {\theta}^{\bar{w}}\, \xi ( \bar{w}_\alpha,  w_\alpha)
\ee
we find from \eqref{D0chi=} that the components of these superfields are bosonic and fermionic functions of complex normalized bosonic spinors (functions on SU$(1,1)\approx \text{SL}(2,{\bb R})$) with definite U$(1)$ charges
$q$ and $(q-1)$,

\bea\label{D0phi=}
\phi =\phi^q ( \bar{w}_\alpha,  w_\alpha) \; , \qquad \left({\bb D}^{(0)} -q\right) \phi^q ( \bar{w}_\alpha,  w_\alpha)
=0 \; .  \qquad \\ \nonumber \\ \label{D0xi=} \xi =\xi^{(q-1)}(\bar{w}_\alpha,  w_\alpha)  \; , \qquad \left({\bb D}^{(0)} -(q-1)\right) \xi^{(q-1)}(\bar{w}_\alpha,  w_\alpha)
=0 \; .  \qquad
\eea

\noindent
These definite U$(1)$ charges make these fields to be  functions on the coset
SU$(1,1)/\text{U}(1)$ \footnote{Actually they are sections of a fiber bundle $\pi : \text{SU}(2) \mapsto \text{SU}(2)/\text{U}(1)$ with structure group U$(1)$, which can be identified with Hopf fibration of ${\mathbb S}^3=\text{SU}(2)$ over ${\mathbb S}^2=\text{SU}(2)/\text{U}(1)$ with fiber ${\mathbb S}^1=\text{U}(1)$. However, such mathematical subtleties are beyond the scope of this paper. }.

As far as $\theta^{\bar{w}}$ carries charge one with respect to U$(1)$ symmetry generated by the operator in \eqref{D0chi=}, this latter equation implies that the superfield $\chi$ has definite charge $q$ with respect to
this U$(1)$ symmetry,

\be
\chi=\chi^q( {\theta}^{\bar{w}}, \bar{w}_\alpha,  w_\alpha) \; .
\ee

Eqs. \eqref{D0phi=} and \eqref{D0xi=} imply the following homogeneity properties of the bosonic and fermionic function under gauge transformations

\be
\phi^q ( e^{-i\vartheta}\bar{w}_\alpha, e^{i\vartheta} w_\alpha)=e^{-iq\vartheta}\phi^q ( \bar{w}_\alpha,  w_\alpha)\; , \qquad  \xi^{(q-1)}( e^{-i\vartheta}\bar{w}_\alpha, e^{i\vartheta} w_\alpha)=e^{-i(q-1)\vartheta}\xi^{(q-1)}(\bar{w}_\alpha,  w_\alpha) \; . \qquad
\ee

\subsubsection{Spacetime interpretation I}

To clarify the physical meaning of these functions, let us notice that, as a result of the constraints \eqref{bww=i} and \eqref{p0=An}-\eqref{bpA=0}, which implies $p_a= mu_a^{(0)}$ and $p_a\gamma^a_{\alpha\beta}=mu_a^{(0)}\gamma^a_{\alpha\beta}=2m\bar{w}_{(\alpha} w_{\beta )}$, we can write
\be
\bar{w}_{\alpha} w_{\beta }=\frac 1 {2m}p_a\gamma^a_{\alpha\beta}+\frac i {2}\epsilon_{\alpha\beta}\; , \qquad p_ap^a=m^2\; .
\ee
As a result, the dependence on the product $\bar{w}_{\alpha} w_{\beta }$ can be identified as dependence on the on-shell momentum of the massive particle, and for integer $q\geq 1$ we can write the following solution of Eqs. \eqref{D0phi=} and \eqref{D0xi=}
\be\label{phiq=+}
\phi^q(\bar{w},w)=\bar{w}^{\alpha_1} \ldots \bar{w}^{\alpha_q} \phi_{\alpha_1\ldots \alpha_q}(p)\vert_{p^2=m^2}\; , \qquad
\xi^{q-1}(\bar{w},w)=\bar{w}^{\alpha_1} \ldots \bar{w}^{\alpha_{q-1}} \xi_{\alpha_1\ldots \alpha_{q-1}}(p)\vert_{p^2=m^2}\; , \qquad q \geq 1\; .
\ee
To be in consonance with the generic spin-statistic theorem\footnote{Notice that in 3D the set of possible statistics is not exhausted by  the  Bose-Einstein and Fermi-Dirac options and includes also anyons  \cite{Wilczek:1981du,Wilczek:1981dr,Iengo:1991zbc}, in particular quartions \cite{Volkov:1989qa}, which can be collected in exotic  supermultiplets \cite{Sorokin:1993ua}. We will not address these possibilities in the present paper.}, for even $q$,  the field $\phi^q(\bar{w},w)$ should be bosonic and $\xi^{q-1}(\bar{w},w)$ should be fermionic while, for  odd $q$, it should be the other way around. For $q\leq 0$ the similar solution is
\be\label{phiq=-}
\phi^q(\bar{w},w)={w}^{\alpha_1} \ldots {w}^{\alpha_{-q}} \phi_{\alpha_1\ldots \alpha_{-q}}(p)\vert_{p^2=m^2}\; , \qquad
\xi^{q-1}(\bar{w},w)={w}^{\alpha_1} \ldots {w}^{\alpha_{-q-1}} \xi_{\alpha_1\ldots \alpha_{-q-1}}(p)\vert_{p^2=m^2}\; , \qquad q \geq 1\; .
\ee
Finally, for $q=1$ we have
\be\label{phiq=0}
\phi^{(1)}(\bar{w},w)=\bar{w}^{\alpha} \phi_{\alpha}(p)\vert_{p^2=m^2}\; , \qquad
\xi^{(0)}(\bar{w},w)= \xi(p)\vert_{p^2=m^2}\; , \qquad q \geq 1\; .
\ee
with fermionic $ \phi_{\alpha}(p)$ and bosonic $\xi(p)$.

However, as it is not manifestly clear how/whether one can normalize the above analytic solutions, below we will describe the solution in terms of different type of functions.

\bigskip

\subsubsection{Spacetime interpretation II}

To pass to a more standard description in terms of central basis coordinates, let us first resume  the above discussion by stating that the state vector superfield of single massive 3D ${\cal N}=2$  superparticle (3D counterpart of D$0$-brane) has the form

\be\label{Xiq=phiqexp}
\Xi_0=\Xi_0^q= e^{im{\rm x}^{(0)}}\left(\phi^q(\bar{w}_\alpha,  w_\alpha) +i {\theta}^{\bar{w}}\xi^{q-1}(\bar{w}_\alpha,  w_\alpha) - 2m{\theta}^{\bar{w}}\bar{\theta}^{{w}}{\phi}{}^q(\bar{w}_\alpha,  w_\alpha)\right)\; .
\ee
with $\phi^q(\bar{w}_\alpha,  w_\alpha)$ and $\xi^{q-1}(\bar{w}_\alpha,  w_\alpha)$ obeying \eqref{D0phi=} and \eqref{D0xi=}.

Now, using \eqref{x0=}, \eqref{thw=}, \eqref{bthw=} (or \eqref{ddxa=an}-\eqref{bDal=an}) and  acting by covariant derivative in central basis on this superfield, we obtain

\be\label{(bD-imth)Xiq=0}
(\bar{\text{D}}_\alpha +im\theta_\alpha)\Xi_0^q=0\; ,
\ee
which can be obtained from a (generalized) dimensional reduction of the chirality constraint in $\text{D}=4$
\footnote{The fermionic operator in the l.h.s. of \eqref{(bD-imth)Xiq=0} results from the  dimensional reduction of the 4d fermionic covariant derivative $\bar{D}_{\dot{\alpha}}=\bar{\partial}_{\dot{\alpha}}+i(\theta\sigma^{\underline{a}})_{\dot{\alpha}}{\partial}_{\underline{a}}$
assuming the dimensional reduction  implies ${\partial}_{x^\perp}:={\partial}_{x^2}=im$.} (see
\cite{Bandos:2021vrq}),
as well as
\be\label{(d+imu0)Xiq=0}
(\partial_a-imu_a^{(0)}) \Xi_0^q=0
\ee
which implies the Klein-Gordon equation
\be\label{(d*d+m2)Xiq=0}
(\partial_a\partial^a +m^2) \Xi_0^q=0 \; . \ee

\bigskip

Let us consider the leading component of the superfield \eqref{Xiq=phiqexp}, $e^{im{\rm x}^{(0)}}\phi^q(\bar{w}_\alpha,  w_\alpha)$. Roughly speaking (which is to say, following the line of \eqref{phiq=+}),
for the case $q=0$ this can be written in the form

\be \Xi_0^{(0)}\vert_{_{\theta =0}} = e^{im{\rm x}^{(0)}}\phi^{(0)}(\bar{w}_\alpha,  w_\alpha)=  e^{ip_a {x}^a} \phi(p)\vert_{p^2=m^2}  \; , \qquad \ee
where we have used \eqref{p0=An}-\eqref{bpA=0}. In this form it becomes evident that our state vector depends on both momenta and coordinates. To have a usual state vector field (``wavefunction'') in the coordinate representation, we should integrate over on-shell momentum.
In the standard textbook notation this reads

\be
\phi(x)=\left. \int \frac {d^2\vec{p}} {(2\pi)^2 p^0} \left(e^{ip_a {x}^a} \phi(p)+ e^{-ip_a {x}^a} \phi^*(p)\right)\right|_{p^0=\sqrt{\vec{p}^2+m^2}}\; .
\ee

Our twistor-like representation describes only one of two terms in the integrand of the above equation as the sign of $p_0$ is fixed by
\eqref{Phi0=0}.
This latter relation also allows to find the expression for integrating measure in terms of moving frame and spinor moving frame variables. Indeed, it implies (see \eqref{du0})

\be\label{dpa=}
\text{d}p_a=im f\bar{u}_a -im\bar{f}u_a \; , \qquad
\ee
and, after the use of \eqref{euuu=-2i},

\be
\frac 1 2 \epsilon^{abc}\text{d}p_b\wedge \text{d}p_c=-2i u^{a(0)} \bar{f}\wedge f \; .
\ee
Hence, the natural correspondence is

\be
{\text{d}^2\vec{p}} \; \leftrightarrow \;  \bar{f}\wedge f
\ee
and the standard coordinate representation of the state vector field with $q=0$ is

\be\label{phix=0}
\phi (x)= \int \bar{f}\wedge f  \, e^{imx^au_a^{(0)}} \phi^{(0)}(\bar{w},w)\; , \qquad q=0 \; , \qquad
\ee
where (let us recall)

\be
 u_a^{(0)}=\bar{w}\gamma_aw\; , \qquad f=w^\alpha \text{d}w_\alpha =(\bar{f})^*\; , \qquad \bar{w}^\alpha w_\alpha=i\; .
\ee
All this line can be easily generalized for the case of nonvanishing $q$ giving

\be\label{phix=+q}
\phi_{\alpha_1\ldots \alpha_q} (x)= \int \bar{f}\wedge f  \, w_{\alpha_1}\ldots  w_{\alpha_q} e^{im x^au_a^{(0)}} \phi^{q}(\bar{w},w)\; , \qquad q> 0 \; , \qquad
\ee
and

\be\label{phix=-q}
\phi_{\alpha_1\ldots \alpha_{-q}} (x)= \int \bar{f}\wedge f  \, \bar{w}_{\alpha_1}\ldots  \bar{w}_{\alpha_{-q}} e^{im x^au_a^{(0)}} \phi^{q}(\bar{w},w)\; , \qquad q< 0 \; . \qquad
\ee

Notice that generically, when we use equations involving $\phi^{q}(\bar{w},w)$, we have to be careful about the class of functions we choose. In particular, we should not substitute \eqref{phiq=+}-\eqref{phiq=0} for $\phi^{q}(\bar{w},w)$ in Eq. \eqref{phix=-q}. We will not work out the precise mathematical formulation of the problem (see \cite{Bandos:1990ji} and \cite{Fedoruk:1994ij,Zima:1995db} for a more rigorous treatment of $\text{D}=4$ case) but rather state that Eqs. \eqref{phix=+q} and  \eqref{phix=-q} describe the solutions of the wave equation assuming that $\phi^{q}(\bar{w},w)$ are such that the integrals converge.

Similarly, the standard superfield representation of the state vector can be obtained as

\bea\label{XiZ=q}
\Xi_{\alpha_1\ldots \alpha_{|q|}} (x,\theta,\bar{\theta})= \begin{cases} \int \bar{f}\wedge f  \, w_{\alpha_1}\ldots  w_{\alpha_q} \; e^{im x^au_a^{(0)}-2m\theta{\bar{w}}\,\bar{\theta}w} \; \chi^{q}(\bar{w},w, \theta \bar{w})\; , \qquad q\geq 0 \; , \qquad \cr \cr   \int \bar{f}\wedge f  \, \bar{w}_{\alpha_1}\ldots  \bar{w}_{\alpha_{-q}} e^{im x^au_a^{(0)}-2m\theta{\bar{w}}\,\bar{\theta}w}\; \chi^{q}(\bar{w},w, \theta{\bar{w}})\; , \qquad q< 0 \; , \qquad \end{cases} \\ \nonumber \\  \text{where}\qquad \theta \bar{w}= \theta^\alpha \bar{w}_\alpha =(\bar{\theta}w)^*\; , \qquad u_a^{(0)}= \bar{w}\gamma_aw \; .  \qquad
\eea
These superfields obey Eqs.  \eqref{(bD-imth)Xiq=0} and  \eqref{(d*d+m2)Xiq=0},

\bea\label{(bD-imth)Xi=0}
(\bar{\text{D}}_\alpha +im\theta_\alpha)\;\Xi_{\alpha_1\ldots \alpha_{|q|}} (x,\theta,\bar{\theta})=0\; , \qquad
\eea
\be\label{(d*d+m2)Xi=0}
(\partial_a\partial^a +m^2)\; \Xi_{\alpha_1\ldots \alpha_{|q|}} (x,\theta,\bar{\theta}) =0 \; . \ee
Furthermore, one can easily check that for  $q\not= 0$ superfield \eqref{XiZ=q}  obeys the following Dirac equation

\be\label{DiracXi=mXi}
\partial^{\alpha\alpha_1} \Xi_{\alpha_1\alpha_2\ldots \alpha_{|q|}}=-m\,\frac q {|q|}\Xi^{\alpha}{}_{\alpha_2\ldots \alpha_{|q|}}  \,  .  \ee

Thus the quantum state vector of $\text{D}=3$ ${\cal N}=2$  superparticle can be identified with superfield in the standard $\text{D}=3$~${\cal N}=2$   superspace obeying   Eqs. \eqref{(bD-imth)Xi=0}-\eqref{DiracXi=mXi}. This is equivalent to the description by a function of the coordinates of the analytical basis of Lorentz harmonic superspace \eqref{Xi=cen1N} obeying \eqref{d-dthw1N}-\eqref{D0Xi-=01N}.

Unfortunately, in the generic case of $N>1$, the description of the first type is not available (or, probably better to say, not efficient). This is because the bosonic and fermionic matrix coordinates of the 3D mD0 configurational superspace are not inert under the U$(1)$ symmetry. Thus we have to work with the description of the second type, in terms of superfield depending on the coordinates of the analytical basis of
center of mass superspace \eqref{Xi==} satisfying the set of equations
\eqref{id-xx0}-\eqref{hGXi=0}.

\subsection{Dependence on Grassmann center of mass coordinates for generic case  $N>1$. }

The set of fermionic equations for the state vector \eqref{d-dthw}-\eqref{d-dbthw} can be solved
in the generic case of $N>1$. To this end, let us first notice that  the commutators and anticommutators of the SYM currents simplify essentially when applied to the state vector obeying  Eq. \eqref{hGXi=0},

\begin{eqnarray}\label{hnuhbnu=-H}
&& \hat{\tilde{\nu}}^2 \,\Xi =  0\; , \qquad  \hat{\bar{\tilde{\nu}}}^2 \,\Xi =  0\; , \qquad  \{\hat{\tilde{\nu}} , \hat{\bar{\tilde{\nu}}}  \}\,\Xi   =\hat{\tilde{\mathcal{H}}}\,\Xi~, \\ \nonumber \\
&& {}[\hat{\tilde{\nu}}, \hat{\tilde{\mathcal{H}}}]\, \Xi = 0 \; , \qquad {}[\hat{\bar{\tilde{\nu}}}, \hat{\tilde{\mathcal{H}}}]\, \Xi = 0 \; ,
\qquad
\end{eqnarray}

\noindent
which, in particular, results in $e^{-2\theta^w\hat{\bar{\tilde{\nu}}}}\hat{{\tilde{\nu}}}= \hat{{\tilde{\nu}}}e^{-2\theta^w\hat{\bar{\tilde{\nu}}}} - 2\theta^w\hat{\tilde{\mathcal{H}}}$.

Then it is easy to check that Eqs.  \eqref{d-dthw}-\eqref{d-dbthw} are solved (formally)  by

\bea
\Xi&=& e^{-2\theta^w\hat{\bar{\tilde{\nu}}}}e^{-2\bar{\theta}{}^{\bar{w}}\hat{{\tilde{\nu}}}}  e^{+2\theta^w\bar{\theta}{}^{\bar{w}}\hat{\tilde{\mathcal{H}}}}e^{-2\theta^{\bar{w}}\bar{\theta}{}^{{w}}(m+
\hat{\tilde{\mathcal{H}}})}\;\left({\mathfrak{B}} + \theta^{\bar{w}} {\mathfrak{F}}\right) = \nonumber  \\ \nonumber \\
&=& \left(1- 2\theta^w\hat{\bar{\tilde{\nu}}}- 2\bar{\theta}{}^{\bar{w}}\hat{{\tilde{\nu}}}   +2\theta^w\bar{\theta}{}^{\bar{w}}(\hat{\tilde{\mathcal{H}}}-2\hat{\bar{\tilde{\nu}}}\hat{{\tilde{\nu}}} )-2\theta^{\bar{w}}\bar{\theta}{}^{{w}}(m+
\hat{\tilde{\mathcal{H}}})+\right. \nonumber \\ \nonumber \\ && \left. +4\theta^w\theta^{\bar{w}}\bar{\theta}{}^{{w}}\hat{\bar{\tilde{\nu}}}(m+
\hat{\tilde{\mathcal{H}}})+4\theta^{\bar{w}}\bar{\theta}{}^{{w}}\bar{\theta}{}^{\bar{w}}\hat{{\tilde{\nu}}}(m+
\hat{\tilde{\mathcal{H}}})- \right. \nonumber \\ \nonumber \\ && \left.
-4 \theta^w \bar{\theta}^{\bar{w}} \theta^{\bar{w}} \bar{\theta}^w (\hat{\tilde{\mathcal{H}}} - 2 \hat{\bar{\tilde{\nu}}}\hat{\tilde{\nu}})( m + \hat{\tilde{\mathcal{H}}}) \right)\; \left({\mathfrak{B}} + \theta^{\bar{w}} {\mathfrak{F}}\right)
\; , \qquad
\eea

\noindent
where

\be
{\mathfrak{B}}={\mathfrak{B}}\left(\text{x}^{(0)}, \bar{w}, w; {\mathbb{Z}}, \bar{\mathbb{Z}};\mathbf{\Psi} \right) \; , \qquad
{\mathfrak{F}}={\mathfrak{F}}\left(\text{x}^{(0)}, \bar{w}, w;  {\mathbb{Z}}, \bar{\mathbb{Z}};\mathbf{\Psi} \right)
\ee

\noindent are functions (superfields)  of opposite statistics (one bosonic another fermionic) dependent of spinor moving frame variables  and of matrix coordinates of our enlarged superspace. They should obey

\be
\hat{{\bb G}}_j{}^i  {\mathfrak{B}}= 0 \; , \qquad \hat{{\bb G}}_j{}^i  {\mathfrak{F}}= 0 \;  \qquad
\ee

\noindent with $\hat{{\bb G}}_j{}^i  $ from \eqref{hGXi=0} (or, equivalently, \eqref{hGXi==0}), which implies that they should be constructed from SU$(N)$ invariant combinations of the matrix fields,
and

\be\label{U0fB=qfB}
 \left( \mathbb{D}^{(0)} - 2i \hat{\tilde{\mathcal{B}}} - q \right) {\mathfrak{B}}= 0  \; , \qquad  \left( \mathbb{D}^{(0)} - 2i \hat{\tilde{\mathcal{B}}} - (q-1) \right) {\mathfrak{F}}= 0\; , \qquad
\ee

\noindent which fix the charge of the fields with respect to U$(1)$ group acting on spinor frame variables and matrix fields,

\be\label{fB=fBq}
 {\mathfrak{B}}=  {\mathfrak{B}}^{(q)} \left(\text{x}^{(0)}, \bar{w}, w; {\mathbb{Z}}, \bar{\mathbb{Z}};\mathbf{\Psi} \right)  \; , \qquad  {\mathfrak{F}}=  {\mathfrak{F}}^{(q-1)}\left(\text{x}^{(0)},  \bar{w}, w; {\mathbb{Z}}, \bar{\mathbb{Z}};\mathbf{\Psi} \right)   \; . \qquad
\ee

Furthermore, as a result of \eqref{id-xx0} these functions obey the Schr\"odinger equation with a shifted SYM Hamiltonian in which  $\text{x}^{(0)}$ coordinate playing the role of time,

\begin{equation}\label{id-xx0fB=}
\left( i \partial_{\text{x}^{(0)}} + m + 2 \hat{\tilde{\mathcal{H}}} \right) {\mathfrak{B}}^{(q)}  = 0~, \qquad \left( i \partial_{\text{x}^{(0)}} + m + 2 \hat{\tilde{\mathcal{H}}} \right) {\mathfrak{F}}^{(q-1)}  = 0~.
\end{equation}


To find \eqref{U0fB=qfB} from \eqref{D0Xi-=0} we have used the relations

\begin{eqnarray}\label{D0exp-=}
 i\hat{\tilde{\tilde{U}}}^{(0)} e^{-2\theta^w\hat{\bar{\tilde{\nu}}}}= \left( \mathbb{D}^{(0)} - \bar{\theta}^w \partial_{\bar{\theta}^w} + \theta^{\bar{w}} \partial_{\theta^{\bar{w}}} -4 \theta^{w} \bar{\theta}^{\bar{w}} \hat{\tilde{\mathcal{H}}} + 2 \theta^w \hat{\bar{\tilde{\nu}}}- 2 \bar{\theta}^{\bar{w}} \hat{\tilde{\nu}} - 2i \hat{\tilde{\mathcal{B}}} - q \right) e^{-2\theta^w\hat{\bar{\tilde{\nu}}}}=  \nonumber  \\ \nonumber \\
  = e^{-2\theta^w\hat{\bar{\tilde{\nu}}}}  \left( \mathbb{D}^{(0)} - \bar{\theta}^w \partial_{\bar{\theta}^w} + \theta^{\bar{w}} \partial_{\theta^{\bar{w}}}  + 2 \bar{\theta}^{\bar{w}} \hat{\tilde{\nu}} - 2i \hat{\tilde{\mathcal{B}}} - q \right)  \; , \qquad    \\ \nonumber \\
  \left( \mathbb{D}^{(0)} - \bar{\theta}^w \partial_{\bar{\theta}^w} + \theta^{\bar{w}} \partial_{\theta^{\bar{w}}}  - 2 \bar{\theta}^{\bar{w}} \hat{\tilde{\nu}} - 2i \hat{\tilde{\mathcal{B}}} - q \right) e^{-2\bar{\theta}{}^{\bar{w}}\hat{{\tilde{\nu}}}}  = e^{-2\bar{\theta}{}^{\bar{w}}\hat{{\tilde{\nu}}}}  \left( \mathbb{D}^{(0)} - \bar{\theta}^w \partial_{\bar{\theta}^w} + \theta^{\bar{w}} \partial_{\theta^{\bar{w}}}  - 2i \hat{\tilde{\mathcal{B}}} - q \right)\; , \qquad
\end{eqnarray}

\noindent which follow from the form of  the commutators of the U(1) generators for matrix fields and SYM supercurrents

\be[\hat{{\tilde{\nu}}} , \mathcal{B}]\, = \frac i 2 \mu^6 \hat{{\tilde{\nu}}},\; \qquad [\hat{\bar{\tilde{\nu}}}, \mathcal{B}]\, =  -\frac i 2 \mu^6 \hat{\bar{\tilde{\nu}}}~.\; \qquad
\ee

Of course, we can easily write the formal solution of \eqref{id-xx0fB=}

\bea\label{fB=fBq=}
{\mathfrak{B}}^{(q)} \left(\text{x}^{(0)}, \bar{w}, w; {\mathbb{Z}}, \bar{\mathbb{Z}};\mathbf{\Psi} \right) =e^{i\text{x}^{(0)}(m+
2\hat{\tilde{\mathcal{H}}})} \check{{\mathfrak{B}}}{}^{(q)} \left(\bar{w}, w; {\mathbb{Z}}, \bar{\mathbb{Z}};\mathbf{\Psi} \right)  \; , \qquad \nonumber \\  \nonumber \\   {\mathfrak{F}}^{(q-1)}\left(\text{x}^{(0)}, \bar{w}, w; {\mathbb{Z}}, \bar{\mathbb{Z}};\mathbf{\Psi} \right)=e^{i\text{x}^{(0)}(m+
2\hat{\tilde{\mathcal{H}}})}  \check{{\mathfrak{F}}}{} ^{(q-1)}\left( \bar{w}, w; {\mathbb{Z}}, \bar{\mathbb{Z}};\mathbf{\Psi} \right)  \; , \qquad
\eea

\noindent but its usefulness is restricted: an explicit solution of the Schr\"odinger equation would carry much more information on the system.

\bigskip

\subsection{Solutions with BPS-type configurations of the SYM sector}

Let us consider the generic case $N>1$ but impose the following set of conditions on the state vector superfield
\bea\label{nuXi=0}
\hat{\tilde{\nu}}\, \Xi =0\; , \qquad \\ \nonumber  \\ \label{bnuXi=0} \hat{\bar{\tilde{\nu}}} \,\Xi =0\; , \qquad \\ \nonumber  \\ \label{cHXi=0} \hat{\tilde{\mathcal{H}}} \,\Xi =0\; ,  \qquad
\eea
where $ \hat{{\tilde{\nu}}}$,  $ \hat{\bar{\tilde{\nu}}}$, $ \hat{\tilde{\mathcal{H}}}$ are given in \eqref{hnu=}, \eqref{hbnu=} and
\eqref{hcH=}.
Then Eqs.  \eqref{id-xx0}-\eqref{d-dbthw}  reduce to   \eqref{d-dthw1N}-\eqref{D0Xi-=01N} but imposed on the state vector superfield depending on matrix coordinates,  \eqref{Xi==}.

Furthermore, in this case the center of mass part of the equations of motion is separated from the relative motion equations the set of which is given now by  \eqref{nuXi=0}-\eqref{cHXi=0} and
\eqref{hGXi=0}.

Hence we can search for a solution in the form with factorized center of mass motion, i.e. in the form of sum of products

\be
\Xi^q=\sum\limits{q^\prime} {\Xi}_0^{q-q^\prime}(\text{x}^{(0)}, \bar{w},w;\, \theta^{\bar{w}}, \bar{\theta}{}^w )\; \underline{\Xi}{}^{q^\prime}  (Z^I,\bar{Z}^I,\Psi^I)
\ee

\noindent of functions of center of mass variables $ {\Xi}_0^{q-q^\prime}(\text{x}^{(0)}, \bar{w},w;\, \theta^{\bar{w}}, \bar{\theta}{}^w )$ obeying \eqref{d-dthw1N}, \eqref{d-dbthw1N}-\eqref{id-xx01N} and

\begin{equation}\label{D0Xi-=01Nq'}
\left( \mathbb{D}^{(0)} - \bar{\theta}^w \partial_{\bar{\theta}^w} + \theta^{\bar{w}} \partial_{\theta^{\bar{w}}} - (q-q^\prime) \right){\Xi}_0^{q-q^\prime} = 0~,
\end{equation}

\noindent and of functions of relative motion  variables  $\underline{\Xi}{}^{q^\prime}  (Z^I,\bar{Z}^I,\Psi^I)$ obeying
 \eqref{nuXi=0}-\eqref{cHXi=0}, \eqref{hGXi=0} and

\begin{equation}\label{cBXi=q'}
\left(
 \hat{\tilde{\mathcal{B}}}-\frac i 2 \, q^\prime \right)\,\underline{\Xi}{}^{q^\prime}  (Z^I,\bar{Z}^I,\Psi^I)=0 \; ,
\end{equation}

\noindent with operators defined in \eqref{hGXi==0}-\eqref{tbnu=}.

The set of Eqs. \eqref{tnu=}-\eqref{htcH=}  and  \eqref{hGXi=0} (with \eqref{hGXi==0})  coincides with  the equations for supersymmetric vacua of 1d reduction of 3D SYM model in the Schr\"odinger representation (see \cite{Nair:2023hsy} for study of 3D YM in this representation).
The related studies of the ground state of YM, SYM, and BFSS  Matrix model were carried out in \cite{Frohlich:1999zf,Lin:2014wka,Hoppe:2023vph} and in refs. therein.

The solutions of the equations for $ {\Xi}_0^{q-q^\prime}$ are the solution of the equations for single D$0$-brane and we have already discussed these in the previous (sub)section.
Here we consider Eqs. \eqref{htcH=}, \eqref{tnu=},  \eqref{tbnu=} and  \eqref{hGXi=0} for the simplest case of $N=2$.

\subsubsection{$N=2$ case: field theory  of simplest nontrivial 3D counterpart of mD$0$ system  }

Let us search for a solution of mD$0$ equations which obey the additional conditions \eqref{nuXi=0}
for the case of $N=2$, i.e. for the system of two D$0$-branes and strings ending on the same and on different D$0$-branes.

In this case the decomposition of $ \underline{\Xi}{}^{q^\prime}  (Z^I,\bar{Z}^I,\Psi^I)$ superfield on the fermionic coordinates   $ \Psi^I = (\Psi^1 ,  \Psi^2 , \Psi^3)$ reads

\be\label{Xq'=}
\underline{\Xi}{}^{q^\prime}  (Z^I,\bar{Z}^I,\Psi^I)=\mathfrak{f}{}^{q^\prime}  (Z,\bar{Z})+ \Psi^I \mathfrak{f}{}_I^{q^\prime -1} (Z,\bar{Z})+\frac 1 2  \Psi^I  \Psi^J\mathfrak{f}{}_{IJ}{}^{q^\prime -2} (Z,\bar{Z}) +\Psi^{\wedge 3}\mathfrak{f}{}{}^{q^\prime -3}  (Z,\bar{Z})
 \ee

\noindent where

\be
\Psi^{\wedge 3}:= \frac 1 {3!}\epsilon_{IJK} \Psi^I  \Psi^J \Psi^K = \Psi^1  \Psi^2 \Psi^3
\ee

\noindent is the basic monomial  of maximal power constructed from three fermionic variables.

Let us substitute this decomposition in Eqs. \eqref{cBXi=q'}, \eqref{cHXi=0} , \eqref{nuXi=0}, \eqref{bnuXi=0} and  \eqref{hGXi=0} with \eqref{hGXi==0}.

First of all,  Eq.  \eqref{cBXi=q'} clearly implies that the  coefficient functions in the decomposition \eqref{cBXi=q'} are  the eigenfunctions of the operator

\begin{equation}\label{hcB0==}
-2i\hat{\tilde{\mathcal{B}}}_0 = \left(
2\bar{{Z}}^I\dfrac{\partial}{\partial \bar{{Z}}^I}-2 {Z}^I \dfrac{\partial}{\partial {Z}^I}\right)~,
\end{equation}

\noindent with eigenvalues (having the meaning of U$(1)$ charges) written as superindices of these component fields.

\bigskip

Now, let us observe that the anticommutators of the operators  in \eqref{nuXi=0}  and   \eqref{bnuXi=0} are expressed in terms of the operators  \eqref{cHXi=0}  and  \eqref{hGXi=0}, so that it is sufficient to solve two  former equations.

Substituting \eqref{Xq'=} into   Eq.  \eqref{nuXi=0} (and re-denoting $q'$ by $q$) we find that this superfield equation implies the following set of four equations for the component functions

\bea\label{Psi0}
\epsilon_{IJK} Z^J\bar{Z}^K {\mathfrak{f}}_I^{(q-1)}=0\; , \qquad
\\ \nonumber \\ \label{Psi1}
\bar{\partial}_{I} {\mathfrak{f}}^{(q)}=4\sqrt{2}
\epsilon_{JKL} Z^J\bar{Z}^K  {\mathfrak{f}}^{(q-2)}_{LI}\; , \qquad
\\ \nonumber \\ \label{Psi2}
{}  \bar{\partial}_{[I} {\mathfrak{f}}_{J]}{}^{(q-1)}= 4\sqrt{2}
 Z^{[I}\bar{Z}^{J]}  {\mathfrak{f}}^{(q-3)} \; , \qquad
\\ \nonumber \\  \label{Psi3}
{}  \bar{\partial}_{[I} {\mathfrak{f}}_{JK]}{}^{(q-2)}=0  \; , \qquad
\eea

\noindent where (and below)

\be
\label{bd=ddbZ}
\bar{\partial}_I := \frac \partial {\partial \bar{Z}{}^{I}}\; ,  \qquad {\partial}_I := \frac \partial {\partial {Z}{}^{I}}\; . \qquad
\ee

\noindent By Poincar\'e lemma, Eq. \eqref{Psi3} is solved by

\be\label{fIJ=dIfJ}
 {\mathfrak{f}}^{(q-2)}_{IJ}=
{}  \bar{\partial}_{[I} {\mathfrak{f}}_{J]}{}^{(q)}~,
\ee

\noindent so that \eqref{Psi1} acquires the form

\be\label{Psi1==}
\bar{\partial}_{I} {\mathfrak{f}}^{(q)}=4\sqrt{2}
\epsilon_{JKL} Z^J\bar{Z}^K   \bar{\partial}_{[L} {\mathfrak{f}}_{I]}{}^{(q)} \qquad \Leftrightarrow \qquad \bar{\partial}_{I} \left( {\mathfrak{f}}^{(q)}+2\sqrt{2}
\epsilon_{JKL} Z^J\bar{Z}^K   {\mathfrak{f}}_{L}{}^{(q)}\right) =2\sqrt{2}
\epsilon_{JKL} Z^J\bar{Z}^K   \bar{\partial}_{L} {\mathfrak{f}}_{I}{}^{(q)} \; . \qquad
\ee

\noindent

Similarly one can find that Eq. \eqref{bnuXi=0} for superfield \eqref{Xq'=} with \eqref{fIJ=dIfJ} implies the following three equations (since  the superfield equation \eqref{bnuXi=0} has the highest component identically equal to zero)

\bea \label{Psi0=}
\mu^6\, \partial_{I}  {\mathfrak{f}}^{(q-1)}_{I} =0 \; , \qquad
\\ \nonumber \\ \label{Psi1=}
\mu^6\, \partial_{J}  \bar{\partial}_{[I} {\mathfrak{f}}^{(q)}_{J]}=- \frac 1 {2\sqrt{2} \mu^{6}}\epsilon_{IJK}Z^J \bar{Z}{}^{K}f^{(q)}\; ,
\\
 \nonumber \\
 \label{Psi2=}  \mu^6\,  \partial_{I}  {\mathfrak{f}}^{(q-3)} = - \frac 1 {\sqrt{2} \mu^{6}} Z^{[I}\bar{Z}{}^{J]} {\mathfrak{f}}^{(q-1)}_{J}\; .  \qquad
\eea

In appendix \ref{AppBornOppenhemer} we discuss the Born-Oppenheimer-like method to search for
 asymptotic form of the solution of this system of equations in the line of \cite{Frohlich:1999zf} (see also \cite{Lin:2014wka}).

\bigskip

\newpage

\setcounter{equation}{0}

\section{Conclusion and outlook}
\label{section:Conclusion}

In this paper we have initiated the program of construction of relativistic invariant (super)field theory of multiple-D$0$-brane (mD$0$) system by studying the quantization of its simplest 3D counterpart which we roughly call 3D mD$0$. This simplest system is described by the action which is the 3D counterpart of the 10D action constructed in \cite{Bandos:2018ntt} which was  considered as a candidate action for mD$0$ since i) it is formulated in terms of variables which are expected to describe mD$0$ and ii) it possesses local  fermionic $\kappa$-symmetry so that its ground state preserves a part (one half) of spacetime (superspace) supersymmetry.

Actually, in \cite{Bandos:2022uoz} we found a family of the actions possessing these properties which includes an arbitrary positive definite function ${\cal M}({\cal H})$ of the  SYM Hamiltonian ${\cal H}$ constructed from the matrix fields of the model; the model of \cite{Bandos:2018ntt} corresponds to the choice  ${\cal M}({\cal H})=m$  with the constant $m$ entering the center of mass part of the action.
As we have shown in \cite{Bandos:2022dpx}, the supersymmetric ground state configuration is the same for all the models with positive definite  ${\cal M}({\cal H})$, however the most promising  candidate is given by ${\cal M}({\cal H})$ of the form \eqref{cM=m+} since this can be obtained by dimensional reduction of 11D multiple M-wave (multiple M$0$ or mM$0$) model proposed in \cite{Bandos:2012jz}. The 3D counterpart of the 10D model with arbitrary positive definite ${\cal M}({\cal H})$ was constructed in \cite{Bandos:2021vrq} where it was shown that the model with  ${\cal M}({\cal H})$ of the form \eqref{cM=m+} can be obtained by dimensional reduction from the 4D counterpart of the 11D mM$0$ model.

As we discussed in the Introduction, the quantization of mD$0$-brane system in a coordinate representation should result in  equations of field theory of  mD$0$ which might provide us with new insights in String/M-theory. However, as such a quantization appeared to be quite complicated, we decided to begin with a toy 3D model which has the properties characteristic of mD$0$ system, moreover we restricted ourselves
by simplest of such models, with  ${\cal M}({\cal H})=m$. As one can see from the main text, already the quantization   of this toy model, which was the aim of this paper, was quite complex. It indicated some of the problems we will meet in quantization of 10D mD$0$ model(s), suggested the way to resolve these and allowed to develop and to check the methods for such resolution.

The dynamical variables of our 3D mD$0$  system is split on the relative motion and center of mass sectors.
The former is described by traceless $N\times N$ matrix fields of 3-dimensional ${\cal N}=2$ supersymmetric SU$(N)$ Yang-Mills model dimensionally reduced to $\text{d}=1$: bosonic ${\mathbb Z}(\tau)= (\bar{{\mathbb Z}}(\tau))^\dagger$ and fermionic $\mathbf{\Psi}(\tau)=(\bar{\mathbf{\Psi}}(\tau))^\dagger$, as well as auxiliary momenta ${\mathbb P}(\tau)= (\bar{{\mathbb P}}(\tau))^\dagger$ and 1-form gauge field ${\mathbb A}=\text{d}\tau {\mathbb A}_\tau (\tau)$.
The center of mass sector contains bosonic and fermionic coordinate functions $x^\mu (\tau)$,
$\theta^\alpha (\tau)$, $\bar{\theta}{}^\alpha (\tau)$, describing the embedding of the center of mass of the system  in ${\cal N}=2$ 3D flat superspace, as well as spinor moving frame variables, bosonic spinors  $w_\alpha$ and $\bar{w}_\alpha$ constrained by $\bar{w}{}^\alpha w_\alpha=i$ and  parametrizing SL$(2,{\bb R})/\text{U}(1)$ coset. Although auxiliary, these bosonic spinor  variables are essential to couple the  center of mass  sector to the relative motion one. In particular, the matrix fields of the latter are charged with respect to U$(1)$ symmetry which is a defining symmetry of  the spinor moving frame sector. As a result, their  time derivative in the action are completed to the covariant time derivatives containing, besides the 1d SU$(N)$ gauge fields, also U$(1)$  connection composed from spinor frame variables.

We have constructed Hamiltonian mechanics of this model, one of the subtleties of which is the presence of manifestly constrained spinor frame variables which we treat using the method of Cartan forms and introducing the so-called covariant momenta conjugate to these Cartan forms.
We discussed in detail the Hamiltonian approach to complex matrix fermionic fields $\mathbf{\Psi}$ and  $\bar{\mathbf{\Psi}}$ and the Dirac brackets which allows to treat $\bar{\mathbf{\Psi}}$ as momentum for ${\mathbf{\Psi}}$. The calculation of canonical momenta for center of mass  variables (including covariant momenta for spinor frame variables) and for the SU$(N)$ gauge field has lead to identification of primary constraints.
 The SU$(N)$ gauge field does not carry physical degrees of freedom, but generate the non-Abelian counterpart of the  Gauss law constraint which appears as the only secondary constraint in our system. The canonical Hamiltonian vanishes in the weak sense, i.e. when constraints are taken into account. This reflects the reparametrization invariance of our dynamical system.

We have separated constraints into first class ones, generating gauge symmetries on Poisson brackets, and  second class ones, the set of which can be split on  canonically conjugate pairs. The Poisson brackets of these first and second class constraints represent a quite nontrivial algebra which makes the application of the procedure of the Abelization of the second class constraints
quite difficult and thus hampered the way to Hamiltonian BRST quantization by Batalin-Fradkin-Fradkina method \cite{Batalin:1989dm}. Since  the Gupta-Bleuler method also failed in the sector of bosonic second class constraints (quite unexpectedly, at least for us, and even in $N=1$ case corresponding to single D-particle; see Appendix \ref{GB-failed}), the original Dirac brackets method seemed to be the only way to quantization of our simplest 3D mD$0$ system.

We  simplified the quantization by passing to the so-called analytical coordinate basis of the center of mass superspace, in which the bosonic second class constraints appeared to be resolved  thus allowing us to remove ``unphysical'' coordinates  from the phase space without passing to the Dirac brackets. To deal with the fermionic second class constraints we  applied the Gupta-Bleuler or, better to say, generalized conversion method.
In such a way we arrive at a set of effective first class constraints which, upon quantization, can be imposed on the state vector.

We used the (generalized) coordinate representation in which  momenta are represented by  differential operators
over the configuration (super)space coordinates. Some subtleties appear in the spinor moving frame sector, where one should deal with covariant momenta and a kind of covariant derivatives, as well as in the case of two Hermitian conjugate fermionic matrix fields, where we have used the so-called holonomic representation in which $\hat{\bar{\mathbf{\Psi}}}$ operator is represented by  derivative with respect to  ${\mathbf{\Psi}}$.

Imposing the quantum constraints thus represented on the state vector $\Xi$ we obtained equations of the field theory of simplest 3D counterpart of mD$0$ brane system. This field theory is defined on the superspace with additional matrix coordinates given by traceless $N\times N$ matrices ${\mathbb Z}= (\bar{{\mathbb Z}})^\dagger$ and fermionic $\mathbf{\Psi}$ and $\bar{\mathbf{\Psi}}$. The superspace also contains the center of mass sector which can be identified with Lorentz harmonic  $\text{D}=3$ ${\cal N}=2$ superspace. That is parametrized by  spinor moving frame variables and by the coordinate of usual $\text{D}=3$ ${\cal N}=2$ superspace $x^a$, $\theta^\alpha$, $\bar{\theta}{}^\alpha$, but taken in the so-called analytical basis, e.g. ${\rm x}^{(0)}= x^a \bar{w}\gamma_a w\, $, $\, \bar{\theta}{}^w=\bar{\theta}{}^\alpha w_\alpha $, etc.

Even in the case of $N=1$ corresponding to the simplest 3D counterpart of single D$0$-brane field theory, such a field theory looks unusual. We have discussed this case in detail and show how the equations for the  state vector superfield looks like in the central basis of the Lorentz harmonic superspace, in which $\Xi=\Xi(x^a,\theta,\bar{\theta};\bar{w}, w)$, as well as how one can pass to the standard 3d field theory equations in spacetime and to their solutions.

For the generic $N>1$ case we have presented the formal solution of the system of the  equations specifying dependence of the state vector on fermionic coordinates of the center of mass sector. We have also discussed the solutions with BPS-like configuration of the relative motion sector in which
this sector is decoupled from the center of mass one. Such type equations describe also the ground state of the
BFSS matrix model \cite{Banks:1996vh} at finite $N$ which was the subject of study in
\cite{Halpern:1997fv,Graf:1998bm,Frohlich:1999zf,Hoppe:2000tj,Hasler:2002wt,Lin:2014wka,Hoppe:2023vph}.

For the simplest $N=2$ case with SU$(2)$ gauge invariance we have obtained the set of field theory equations which describes the ground state for the $\text{D}=3$
${\cal N}=2$ SU$(2)$ SYM theory in Schrödinger-type representation \footnote{See \cite{Nair:2023hsy} and refs. therein for studies of 3D YM model in Schr\"{o}dinger representation.}. We have also discussed, following the method of \cite{Frohlich:1999zf}, the asymptotic form of the solution of these equations.

One of the natural directions for further study is to search for the solution of our system of (super)field equations
with coupled center of mass and relative motion sectors.

Another  interesting problem is the construction of Hamiltonian mechanics and quantization of the most general 3D counterpart of the mD$0$ system the action of which \eqref{eq:3DmD0_L} includes an arbitrary positive definite function ${\cal M}({\cal H})$ of the SYM Hamiltonian ${\cal H}$. The presence of such a function makes the straightforward elaboration of the Hamiltonian formalism quite technically involved. Probably a shortcut passes through first quantizing the 4D counterpart of multiple M-wave (mM$0$) system of \cite{Bandos:2012jz}, which was constructed in \cite{Bandos:2021vrq}. Indeed, as the dimensional reduction of this system to $\text{D}=3$ was shown to produce generic 3D counterpart of the mD$0$ system described by the Lagrangian \eqref{eq:3DmD0_L}, it is natural to expect that the dimensional reduction of the field theory of this 4D mM$0$ would give the field theory for such a generic 3D mD$0$ model \footnote{To avoid confusion, let us stress that the problem of quantization of  4D mM$0$ system is more complicated with respect to our study in the present paper as we have addressed just
the simplified version of the model obtained from  4D mM$0$  by dimensional reduction in which the positive definite function ${\cal M}({\cal H})$ \eqref{cM=m+} is chosen to be equal to the constant $m$.}.

Of course, all these studies, as well as the results  of the present paper, will be  preparatory steps toward the quantization of the 10D  mD$0$ model proposed in \cite{Bandos:2022dpx}. Such a quantization will result in a field theory in 10D superspace extended by 9 bosonic and 16 fermionic matrix coordinates the study of which might give new insights in the structure of String/M-theory.

\section*{Acknowledgments}
We are thankful to Dmitri Sorokin for useful comments and discussions, and to Padova Section of INFN  for hospitality at Padova University where these discussion took place. IB thanks the Theoretical Physics Department of CERN for hospitality during his  visit at the  intermediate period of this work. The partial support by grant PID2021-125700NB-C21 funded by Spanish MCIN/ AEI and by EU ERDF, and by the Basque Government Grant IT-1628-22 is greatly acknowledged.

\newpage

\appendix

\setcounter{equation}{0}

\def\theequation{A.\arabic{equation}}
\section{Useful Poisson bracket relations }
\label{App=PB}

Let us begin by presenting the Poisson brackets of the Gauss law constraint
$\mathbb{G}$  \eqref{bbG:=}
with the matrix variables. It is convenient  to encode these in the brackets of the trace of the product of $\mathbb{G}$ with some ``reference'' traceless matrix $\mathbb{Y}$, i.e. of $\text{tr}(\mathbb{Y}\mathbb{G}) = \mathbb{Y}_i^j \mathbb{G}^i_j$. In such a way we arrive at simple and transparent expressions
\begin{equation}
\begin{array}{ccc}
[\text{tr}(\mathbb{Y}\mathbb{G}), \mathbb{Z}]_{\text{PB}} = [\mathbb{Z}, \mathbb{Y}]~,&~&[\text{tr}(\mathbb{Y}\mathbb{G}), \bar{\mathbb{Z}}]_{\text{PB}} = [\bar{\mathbb{Z}}, \mathbb{Y}]~,
\end{array}
\end{equation}
\begin{equation}
\begin{array}{ccc}
[\text{tr}(\mathbb{Y}\mathbb{G}), \mathbb{P}]_{\text{PB}} = [\mathbb{P}, \mathbb{Y}]~,&~&[\text{tr}(\mathbb{Y}\mathbb{G}), \bar{\mathbb{P}}]_{\text{PB}} = [\bar{\mathbb{P}}, \mathbb{Y}]~,
\end{array}
\end{equation}
\begin{equation}
\begin{array}{ccc}
[\text{tr}(\mathbb{Y}\mathbb{G}), \mathbf \Psi]_{\text{PB}} = [\mathbf \Psi, \mathbb{Y}]~,&~&[\text{tr}(\mathbb{Y}\mathbb{G}), \bar{\mathbf \Psi}]_{\text{PB}} = [\bar{\mathbf \Psi}, \mathbb{Y}]
\end{array}
\end{equation}
which clearly represent the infinitesimal SU$(N)$ transformations.

These relations allow, in particular, to obtain the $\mathfrak{su}(N)$ algebra \eqref{GG=G} generated by the Gauss constraints

\begin{equation}
[\text{tr}(\mathbb{Y}\mathbb{G}), \text{tr}(\mathbb{Y}^\prime \mathbb{G})]_{\text{PB}} = \text{tr}\left([\mathbb{Y}, \mathbb{Y}^\prime]\mathbb{G}\right)~.
\end{equation}

It is also easy to check that the Poisson brackets of the Gauss law with different currents that appear in our constraints,
\begin{equation}
\begin{array}{ccccc}
{\nu} := \text{tr}(\mathbf \Psi \mathbb{P} + \bar{\mathbf \Psi}[\mathbb{Z}, \bar{\mathbb{Z}}])~,&~&{\bar{\nu}} := \text{tr}( \bar{\mathbf \Psi} \bar{\mathbb{P}} + \mathbf \Psi[\mathbb{Z}, \bar{\mathbb{Z}}])~,&~& \mathcal{B}:= \text{tr}\left(\bar{\mathbb{P}}\mathbb{Z} - \mathbb{P}\bar{\mathbb{Z}} + \dfrac i {8}{\mathbf \Psi} \bar{\mathbf \Psi} \right),
\label{eq:currentsMatrix}
\end{array}
\end{equation}
and
\begin{equation}\label{cH==B}
\begin{array}{ccc}
\mathcal{H} =    {\rm tr}\left( {\bb P} \bar{\bb P} +  [{\bb Z},  \bar{\bb Z}]^2 -
\dfrac{i}{2} {\bb Z}{ \mathbf \Psi}{ \mathbf \Psi} + \dfrac{i}{2} \bar{\bb Z} \bar{\mathbf \Psi}  \bar{\mathbf \Psi} \right)~
\end{array}
\end{equation}
vanish,

\begin{eqnarray}
&& [\text{tr}(\mathbb{Y}\mathbb{G}), \nu]_{\text{PB}} = 0~, \qquad [\text{tr}(\mathbb{Y}\mathbb{G}), \bar{\nu}]_{\text{PB}} = 0~,\qquad [\text{tr}(\mathbb{Y}\mathbb{G}), \mathcal{B}]_{\text{PB}} = 0~, \qquad  [\text{tr}(\mathbb{Y}\mathbb{G}), \mathcal{H}]_{\text{PB}} = 0~,\qquad
\end{eqnarray}
which is just the reflection of SU$(N)$ invariance of these objects.

The Poisson brackets of the fermionic and bosonic currents \eqref{eq:currentsMatrix} among themselves and with the relative motion Hamiltonian \eqref{cH==B} are

\begin{equation}
\begin{array}{ccccc}
\{\nu, \nu \}_{\text{PB}} =  -8i(\mu^{6})^2 {\rm tr} ({\bb Z}{\bb G})~,&~& \{\bar{\nu}, \bar{\nu} \}_{\text{PB}} =8i(\mu^{6})^2 {\rm tr} (\bar{{\bb Z}}{\bb G})~,&~& \{\nu, \bar{\nu} \}_{\text{PB}} =-4i \mu^6 \mathcal{H}~, \\ \\ {}
[\nu, \mathcal{B}]_{\text{PB}}= \dfrac 1 2 \mu^6 \nu,&~&[\bar{\nu}, \mathcal{B}]_{\text{PB}}=  -\dfrac 1 2 \mu^6 \bar{\nu}~,&~&
\end{array}
\end{equation}

\noindent and

\begin{equation}
\begin{array}{ccccc}
[\nu, \mathcal{H}]_{\text{PB}} = (\mu^{6})^2 {\rm tr}(\bar{\mathbf{\Psi} }{\bb G})~,&~&[\bar{\nu}, \mathcal{H}]_{\text{PB}} = -(\mu^{6})^2 {\rm tr}(\mathbf{\Psi }{\bb G})~,&~&[\mathcal{B}, \mathcal{H}]_{\text{PB}} = 0~. \\ {}
\end{array}
\end{equation}

In terms of renormalized currents and matrix fields \eqref{tcH:=}, \eqref{tB=} and \eqref{tG=G/m} the above relations simplify to

\begin{equation}
\begin{array}{ccccc}
\{\tilde{\nu}, \tilde{\nu} \}_{\text{PB}} =  -{2i}{\rm tr} ({\bb Z}\tilde{{\bb G}})~,&~& \{\bar{\tilde{\nu}}, \bar{\tilde{\nu}} \}_{\text{PB}} =2i{\rm tr} (\bar{{\bb Z}}\tilde{{\bb G}})~,&~& \{\tilde{\nu}, \bar{\tilde{\nu}} \}_{\text{PB}} =-i \tilde{\mathcal{H}}~, \\ \\ {}
[\tilde{\nu}, \tilde{\mathcal{B}}]_{\text{PB}}= \dfrac 1 2\tilde{ \nu},&~&[\bar{\tilde{\nu}}, \tilde{\mathcal{B}}]_{\text{PB}}=  -\dfrac 1 2 \bar{\tilde{\nu}}~,&~&[\tilde{\mathcal{B}}, \tilde{\mathcal{B}}]_{\text{PB}} \equiv 0~,
\\ \\
{}[\tilde{\nu}, \tilde{\mathcal{H}}]_{\text{PB}} = \dfrac 1 2  {\rm tr}(\tilde{\bar{\mathbf{\Psi} }}\tilde{{\bb G}})~,&~&[\bar{\tilde{\nu}}, \tilde{\mathcal{H}}]_{\text{PB}} = -\dfrac 1 2 {\rm tr}(\tilde{{\mathbf{\Psi} }}\tilde{{\bb G}})~,&~&[\tilde{\mathcal{B}}, \tilde{\mathcal{H}}]_{\text{PB}} = 0~. \\ {}
\end{array}
\end{equation}

The above relations were obtained by using the brackets of currents with matrix fields which are

\begin{equation}
\begin{array}{lllll}
[\nu, \mathbb{Z}]_{\text{PB}}=0~,&~&[\nu, \mathbb{P}]_{\text{PB}}=0~,&~&\{ \nu, \mathbf \Psi \}_{\text{PB}}= -{4i \mu^6 [\mathbb{Z}, \bar{\mathbb{Z}}]}~,
\\ \\ {}
[\nu, \bar{\mathbb{Z}}]_{\text{PB}}=- {\mu^6 \mathbf \Psi}~,&~&[\nu, \bar{\mathbb{P}}]_{\text{PB}}=0~,&~& \{\nu, \bar{\mathbf \Psi} \}_{\text{PB}}=-4i \mu^6 \mathbb{P},
\\ \\ {}
[\bar{\nu}, \mathbb{Z}]_{\text{PB}}=-{\mu^6 \bar{\mathbf \Psi}}~,&~&[\bar{\nu}, \mathbb{P}]_{\text{PB}}=0~,&~&\{\bar{\nu}, \mathbf \Psi \}_{\text{PB}}=-4i \mu^6 \bar{\mathbb{P}} ~,
\\ \\ {}
[\bar{\nu}, \bar{\mathbb{Z}}]_{\text{PB}}=0~,&~&[\bar{\nu}, \bar{\mathbb{P}}]_{\text{PB}}=0~,&~&\{\bar{\nu}, \bar{\mathbf{\Psi}}\}_{\text{PB}}= {-4i \mu^6 [\mathbb{Z}, \bar{\mathbb{Z}}]}~,
\\ \\ {}
[\mathcal{B}, \mathbb{Z}]_{\text{PB}}=-\mu^6 \mathbb{Z}~,&~&[\mathcal{B}, \mathbb{P}]_{\text{PB}}=-\mu^6 \mathbb{P}~,&~&[ \mathcal{B}, \mathbf{\Psi} ]_{\text{PB}}=\;  \dfrac 1 2 \mu^6 \mathbf \Psi ~,
\\ \\ {}
[\mathcal{B}, \bar{\mathbb{Z}}]_{\text{PB}}=\mu^6 \bar{\mathbb{Z}}~,&~&[\mathcal{B}, \bar{\mathbb{P}}]_{\text{PB}}=\mu^6 \bar{\mathbb{P}}~,&~&[ \mathcal{B}, \bar{\mathbf \Psi} ]_{\text{PB}}= -\dfrac 1 2 \mu^6 \bar{\mathbf \Psi} ~.
\end{array}
\end{equation}

\setcounter{equation}{0}

\def\theequation{B.\arabic{equation}}
\section{Problems with Gupta-Bleuler quantization scheme for massive particle in moving frame formulation }
\label{GB-failed}

Let us consider the system of the first class constraints \eqref{Phi0=0}, \eqref{U0=0} and second class constraints  \eqref{eqs:frak_d}, \eqref{eqs:Phi} and set to zero all the fermionic and matrix fields. Then such a system of constraints describes, in the frame of Hamiltonian approach, just a massive bosonic particle in its spinor moving frame formulation.

In this appendix we show that, surprisingly, the quantization of such a simple system by the Gupta-Bleuler method in its canonical form fails \footnote{Of course, the quantization of the same system in the frame of standard formulation, with the action given by the bosonic limit of \eqref{SD0=A+L}, can be easily performed and gives an expected result: theory of free scalar field obeying the Klein-Gordon equation. We however are interested in quantization of this simple system  in its moving frame formulation as this arises as a pure bosonic limit of our 3D mD0 system when we set to zero all the matrix fields. }.

Imposing the quantum versions of the first class constraints
\eqref{Phi0=0}, \eqref{U0=0} and two of second class constraints  \eqref{eqs:frak_d}, \eqref{eqs:Phi} which commute  and are not related by Hermitian  conjugation among themselves, and choosing coordinate representation for the state vector, $\Xi_0=\Xi_0(x^a,\bar{w},w)$, we can arrive at the following system of differential equations

\bea
\label{u0dxXi0=}
(iu^{(0)a}\partial_{a}+m)\Xi_0 =0  \; , \qquad  \\ \nonumber \\
\label{D0Xi0=}
\left({\bb D}^{(0)}  -q\right) \Xi_0=0 \; , \qquad
\\ \nonumber \\
\label{barbbDXi0=}
\bar{{\bb D}}  \Xi_0=0 \; , \qquad
\\ \nonumber \\ \label{budxXi0=0}
\bar{u}{}^{a}\partial_{a}\Xi_0 =0  \; . \qquad  \\ \nonumber
\eea

The $x^a$-dependence of the state vector is fixed by Eqs. \eqref{u0dxXi0=} and \eqref{budxXi0=0} to be

\be\label{Xi0=exp.chi}
\Xi_0 = e^{imx^a{u}_a^{(0)} } \, \chi_0 ( x^a\bar{u}_a, \bar{w}_\alpha,  w_\alpha) \; . \qquad
\ee
Using this we find from \eqref{barbbDXi0=} that $\chi_0 ( x^a\bar{u}_a, \bar{w}_\alpha,  w_\alpha)$ obeys

\be
\bar{{\mathbb D}} \chi_0 ( x^a\bar{u}_a, \bar{w}_\alpha,  w_\alpha):=
\bar{w}_\alpha \frac {\partial} {\partial {w}_\alpha}
\chi_0 ( x^a\bar{u}_a, \bar{w}_\alpha,  w_\alpha) \; =- imx^a\bar{u}_a\, \chi_0 ( x^a\bar{u}_a, \bar{w}_\alpha,  w_\alpha) \; .
\ee
The second form of the l.h.s. of this equation makes manifest that its nontrivial solution, if existed, would be
$\chi_0 = e^{-imx^a{u}_a^{(0)}}\tilde{\chi}_0 ( x^a\bar{u}_a, \bar{w}_\alpha, w_\alpha)$, but this form contradicts to the fact that in \eqref{Xi0=exp.chi} $\chi_0 = \chi_0 ( x^a\bar{u}_a, \bar{w}_\alpha,  w_\alpha)$ is independent on $x^a{u}_a^{(0)}$.

This example (actually the only presently known to the authors) indicates that the selfconsistency of the Gupta-Bleuler quantization method is not guaranteed and thus its should be used  with precaution.

Curiously, if we modify the above procedure by imposing besides \eqref{u0dxXi0=} and \eqref{D0Xi0=} the complex (Hermitian)  conjugate but not canonically conjugate pair of second class constraints
 \eqref{eqs:frak_d}, which commute among themselves, we  arrive at the system

\bea
\label{u0dxXi0==}
(iu^{(0)a}\partial_{a}+m)\Xi_0 =0  \; , \qquad  \\ \nonumber \\
\label{D0Xi0==}
\left({\bb D}^{(0)}  -q\right) \Xi_0=0 \; , \qquad
\\ \nonumber \\
\label{udxXi0==0}
u^{a}\partial_{a}\Xi_0 =0  \; , \qquad
\\ \nonumber \\ \label{budxXi0==0}
\bar{u}{}^{a}\partial_{a}\Xi_0 =0  \; \qquad  \\ \nonumber
\eea
which does have a nontrivial solution which is equivalent to the solution of the massive Klein-Gordon equation.

Such a modification clearly does not feet in the standard Gupta-Bleuler prescription. However, as we show in the main text and also in Appendix \ref{App=D0}, its result is equivalent to the quantization with the method of Dirac brackets, which is further simplified and reduced to the explicit resolution of the constraints in the analytical basis. Such Dirac brackets/anaytical basis quantization  gives a consistent field theory with nontrivial solution which, in the simplest case considered in this Appendix, coincides with the result of quantization in the standard formulation of the massive particle.


\bigskip

\setcounter{equation}{0}
\def\theequation{C.\arabic{equation}}
\section{Hamiltonian formalism of the 3D single D0-brane in the so-called analytical basis}
\label{App=D0}

The Lagrangian of 3D counterpart of single D0-brane in spinor moving frame formulation reads
({\it cf. } \eqref{SD0=A+L})

\begin{equation}\label{LD0=}
 \mathcal{L}_{\text{D}0} =\text{d} \tau {L}_{\text{D}0} = \, m  {\rm E}^{(0)} + m (\text{d}\theta^\alpha \bar{\theta}_\alpha - \text{d}\bar{\theta}^\alpha \theta_\alpha)
\end{equation}
where ${\rm E}^{(0)}$ is defined in \eqref{E0=} with  \eqref{Ea=}. The dynamical variables in this Lagrangian 1-form  are coordinate functions which define parametrically  the embedding of worldline in flat superspace
\begin{equation}
\begin{array}{ccccccc}
\mathcal{W}^1  \subset \Sigma^{\left(3\left. \right| 4 \right)}: &~& x^a = x^a (\tau), &~&  \theta^\alpha= \theta^\alpha(\tau),&~& \bar{\theta}^\alpha =  \bar{\theta}^\alpha(\tau)
\end{array}~
\end{equation}
and spinor frame variables \eqref{w=}, $w_\alpha(\tau)$ and its c.c. $\bar{w}_\alpha(\tau)$ obeying \eqref{bww=i}.

In terminology of \cite{Bandos:1990ji}, originating in the seminal papers on Harmonic superspaces \cite{Galperin:1984av,Galperin:1984bu,Sokatchev:1985tc,Sokatchev:1987nk}, these latter variables can be called Lorentz harmonics thus reflecting the fact that they parametrize the double covering of 3D Lorentz group SL$(2,{\bb R})$ and, if the requirement of the invariance under  U$(1)$ gauge transformations acting on $w$ and $\bar{w}$ is imposed, as homogeneous coordinates of the coset  SL$(2,{\bb R})/\text{U}(1)$.
Furthermore, the configuration space of our dynamical system, this is to say the target space of the spinor moving frame formulation of the 3D D$0$-brane,   can be called, following the above terminology, {\it Lorentz harmonic superspace}, $\Sigma^{(3+3|4)}$ . This is the superspace with coordinates
\be
\Sigma^{(3+3|4)} = \{ (x^a,\theta^\alpha , \bar{\theta}^\alpha , w_\alpha , \bar{w}_\alpha )\} =: \{ (Z^M , w_\alpha , \bar{w}_\alpha )\} \; , \qquad \bar{w}^\alpha{w}_\alpha =i\;
\ee
and in the spinor moving frame approach we are considering D$0$-brane as a  particle moving in this extended and enlarged $\text{D}=3$ superspace. Here ``extended'' refers to ${\cal N}=2$ supersymmetry and ``enlarged'' reflects the presence of additional bosonic spinor (spinor moving frame or Lorentz harmonic) coordinates.

This perspective is convenient, in particular, because it suggests the possibility to choose an alternative coordinate basis of the Lorentz harmonic superspace, the analytical coordinate  basis
\be
\Sigma^{(3+3|4)} = \{ ({\rm x}{}^{(0)},{\rm x}_A,\bar{{\rm x}}_A\; ,  \theta^w , {\theta}^{\bar{w}} , \bar{\theta}^{{w}} ,\bar{\theta}^{\bar{w}}  ; w_\alpha , \bar{w}_\alpha )\} =:\{ ({\rm Z}{}^{(M)}_{An}, \; w_\alpha , \bar{w}_\alpha )\} \; , \qquad \bar{w}^\alpha{w}_\alpha =i\;
\ee
defined  by (see \cite{Sokatchev:1985tc,Sokatchev:1987nk,Bandos:1990ji})
\bea\label{xA=D0}
{\rm x}{}^{(0)}:=x^au_a^{(0)}\; , \qquad & \fbox{${\rm x}_A := {\rm x}-2i \theta^w \bar{\theta}^{{w}}$}=x^au_a-2i \theta^w \bar{\theta}^{{w}}\; , \qquad & \fbox{$\bar{{\rm x}}_A:= \bar{{\rm x}} +2i{\theta}^{\bar{w}} \bar{\theta}^{\bar{w}}$}=x^a \bar{u}_a+2i {\theta}^{\bar{w}} \bar{\theta}^{\bar{w}}\; , \\
\nonumber
\\ \label{thw=D0}
& \theta^w := \theta^\alpha w_\alpha~, \qquad & \theta^{\bar{w}}:= \theta^\alpha \bar{w}_\alpha~, \qquad \\
\nonumber \\ \label{bthw=D0}
& \bar{\theta}^w := \bar{\theta}^\alpha w_\alpha~, \qquad & \bar{\theta}^{\bar{w}}:= \bar{\theta}^\alpha \bar{w}_\alpha~. \qquad
\eea


The definition of ${\rm x}_A=(\bar{{\rm x}}_A)^*$ is designed in such a way that in the analytical basis the Lagrangian form \eqref{LD0=} simplifies to

\begin{eqnarray}
 \mathcal{L}_{\text{D}0} &=& m \text{d}{\rm x}^{(0)} - im f\; \bar{{\rm x}}_A  + im \bar{f}\; {\rm x}_A  -4ma\theta^{\bar{w}} \bar{\theta}^w - 2im (\text{d}\theta^{\bar{w}} \bar{\theta}^{w} + \text{d}\bar{\theta}^{w} \theta^{\bar{w}})= \\ \nonumber \\
 &=&  m \text{d}{\rm x}^{(0)} - 2im (\text{d}\theta^{\bar{w}}-ia\theta^{\bar{w}})\, \bar{\theta}^{w}  - 2im (\text{d}\bar{\theta}^{w}+ia\bar{\theta}^{w} )\, \theta^{\bar{w}}  - im f\; \bar{{\rm x}}_A  + im \bar{f}\; {\rm x}_A \; .
\end{eqnarray}

In the main text we will need also separate expressions for the kinetic and WZ terms of single 3D D$0$ action,

\begin{eqnarray}\label{E0=an}
& {\rm E}^{0} =&  \text{d}{\rm x}^{(0)} - i(\text{d}\theta^{{w}}+ia\theta^{{w}})\, \bar{\theta}^{\bar{w}}- i(\text{d}\theta^{\bar{w}}-ia\theta^{\bar{w}})\, \bar{\theta}^{w}
  -i (\text{d}\bar{\theta}^{w}+ia\bar{\theta}^{w}) \theta^{\bar{w}} -i (\text{d}\bar{\theta}^{\bar{w}}-ia\bar{\theta}^{\bar{w}}) \theta^{w}+i f\; \bar{{\rm x}} -i \bar{f}\; {\rm x}  \; ,  \qquad  \\  && \nonumber \\ \label{WZ=an}
&\text{d}\theta^\alpha \bar{\theta}_\alpha& -\text{d}\bar{\theta}^\alpha {\theta}_\alpha   = 2f\;\theta^{\bar{w}} \bar{\theta}^{{\bar{w}}}  + 2 \bar{f}\; \theta^{{w}} \bar{\theta}^{{w}} + i(\text{d}\theta^{{w}}+ia\theta^{{w}})\, \bar{\theta}^{\bar{w}}- i(\text{d}\theta^{\bar{w}}-ia\theta^{\bar{w}})\, \bar{\theta}^{w} - \nonumber \\ \nonumber \\  && {}\hspace{4.8cm}
  -i (\text{d}\bar{\theta}^{w}+ia\bar{\theta}^{w}) \theta^{\bar{w}} +i (\text{d}\bar{\theta}^{\bar{w}}-ia\bar{\theta}^{\bar{w}}) \theta^{w} \; .  \qquad
\end{eqnarray}

The canonical Hamiltonian $H_0$ in terms of the coordinates of the analytical basis is defined by

\begin{equation}
\text{d}\tau H_0 = \text{d}\text{x}^{(0)}p^{(0)} + \text{d}{\rm x}_A \bar{p} + \text{d}\bar{\text{x}}_A p + \text{d}\theta^w \Pi^\theta_w + \text{d}\theta^{\bar{w}} \Pi^\theta_{\bar{w}} + \text{d} \bar{\theta}^w \bar{\Pi}^{\bar{\theta}}_w + \text{d} \bar{\theta}^{\bar{w}} \bar{\Pi}^{\bar{\theta}}_{\bar{w}} + ia \tilde{\mathfrak{d}}^{(0)} + i f \bar{\tilde{\mathfrak{d}}} - i \bar{f} \tilde{\mathfrak{d}} - \mathcal{L}_{\text{D}0}
\end{equation}
where the momenta conjugate to  the coordinate functions, i.e. having the nonvanishing Poisson brackets

\begin{equation}
\begin{array}{ccccc}
\left[p^{(0)}, \text{x}^{(0)} \right]_{\text{PB}} = -1~,&~&\left[p, \bar{{\rm x}}_A \right]_{\text{PB}} = -1~,&~&\left[\bar{p}, {\rm x}_A \right]_{\text{PB}} = -1~.
\end{array}
\end{equation}

\noindent
are defined by \eqref{up=-2p},

\begin{equation}\label{p-pAnA}
\begin{array}{ccccc}
p^{(0)}:= u^{a(0)}p_a~,&~&p := -\dfrac{1}{2}u^a p_a~,&~&\bar{p} := -\dfrac{1}{2}\bar{u}^a p_a
\end{array}
\end{equation}

Similarly, the momenta conjugate to the fermionic coordinate functions of the analytical basis, i.e. obeying

\begin{equation}
\begin{array}{ccccccc}
{}\{\Pi^\theta_w , \theta^w\}_{\text{PB}} = -1~,&~& \{\Pi^\theta_{\bar{w}} , \theta^{\bar{w}}\}_{\text{PB}} = -1~, &~& \{\bar{\Pi}^{\bar{\theta}}_w , \bar{\theta}^w\}_{\text{PB}} = -1~, &~& \{\bar{\Pi}^{\bar{\theta}}_{\bar{w}} , \bar{\theta}^{\bar{w}}\}_{\text{PB}} = -1~.
\end{array}
\end{equation}

\noindent are related to the ones of the central basis by \eqref{Pi-PiAn}

\begin{equation}\label{Pi-PiAnA}
\begin{array}{ccc}
\Pi^\theta_{w}:= -i\bar{w}^\alpha \Pi_\alpha+2i \bar{\theta}^w\bar{p}~,&~& \Pi^\theta_{\bar{w}}:= iw^\alpha \Pi_\alpha - 2i \bar{\theta}^{\bar{w}}p~,\\
~\\
\bar{\Pi}^{\bar{\theta}}_{w}:= -i\bar{w}^\alpha \bar{\Pi}_\alpha -2i \theta^w \bar{p} ~,&~& \bar{\Pi}^{\bar{\theta}}_{\bar{w}}:= iw^\alpha \bar{\Pi}_\alpha+ 2i \theta^{\bar{w}}p~.
\end{array}
\end{equation}

The covariant momenta of the analytical basis are related to these of the central basis by

 \begin{equation}
\begin{array}{lll}
&~& \mathfrak{d}  := \tilde{\mathfrak{d}} + (\text{x}_A+2i \theta^{w}  \bar{\theta}^{\bar{w}} ) p^{(0)} + 2(\text{x}^{(0)}+i \theta^{w}  \bar{\theta}^{\bar{w}}+i \theta^{\bar{w}}  \bar{\theta}^{{w}}) p + \theta^{w} \Pi_{\bar{w}}^\theta +  \bar{\theta}^{w} \bar{\Pi}_{\bar{w}}^{\bar{\theta}}~,\\
~\\ &~& \bar{\mathfrak{d}} := \bar{\tilde{\mathfrak{d}}} +(\bar{\text{x}}_A -2i{\theta}^{\bar{w}}\bar{\theta}^{\bar{w}}) p^{(0)} + 2(\text{x}^{(0)}-i{\theta}^{{w}}\bar{\theta}^{\bar{w}}-i{\theta}^{\bar{w}}\bar{\theta}^{{w}})\bar{p} + \theta^{\bar{w}} \Pi_{w}^\theta +  \bar{\theta}^{\bar{w}} \bar{\Pi}_{w}^{\bar{\theta}}~,\\
~\\
&~&{\mathfrak{d}}^{(0)} := \tilde{\mathfrak{d}}^{(0)}- 2\text{x}_A \bar{p} + 2\bar{\text{x}}_Ap+ \theta^{\bar{w}} \Pi_{\bar{w}}^\theta + \bar{\theta}^{\bar{w}} \bar{\Pi}_{\bar{w}}^{\bar{\theta}}- \theta^{w} \Pi^{\theta}_w - \bar{\theta}^w \bar{\Pi}^{\bar{\theta}}_w~.
\end{array}
\end{equation}
This implies that
\begin{equation}
\begin{array}{lll}
{}[\tilde{\mathfrak{d}}^{(0)} , {\rm Z}_{An}^{(M)}]_{\text{PB}} =0 \; , \qquad {}[\tilde{\mathfrak{d}}, {\rm Z}_{An}^{(M)}]_{\text{PB}} =0 \; , \qquad  {}[\bar{\tilde{\mathfrak{d}}}, {\rm Z}_{An}^{(M)}]_{\text{PB}} =0 \; , \qquad
\end{array}
\end{equation}
while
\begin{equation}
\begin{array}{lll}
{}[{\mathfrak{d}}^{(0)} , Z^{M}]_{\text{PB}} =0 \; , \qquad {}[{\mathfrak{d}},  Z^{M}]_{\text{PB}} =0 \; , \qquad  {}[\bar{{\mathfrak{d}}}, Z^{M}]_{\text{PB}} =0 \; . \qquad
\end{array}
\end{equation}

The calculation of all the canonical and covariant momenta result in the (primary) constraints

\begin{eqnarray}\label{Phi=An}
\Phi^{(0)}:= p^{(0)} - m \approx 0\; ,\qquad & \qquad    \bar{\Phi}:= \bar{p}  \approx 0\; ,\qquad & \Phi:= p  \approx 0\; ,\qquad \\ \nonumber  \\ \label{frakd=An}
\tilde{\mathfrak{d}}^{(0)} -4im \theta^{\bar{w}} \bar{\theta}^w   \approx 0~, \qquad   &  \qquad
\tilde{\mathfrak{d}}+ m {\rm x}_A  \approx 0~,   \qquad
 & \bar{\tilde{\mathfrak{d}}}+ m \bar{{\rm x}}_A  \approx 0~,
 \\ \nonumber
 \\ \label{dw=An} d_w:= \Pi^\theta_w \approx 0~, \qquad & \qquad d_{\bar{w}}:= \Pi^\theta_{\bar{w}}+2i m \bar{\theta}^w \approx 0~, \qquad & \\ \nonumber
 \\ \label{bdw=An} \bar{d}_{\bar{w}} := \bar{\Pi}^{\bar{\theta}}_{\bar{w}} \approx 0~,  \qquad
  & \qquad \bar{d}_w := \bar{\Pi}^{\bar{\theta}}_w + 2im\theta^{\bar{w}} \approx 0~. \qquad  &  \\ \nonumber
\end{eqnarray}
We have written these in three columns in such a way that the lines correspond to different sectors of dynamical  variables, from  the second and third columns one can read the pairs of conjugate  second class constraints, which in the case of bosonic constraints are explicitly solved, while the first column contains a prototypes of the first class constraints.

The canonical Hamiltonian vanishes on the surface of primary constraints, this is to say
\be
H_0\approx 0\; .
\ee

The true first class constraints,

\be\label{1st=An}
 \Phi^{(0)}\approx 0, \qquad d_w,  \approx 0, \qquad \bar{d}_{\bar{w}} \approx 0, \qquad  \text{and}\qquad \tilde{\tilde{U}}{}^{(0)} \approx 0, \qquad
\ee
are given by the first equations in \eqref{Phi=An}, \eqref{dw=An}, \eqref{bdw=An} and by the sum of the first equation in \eqref{frakd=An} with certain linear combination of the second class constraints,

\bea\label{U0=An}
\tilde{\tilde{U}}{}^{(0)} = \tilde{\mathfrak{d}}^{(0)} -4im \theta^{\bar{w}} \bar{\theta}^w  -2\text{x}_A\bar{p}+2\bar{\text{x}}_Ap-\bar{\theta}^{w}\bar{d}_w+ {\theta}^{\bar{w}}d_{\bar{w}} =\qquad  \nonumber \\ \nonumber \\ = \tilde{\mathfrak{d}}^{(0)}   -2\text{x}_A\bar{p}+2\bar{\text{x}}_Ap-\bar{\theta}^{w}\bar{\Pi}^{\bar{\theta}}_w+ {\theta}^{\bar{w}}{\Pi}^{{\theta}}_{\bar{w}} \approx 0 .  \;
\eea

Also the set of second class constraints, to put in the canonical form their algebra, should  be redefined a bit and written as
\be\label{2nd=class}
{\text{2nd class}}: \qquad \qquad \begin{cases} \bar{\Phi}=\bar{p}\approx 0 \cr \tilde{\tilde{\mathfrak{d}}}\approx 0\end{cases}\; , \qquad \begin{cases} {\Phi}={p}\approx 0 \cr \bar{\tilde{\tilde{\mathfrak{d}}}} \approx 0\end{cases}\; ,
\qquad \begin{cases} d_{\bar{w}}\approx 0 \cr \bar{d}_w\approx 0\end{cases}\; , \qquad  \ee
 where
\be\label{ttfrakd=}
\tilde{\tilde{\mathfrak{d}}}:= \tilde{\mathfrak{d}}+ m {\rm x}_A+2i {\theta}^{\bar{w}} \bar{\theta}^{w}p\; , \qquad \bar{\tilde{\tilde{\mathfrak{d}}}}:= \bar{\tilde{\mathfrak{d}}}+ m \bar{{\rm x}}_A -2i {\theta}^{\bar{w}} \bar{\theta}^{w}\bar{p} ~.   \qquad
\ee

The algebra of the first class constraints \eqref{1st=An} is Abelian (this is to say all the Poisson brackets of the first class constraints vanish). The only nonvanishing brackets of the first class constraints with the second class constraints appear when the first class constraint is given by the U(1) generator \eqref{U0=An},
\be
{}[ \tilde{\tilde{U}}{}^{(0)} \, , \, \left(\begin{matrix}\bar{p}\cr \tilde{\tilde{\mathfrak{d}}}\cr  {p} \cr \bar{\tilde{\tilde{\mathfrak{d}}}}
\cr  d_{\bar{w}} \cr \bar{d}_w \end{matrix}\right)]_{{\text{PB}}} =  \, \left(\begin{matrix}-2\bar{p}\cr 2\tilde{\tilde{\mathfrak{d}}}\cr  2{p} \cr -2\bar{\tilde{\tilde{\mathfrak{d}}}}
\cr d_{\bar{w}} \cr -\bar{d}_w \end{matrix}\right)\, .
\ee
The coefficients in the r.h.s. represent the charges of second class constraints with respect to U$(1)$ gauge symmetry.

Finally, the nonvanishing brackets of the second class constraints are
\be
 {}[\bar{p}\, , \,  \tilde{\tilde{\mathfrak{d}}} ]_{{\text{PB}}}= m\; , \qquad  [ {p} \, , \,  \bar{\tilde{\tilde{\mathfrak{d}}}}
]_{{\text{PB}}}=m \; , \qquad \{ d_{\bar{w}},  \bar{d}_w \}_{{\text{PB}}}=4im\; ,
\ee
reflecting their second class nature, and weakly vanishing
\bea
 {}[ \tilde{\tilde{\mathfrak{d}}}\, , \,  \bar{\tilde{\tilde{\mathfrak{d}}}} ]_{{\text{PB}}}= \tilde{\tilde{U}}{}^{(0)} +2\text{x}_A\bar{p}-2\bar{\text{x}}_Ap+\bar{\theta}^{w}\bar{d}_w- {\theta}^{\bar{w}}d_{\bar{w}}\;\approx 0 ,  \qquad \\ \nonumber \\
  {}[ \tilde{\tilde{\mathfrak{d}}}\, , \, d_{\bar{w}} ]_{{\text{PB}}}= 2i\bar{\theta}^{w}p  \approx 0 \; , \qquad {}[ \tilde{\tilde{\mathfrak{d}}}\, , \, d_{\bar{w}} ]_{{\text{PB}}}= -2i {\theta}^{\bar{w}}{p}  \approx 0  \; , \qquad
   \\ \nonumber \\
  {}[ \bar{\tilde{\tilde{\mathfrak{d}}}}\, , \, d_{\bar{w}} ]_{{\text{PB}}}= -2i\bar{\theta}^{w}\bar{p}  \approx 0 \; ,  \qquad  {}[ \bar{\tilde{\tilde{\mathfrak{d}}}}\, , \, d_{\bar{w}} ]_{{\text{PB}}}= 2i {\theta}^{\bar{w}}\bar{p}  \approx 0 \; . \qquad
\eea

Resuming, the algebra of the constraints is represented in the following Table \ref{table:singleD0}

\begin{table}[h!]
\resizebox{\textwidth}{!}{
\begin{tabular}{c||cccc||cccccc}
 $[...,... \}_{\text{PB}}$
& $\Phi^{(0)}$&  $\tilde{\tilde{U}}{}^{(0)}$ & $d_w$ & $\bar{d}_{\bar{w}}$ &
$\bar{p}$ & $\tilde{\tilde{\mathfrak{d}}}$ & ${p}$ & $\tilde{\bar{\tilde{\mathfrak{d}}}}$ & $d_{\bar{w}}$ & $ \bar{d}_w$ \\
 \hline \hline
$\Phi^{(0)}$& 0 & 0 & 0 & 0 & 0 & 0 & 0 & 0 & 0 & 0 \\
 $\tilde{\tilde{U}}{}^{(0)}$ & 0 & 0 & 0 & 0 & $-2\bar{p}$ & $2\tilde{\tilde{\mathfrak{d}}}$ & $2{p}$ & $-2\bar{\tilde{\tilde{\mathfrak{d}}}}$ & $d_{\bar{w}}$ & $- \bar{d}_w$\\
 $d_w$ &  0 & 0 & 0 & 0 & 0 & 0 & 0 & 0 & 0 & 0 \\ $\bar{d}_{\bar{w}}$ &  0 & 0 & 0 & 0 & 0 & 0 & 0 & 0 & 0 & 0 \\
 \hline \hline
$\bar{p}$ &  0 & 2$\bar{p}$ & 0 & 0 & 0 & $m$ & 0 & 0 & 0 & 0 \\
$\tilde{\tilde{\mathfrak{d}}}$  &  0 & $-2\tilde{\tilde{\mathfrak{d}}}$ & 0 & 0 & $-m$  & 0 & 0 & \fbox{$\begin{matrix}\tilde{\tilde{U}}{}^{(0)} +\bar{\theta}^{w}\bar{d}_w - {\theta}^{\bar{w}}d_{\bar{w}}+ \cr+2 \text{x}_A{p} -2\bar{\text{x}}_Ap\end{matrix}$} & $2i\bar{\theta}{}^w\bar{p}$ & $-2i{\theta}{}^{\bar{w}}{p} $
\\ ${p}$ &  0 & -2$p$ & 0 & 0 & 0 & 0 & 0 & m & 0 & 0  \\ $\bar{\tilde{\tilde{\mathfrak{d}}}}$ &  0 & $2\bar{\tilde{\tilde{\mathfrak{d}}}}$ &0 & 0 & 0 &  \fbox{$\begin{matrix}-\tilde{\tilde{U}}{}^{(0)} -\bar{\theta}^{w}\bar{d}_w +{\theta}^{\bar{w}}d_{\bar{w}}- \cr -2\text{x}_A\bar{p} +2\bar{\text{x}}_Ap\end{matrix}$} & -m & 0 & $-2i\bar{\theta}{}^w\bar{p}$ & $2i{\theta}{}^{\bar{w}}\bar{p}$ \\ $d_{\bar{w}}$ &  0 & -$d_{\bar{w}}$ & 0 & 0 & 0 & 0 & 0 & 0 & 0 & $-4im$\\ $ \bar{d}_w$ &  0 &  $ \bar{d}_w$& 0 & 0 & 0 & 0 & 0 & 0 & $-4im$ & 0\\
 \hline \hline
\end{tabular}}
\caption{Closed algebra of the first and second class constraints of single D0-brane system.}
\label{table:singleD0}
\end{table}

Now the bosonic second class constraints in \eqref{2nd=class}  are explicitly resolved with respect to ${\rm x}_A$, $\bar{{\rm x}}_A$ and their conjugate momenta. Thus a consistent way is to use these to reduce the phase space of our dynamical system before quantization. An equivalent way is provided by (the classical counterpart of)  the generalized/deformed Gupta-Bleuler procedure which implies that  we impose on quantum system only half of the second class constraints $p\approx 0$ and $\bar{p}\approx 0$. At the classical level this implies that
we omit the conjugate constraints $\bar{\tilde{\tilde{\mathfrak{d}}}}\approx 0$ and $\tilde{\tilde{{\mathfrak{d}}}}\approx 0$ from the consideration thus converting their conjugate $p\approx 0$ and $\bar{p}\approx 0$ into the first class constraints \footnote{Notice that this is a deformed version of Gupta-Bleuler quantization. As we have already written in the main text and shown in Appendix \ref{GB-failed}, the canonical Gupta-Bleuler approach to the bosonic second class constraints of our system fails to produce the correct result equivalent to the Dirac bracket quantization.}.

We prefer this later treatment of the bosonic second class constraints because it will be in consonance with the (canonical) Gupta-Bleuler treatment of the fermionic second class constraints in \eqref{2nd=class}.
These are complex conjugates and, after quantization, we will impose on the state vector the quantum counterpart of only one of two fermionic second class constraints, $\bar{d}_w\approx 0$. This is tantamount to omitting, at the classical level, the conjugate
$d_{\bar{w}}\approx 0$ constraints, which converts $\bar{d}_w\approx 0$ into the first class constraints. The resulting algebra of the effective first class constraints appearing as a result of the  implementation of the above classical counterpart of the generalized/deformed Gupta-Bleuler procedure is presented in Table \ref{table:singleD0eff}.

\begin{table}[h!]
\hspace{11.0em}
\begin{tabular}{c||ccccccc}
 $[...,... \}_{\text{PB}}$
& $\Phi^{(0)}$&  $\tilde{\tilde{U}}{}^{(0)}$ & $d_w$ & $\bar{d}_{\bar{w}}$ &
$\bar{p}$ &  ${p}$ &  $ \bar{d}_w$ \\
 \hline \hline
$\Phi^{(0)}$& 0 & 0 & 0 & 0 &  0 & 0 & 0  \\
 $\tilde{\tilde{U}}{}^{(0)}$ & 0 & 0 & 0 & 0 & $-2\bar{p}$ & $2{p}$ &  $- \bar{d}_w$\\
 $d_w$ &  0 & 0 & 0 & 0 & 0 & 0 & 0 \\ $\bar{d}_{\bar{w}}$ &  0 & 0 & 0 & 0 & 0 & 0 & 0 \\
 \hline \hline
$\bar{p}$ &  0 & 2$\bar{p}$ & 0 & 0 & 0 & 0 & 0  \\
 ${p}$ &  0 &  $-2p$ & 0 & 0 & 0 & 0 & 0 \\  $ \bar{d}_w$ &  0 &  $ \bar{d}_w$& 0 & 0 & 0 & 0 & 0 \\
 \hline \hline
\end{tabular}
\caption{Closed algebra of the effective first  class constraints of single D0-brane system.}
\label{table:singleD0eff}
\end{table}

To quantize our dynamical system in the supercoordinate representation we represent our effective first class constraints by differential operators,

\bea \label{hPhi0=}
\hat{\Phi}^{(0)}= -i\partial_{\text{x}^{(0)}}-m \; , \qquad \\ \nonumber \\ \label{hPhi=}
\hat{\bar{\Phi}}= -i\partial_{\text{x}_A}\; , \qquad  \hat{{\Phi}}= -i\bar{\partial}_{\bar{\text{x}}_A}\; , \qquad \\ \nonumber \\ \label{hU0=}
\hat{\tilde{\tilde{U}}}{}^{(0)}=-i\left({\bb D}^{(0)}+4m\theta^{\bar{w}} \bar{\theta}^w  -2\text{x}_A\partial_{\text{x}_A}+2\bar{\text{x}}_A \partial_{\bar{\text{x}}_A}-\bar{\theta}^{w} \bar{d}_w+ {\theta}^{\bar{w}}d_{\bar{w}} -q\right) = \qquad \nonumber \\ \nonumber \\
=-i\left({\bb D}^{(0)}-2\text{x}_A\partial_{\text{x}_A}+2\bar{\text{x}}_A \partial_{\bar{\text{x}}_A}-\bar{\theta}^{w}\partial_{\bar{\theta}^{{w}}}+ {\theta}^{\bar{w}}\partial_{{\theta}^{\bar{w}}}-q \right) \; , \qquad
\\ \label{hdw=}
\hat{d}_w=-i {\partial}_{{\theta}{}^{w}}  \; , \qquad
\\ \nonumber \\ \label{hbdbw=}
\hat{\bar{d}}_{\bar{w}}=-i \bar{\partial}_{\bar{\theta}{}^{\bar{w}}} \; , \qquad \\ \nonumber \\ \label{hbdw=}
\hat{\bar{d}}_w=-i\left(\bar{\partial}_{\bar{\theta}{}^{w}}  -2m{\theta}^{\bar{w}}\right)
\qquad \eea
and impose them on the state vector.
In \eqref{hPhi0=}-\eqref{hbdw=}  the derivatives and covariant derivatives are defined as follows

\bea
\partial_{\text{x}^{(0)}} =\frac {\partial} {\partial \text{x}^{(0)}} \; , \qquad \partial_{{\text{x}_A}} =\frac {\partial} {\partial {\text{x}_A}} \; , \qquad \partial_{\bar{\text{x}}_A} =\frac {\partial} {\partial \bar{\text{x}}_A} \; , \qquad \\
\nonumber \\
{\bb D}^{(0)}= \bar{w}_\alpha \frac {\partial} {\partial \bar{w}_\alpha} - {w}_\alpha \frac {\partial} {\partial {w}_\alpha}  \; , \qquad
{\bb D} = w_\alpha \frac {\partial} {\partial \bar{w}_\alpha}  \; , \qquad  \bar{{\bb D}}= \bar{w}_\alpha\frac {\partial} {\partial {w}_\alpha}  \; . \qquad \\
\nonumber
\eea
Notice the appearance of the ordering constant in the homogeneous differential operator \eqref{hU0=}.

To understand the structure of the fermionic constraint \eqref{hbdw=} it is useful to keep in mind
the relation of the derivatives and covariant derivatives in the central and analytical basis which
can be obtained from the identity

\bea
\text{d}= \text{d}x^a\partial_a + \text{d}\theta^\alpha \partial_\alpha + \text{d}\bar{\theta}{}^\alpha \bar{\partial}_\alpha + ia {\mathbb{D}}^{(0)} + i f \bar{{\mathbb{D}}} - i \bar{f} {\mathbb{D}}
\;  =\Pi^a\partial_a + \text{d}\theta^\alpha \text{D}_\alpha + \text{d}\bar{\theta}{}^\alpha \bar{\text{D}}_\alpha + ia {\mathbb{D}}^{(0)} + i f \bar{{\mathbb{D}}} - i \bar{f} {\mathbb{D}}= \quad  \\ \nonumber \\
=\text{d}{\rm x}^{(0)} \partial_{{\rm x}^0 } + \text{d}{\rm x}_A \partial_{{\rm x}_A} + \text{d}\bar{{\rm x}}_A \partial_{\bar{{\rm x}}_A } + \text{d}\theta^w \partial_{\theta^w } +\text{d}\theta^{\bar{w}} \partial_{\theta^{\bar{w}} }  + \text{d}\bar{\theta}{}^w \partial_{\bar{\theta}{}^w} + \text{d}\bar{\theta}{}^{\bar{w}} \partial_{\bar{\theta}{}^{\bar{w}}} + ia \tilde{\mathbb{D}}^{(0)} + i f \bar{\tilde{\mathbb{D}}} - i \bar{f} \tilde{\mathbb{D}}\, .\qquad \\ \nonumber
\eea
It implies, in particular,
\bea
\partial_a= u_a^{(0)} \partial_{{\rm x}^{(0)} } + u_a \partial_{{\rm x}_A} + \bar{u}_a \partial_{\bar{{\rm x}}_A}\; , \qquad  \\ \nonumber
\\
\text{D}_\alpha = \partial_\alpha + i(\gamma^a\bar{\theta})_\alpha =
w_\alpha \left(\partial_{\theta^w } +i\bar{\theta}{}^{\bar{w}}  \partial_{{\rm x}^{(0)} }  \right) + \bar{w}_\alpha \left(\partial_{\theta^{\bar{w}} } +4i\bar{\theta}{}^{\bar{w}} \partial_{\bar{\rm x}_A} +i\bar{\theta}{}^{{w}}\partial_{{\rm x}^{(0)} }  \right)\; ,  \\ \nonumber \\ \bar{\text{D}}_\alpha = \bar{\partial}_\alpha + i(\gamma^a{\theta})_\alpha =
w_\alpha \left(\bar{\partial}_{\bar{\theta}{}^w} +4i{\theta}{}^{w} \partial_{{\rm x}_A} +i{\theta}{}^{\bar{w}}\partial_{{\rm x}^{(0)} } \right) +  \bar{w}_\alpha \left(\bar{\partial}_{\bar{\theta}^{\bar{w}}} +i{\theta}{}^{w}  \partial_{{\rm x}^{(0)} }  \right)\; .
\\ \nonumber \eea
The last equation can be used to write \eqref{hbdw=} in the form

\be\label{hbdw==}
\hat{\bar{d}}_w=-\bar{w}\bar{\text{D}}+im{\theta}^{\bar{w}}-4i\theta^w\hat{\Phi}-i\theta^{\bar{w}}\hat{\Phi}{}^{(0)}\; .
\ee

\bigskip

\setcounter{equation}{0}

\def\theequation{D.\arabic{equation}}
\section{Born-Oppenheimer-like approach to  asymptotic form of a solution for $\mathbf{\textit{N}}$~=~2 3D mD$0$ system}

\label{AppBornOppenhemer}

Following \cite{Frohlich:1999zf} we can search for an asymptotic form of the solution of Eqs.  \eqref{Psi0}-\eqref{Psi2=}  using the so-called tabular or  endpoint coordinates (see \cite{Frohlich:1999zf} for references)

\be\label{Z=X+z}
Z^I= X^I e^{i\beta} + |X|^{-1/2} z^I \; , \qquad \bar{Z}{}^I= X^I e^{-i\beta} + |X|^{-1/2} \bar{z}{}^I \; , \qquad
\ee

\noindent  in the neighborhood of the classical vacuum configuration $Z^I= X^I e^{i\beta}$, $\bar{Z}{}^I= X^I e^{-i\beta}$, i.e. of the general solution of the conditions of vanishing of the YM potential which reduces to $\epsilon_{IJK}  Z^J\bar{Z}{}^K=0$.

In \eqref{Z=X+y} $X^I$ is real, $X^I=(X^I)^*$, $|X| =\sqrt{X^IX^I}$,  complex $z^I=(\bar{z}{}^I)^*$ are orthogonal to $X^I$, and, besides, can be restricted by 2 conditions which we can choose to be $\Re{\rm e} (z^I e^{-i\beta})=0$,

\be\label{XIzI=0}
X^I z^I=0 = X^I\bar{z}{}^I \; \; , \qquad z^I e^{-i\beta}+ \bar{z}^I e^{i\beta}=0\; .
\ee

\noindent The asymptotic regime corresponds to large value of $|X|$, formally

\be\label{r-infty}
r=|X| =\sqrt{X^IX^I} \mapsto \infty \; ,
\ee

\noindent
with $z^I\bar{z}{}^I$ kept to be finite.

In our case it is convenient to resolve \eqref{XIzI=0} by $z^I= iy^I  e^{i\beta}$ in terms of real vector $y^I$, so that $\bar{z}^I= -iy^I  e^{-i\beta}$ and the above coordinate system is finally described by

\bea\label{Z=X+y}
Z^I= (rn^I  + ir^{-1/2} y^I)e^{i\beta} \; , \qquad \bar{Z}{}^I=( rn^I-i r^{-1/2} y{}^I)e^{-i\beta} \; , \qquad \\ \nonumber \\ \label{r=|X|}
r:= |X| =\sqrt{X^IX^I}\; , \qquad n^I = X^I/r=X^I/|X|\quad \Rightarrow \quad n^In^I=1 \; , \qquad  \\ \nonumber \\ \label{yn=0} y^I = (y^I)^* \; , \qquad n^I y^I=0\; . \qquad
\eea

To proceed, it is convenient to introduce also the complex null-vectors which complete the unit vector $n^I$ till complete SO$(3)$ frame, i.e. which obey

\be
n^IU^I=0\; , \qquad n^I\bar{U}^I=0\; , \qquad U^IU^I=0\; , \qquad\bar{U}^I\bar{U}^I=0\; , \qquad U^I\bar{U}^I=1\; .
\ee

Then

\be\label{I=nn+}
\delta^{IJ}= n^In^J+U^I\bar{U}^J + \bar{U}^IU^J\; , \qquad
\ee
and the constraint \eqref{yn=0} can be resolved by

\be\label{yI=}
y^I= y_U\bar{U}^I+y_{\bar{U}} U^I\; , \qquad {\text{so that}}\qquad y^Iy^I=|y_U|^2 \; .
\ee

\noindent
The orientation of our SO(3) frame is fixed by setting $\epsilon_{IJK}\bar{U}^IU^Jn^K=i$ which implies

\be\label{eUn=iU}
\epsilon_{IJK}\bar{U}^Jn^K=-i\bar{U}^I\; , \qquad  \epsilon_{IJK}{U}^Jn^K=i{U}^I\; . \qquad
\ee

\noindent Using these relations, we find

\be\label{eZbZ=}
\epsilon_{IJK}Z^J\bar{Z}^K= 2\sqrt{r} (y_U\bar{U}{}^I -y_{\bar{U}}U^I )=2i\sqrt{r} \epsilon_{IJK}y^Jn^K \; .
\ee

The derivatives of the constrained  vectors can be expressed in terms of three Cartan forms

\be\label{Om1=}
\Omega_1=U^I\text{d}n^I\, , \qquad \bar{\Omega}_1 = \bar{U}^I\text{d}n^I \qquad \text{ and} \qquad \Omega_1^{(0)}= \bar{U}^I\text{d}{U}^I\;  \qquad
\ee
by

\be\label{dnI=}
\text{d}n^I= U^I \bar{\Omega}_1 + \bar{U}^I\Omega_1 \; ,  \qquad  \text{d}{U}^I= {U}^I\Omega_1^{(0)} - n^I\Omega_1  \; ,  \qquad  \text{d}\bar{U}^I = -\bar{U}^I\Omega_1^{(0)} -n^I  \bar{\Omega}_1\; .  \qquad
\ee

These can be used to express the derivative of the $y^I$ coordinates as

\be\label{dyI=}
\text{d}y^I=  (\text{d}y_U-y_U\Omega_1^{(0)}) \bar{U}^I+(\text{d}y_{\bar{U}}+y_{\bar{U}}\Omega_1^{(0)}) U^I-n^I (y_U \bar{\Omega}_1 + y_{\bar{U}}\Omega_1)\; .
\ee

Now, using  \eqref{Z=X+y}, \eqref{yI=}, \eqref{dnI=}, \eqref{dyI=} we can calculate d$Z^I$ and d$\bar{Z}{}^I$ and then decompose differential in the space of our bosonic coordinates written in two ways

\be
\text{d}=\text{d}Z^I\partial_I +\text{d}\bar{Z}{}^I\bar{\partial}_I =\text{d}r \partial_r +\text{d}\beta \partial_{(\beta)} + \text{d}y^I\partial^y_I +\text{d}n^I\partial_{n^I }
\ee

\noindent on the basis of 6 independent 1-forms $\text{d}r$, $\text{d}\beta$, $\text{d}y_U$, $\text{d}y_{\bar{U}}$, $\Omega_1$, $\bar{\Omega}_1$ \footnote{$\Omega^{(0)}$ will be also present in the expression, but coefficient for it will not produce independent equations. This fact reflects the U$(1)$ gauge symmetry of our construction. }. In such a way we find the following expression for the derivative with respect to the tabular coordinates

\bea\label{d-dr=}
\partial_r= n^I (e^{i\beta}\partial_I + e^{-i\beta}\bar{\partial}_I ) - \frac i 2\,\frac 1 { r^{3/2}}\, (y_U \bar{U}^I+y_{\bar{U}} U^I)\, (e^{i\beta}\partial_I - e^{-i\beta}\bar{\partial}_I )\; , \qquad
\\ \nonumber \\ \label{d-dbeta=} \partial_{(\beta)}=ir n^I (e^{i\beta}\partial_I - e^{-i\beta}\bar{\partial}_I ) - \frac 1 { r^{1/2}}\, (y_U \bar{U}^I+y_{\bar{U}} U^I)\, (e^{i\beta}\partial_I + e^{-i\beta}\bar{\partial}_I )\; , \qquad
\\ \nonumber \\  \label{bUd-dy=}
 \bar{U}^I\left(\partial^y_{I }-\frac i { r^{1/2}}(e^{i\beta}\partial_I - e^{-i\beta}\bar{\partial}_I )\right) =0\, , \qquad
\\ \nonumber \\  \label{Ud-dy=}  {U}^I\left(\partial^y_{I }-\frac i { r^{1/2}}(e^{i\beta}\partial_I - e^{-i\beta}\bar{\partial}_I )\right) =0\, , \qquad
\\ \nonumber \\ \label{bUd-dn=}
 \bar{U}^I\left(\partial_{n^I }-r(e^{i\beta}\partial_I + e^{-i\beta}\bar{\partial}_I )\right)
- y_{\bar{U}}n^I\left(\partial^y_{I } -\frac i { r^{1/2}}(e^{i\beta}\partial_I - e^{-i\beta}\bar{\partial}_I )\right) =0\, , \qquad
\\ \nonumber \\  \label{Ud-dn=}{U}^I\left(\partial_{n^I }-r(e^{i\beta}\partial_I + e^{-i\beta}\bar{\partial}_I )\right)
- y_{U}n^I\left(\partial^y_{I } -\frac i { r^{1/2}}(e^{i\beta}\partial_I - e^{-i\beta}\bar{\partial}_I )\right) =0\, . \qquad
\\ \nonumber
\eea

These relations are exact. But to solve them  for ${\partial}_I $ and $\bar{\partial}_I $ we need to use the conditions of the asymptotic regime
$r\mapsto  \infty$.
To this end let us first write Eqs. \eqref{d-dr=}-\eqref{Ud-dn=} in the form convenient for perturbative solution of the system for

\be \label{d+-I=}\partial^\pm_I= e^{i\beta}\partial_I \pm e^{-i\beta}\bar{\partial}_I \; . \qquad \ee

Firstly we find from \eqref{d-dbeta=}, \eqref{bUd-dy=} and \eqref{Ud-dy=}

\bea
\label{nId-I=}    n^I \partial^-_I=-\frac i r  \partial_{(\beta)}- \frac i {r\sqrt{r}}  (y_U \bar{U}^I+y_{\bar{U}} U^I)\, \partial^+_I\; , \qquad
\\ \nonumber \\ \label{Ud-I=}\bar{U}^I\partial^-_I = -i\sqrt{r} \bar{U}^I \partial^y_{I }\; , \qquad {U}^I\partial^-_I = -i\sqrt{r} {U}^I \partial^y_{I }\; , \qquad
\eea

which, using \eqref{I=nn+}, can be unified in one equation

\be
\partial^-_I=-i\sqrt{r} (\delta^{IJ} - n^In^J)\partial^y_{J}
-\frac i r  n^I \partial_{(\beta)}- \frac i {r\sqrt{r}} n^I (y_U \bar{U}^J+y_{\bar{U}} U^J)\, \partial^+_J\; . \qquad
\ee

Now, using these equations, one can resolve formally Eqs.  \eqref{d-dr=}, \eqref{bUd-dn=} and  \eqref{Ud-dn=} with respect to projections of $\partial^+_I$:

\bea \label{nId+I=} n^I\partial^+_I=
\partial_r+ \frac 1 {2r}\, (y_U \bar{U}^J+y_{\bar{U}} U^J)\, \partial^y_{J} \; , \qquad
\\ \nonumber \\ \label{bUId+I=} \bar{U}^I \partial^+_I=\frac 1 {r}\,  \bar{U}^I\partial_{n^I }- \frac 1 {r}\,  y_{\bar{U}}  n^I\partial^y_{I } + \frac 1 {r^2\sqrt{r}}  y_{\bar{U}} \, \partial_{(\beta)} + \frac 1 {r^3}  y_{\bar{U}} \, (y_U \bar{U}^J+y_{\bar{U}} U^J)\, \partial^+_{J} \, , \qquad
\\ \nonumber \\ \label{UId+I=} {U}^I \partial^+_I=\frac 1 {r}\,  {U}^I\partial_{n^I }- \frac 1 {r}\,  y_{U}  n^I\partial^y_{I } + + \frac 1 {r^2\sqrt{r}}  y_{{U}} \, \partial_{(\beta)}  + \frac 1 {r^3} y_{{U}}  \, (y_U \bar{U}^J+y_{\bar{U}} U^J)\, \partial^+_{J} \, .\qquad
\\ \nonumber
\eea

Now notice that, according to \eqref{bUId+I=} and \eqref{UId+I=},
$(y_U \bar{U}^J+y_{\bar{U}} U^J)\, \partial^+_J \propto \frac 1 r $ and that the only place where $\partial_r$ appears in the above equations  \eqref{nId-I=}-\eqref{UId+I=} is the first term in the r.h.s. of \eqref{nId+I=}. With this in mind and using \eqref{I=nn+} (several times) we arrive at the following approximate solution

\bea \label{d+I=} \partial^+_I= \frac 1 r \left[ n^I \left(r\, \partial_r  + \frac 12 y^J \partial^y_{J} \right) +(\delta^{IJ}- n^I n^J) \, \partial_{n^J} + y^I n^J\partial^y_{J}
\right] +{\cal O}(r^{-5/2})  \; , \qquad
\\ \nonumber
\\ \label{d-I=} \partial^-_I=-i\sqrt{r} (\delta^{IJ} - n^In^J)\partial^y_{J}
-\frac i r  n^I \partial_{(\beta)}+{\cal O}(r^{-5/2}) \;  \qquad
\eea

\noindent in which ${\cal O}(r^{-5/2})$ terms do not involve $\partial_r$ derivative.

Now from \eqref{d+-I=} we easily find

\bea \label{d-dZI=} \partial_I= \frac 1 2 e^{-i\beta} \Bigg[ -i\sqrt{r} (\delta^{IJ} - n^In^J)\partial^y_{J} +\frac 1 r \left( n^I \left(r\, \partial_r  -i \partial_{(\beta)}  + \frac 12  y^J\partial^y_{J} \right) + \right.  \nonumber \\ \nonumber \\ \left. + (\delta^{IJ}- n^I n^J) \, \partial_{n^J} + y^I n^J\partial^y_{J}
\right) \Bigg] +{\cal O}(r^{-5/2})  \; , \qquad
\\ \nonumber
\\ \label{d-dbZI=} \bar{\partial}_I= \frac 1 2 e^{i\beta} \Bigg[ +i\sqrt{r} (\delta^{IJ} - n^In^J)\partial^y_{J} +\frac 1 r \left( n^I \left(r\, \partial_r   +i \partial_{(\beta)} + \frac 12  y^J \partial^y_{J} \right) + \right.  \nonumber \\ \nonumber \\  \left. + (\delta^{IJ}- n^I n^J) \, \partial_{n^J} + y^I n^J\partial^y_{J}
\right) \Bigg] +{\cal O}(r^{-5/2})  \; . \qquad
\\ \nonumber
\eea

Notice that in the asymptotic region \eqref{r-infty}, $r\mapsto \infty$,  the leading order in the above decomposition of the derivatives is $\propto \sqrt{r}$,

\bea \label{d-dZI=r1-2} \partial_I= - \frac i 2 e^{-i\beta} \sqrt{r} (\delta^{IJ} - n^In^J)\partial^y_{J} +{\cal O}(r^{-1})  \; , \qquad
\\ \nonumber
\\ \label{d-dbZI=r1-2} \bar{\partial}_I=  \frac i 2 e^{i\beta}\sqrt{r} (\delta^{IJ} - n^In^J)\partial^y_{J} +{\cal O}(r^{-1})  \; , \qquad
\\ \nonumber
\eea

\noindent
so that

\bea \label{nId-dZI=0} n^I \partial_I= 0+{\cal O}(r^{-1})  \; , \qquad
\\ \nonumber
\\ \label{nId-dbZI=r1-2} n^I\bar{\partial}_I= 0 +{\cal O}(r^{-1})  \; . \qquad
\\ \nonumber
\eea

\noindent
 According to \eqref{eZbZ=} the multiplier  $\sqrt{r}$ is also present in all the nonvanishing r.h.s.'s  of equations \eqref{Psi0}-\eqref{fIJ=dIfJ} and   \eqref{Psi0=}-\eqref{Psi2=} which can be written in the form

\bea\label{Psi0=1}
&& 2i \sqrt{r}\epsilon_{IJK} y^Jn^K {\mathfrak{f}}_I^{(q-1)}=0\qquad \Leftrightarrow \qquad y_U\bar{U}^I{\mathfrak{f}}_I^{(q-1)}=y_{\bar{U}}U^I{\mathfrak{f}}_I^{(q-1)}\; , \qquad
\\ \nonumber \\ \label{Psi1=1}
&& \bar{\partial}_{I} {\mathfrak{f}}^{(q)}=8i\sqrt{2}\sqrt{r}
\epsilon_{JKL} y^Jn^K  \bar{\partial}_{[L} {\mathfrak{f}}_{I]}{}^{(q)} \; , \qquad
\\ \nonumber \\ \label{Psi2=1}
&&  \bar{\partial}_{[I} {\mathfrak{f}}_{J]}{}^{(q-1)}= 8i\sqrt{2}\sqrt{r}
y^{[I}n^{J]}  {\mathfrak{f}}^{(q-3)} \; \qquad
\eea

\noindent and

\bea \label{Psi0==1}
&& \partial_{I}  {\mathfrak{f}}^{(q-1)}_{I} =0 \; , \qquad
\\ \nonumber \\ \label{Psi1==1}
&& \partial_{J}  \bar{\partial}_{[I} {\mathfrak{f}}^{(q)}_{J]}=- \frac i {\sqrt{2} \mu^{12}}\, \sqrt{r}\, \epsilon_{IJK}y^J n^{K}f^{(q)}\; ,
\\
 \nonumber \\
 \label{Psi2=11}
 &&  \partial_{I}  {\mathfrak{f}}^{(q-3)} = - \frac  {i\sqrt{2}}{\mu^{12}} \, \sqrt{r}\, y^{[I}n{}^{J]} {\mathfrak{f}}^{(q-1)}_{J}\; .  \qquad
\eea

The algebraic equation \eqref{Psi0=1} can be easily solved by

\be\label{ffIq-1=}
{\mathfrak{f}}^{(q-1)}_{I}= n^I {\mathfrak{f}}^{(q-1)}_{(n)}+ y^I {\mathfrak{f}}^{(q-1)}_{(y)}\; . \qquad
\ee

Now we can decompose our ``wavefunctions'' in the inverse powers of $r\sqrt{r}$ and try to solve the equation order by order.  To be consistent we set

\be
 {\mathfrak{f}}^{(q...)}_{...}= r^{-k} \left( {\mathfrak{f}}^{(q...)}_{...[0]} + \frac 1 {r\sqrt{r}}   {\mathfrak{f}}^{(q...)}_{...[1]} +  \frac 1 {r^3} {\mathfrak{f}}^{(q...)}_{...[2]}+...\right)\; ,
\ee

\noindent with some $k$ to be determined, for all but the leading components of the state vector superfield while for this we assume

\be
 {\mathfrak{f}}^{(q)}= \sqrt{r}{\mathfrak{f}}^{(q)}_{[0]} + \frac 1 {r}   {\mathfrak{f}}^{(q)}_{[1]} +  \frac 1 {r^2} {\mathfrak{f}}^{(q)}_{[2]}+...\; .
\ee

Here we restrict our discussion by  zero-th order in which we find (following the terms $\propto \sqrt{r}$) the equations

\bea\label{Psi1=1-0}
&& (\delta^{IJ} - n^In^J)\partial^y_{J} {\mathfrak{f}}^{(q)}_{[0]}=8i\sqrt{2}
\epsilon_{JKL} y^Jn^K(\delta^{[L|P} - n^{[L}n^P)\partial^y_{P} {\mathfrak{f}}_{I][0]}{}^{(q)} \; , \qquad
\\ \nonumber \\ \label{Psi2==1}
&&   (\delta^{[I|K} - n^{[I|}n^K)\partial^y_{K} {\mathfrak{f}}_{J][0]}{}^{(q-1)}=16\sqrt{2}e^{-i\beta}
y^{[I}n^{J]}  {\mathfrak{f}}^{(q-3)}_{[0]} \; ,  \qquad
\\ \nonumber \\  \label{Psi0==1-0}
&&   (\delta^{IJ} - n^{I}n^J)\partial^y_{J}  {\mathfrak{f}}^{(q-1)}_{I[0]} =0 \; , \qquad
\\ \nonumber \\ \label{Psi1==1-0}
&& (\delta^{JK} - n^{J}n^K)\partial^y_{K} \, (\delta^{[I|L} - n^{[I|}n^L)\partial^y_{L}    {\mathfrak{f}}^{(q)}_{|J][0]}=-\frac {2\sqrt{2}\, i }  {(\mu^{6})^2}\,  \,\epsilon_{IJK}y^J n^{K}f^{(q)}_{[0]}\; ,
\\
 \nonumber \\
 \label{Psi2==1-0}
 &&  (\delta^{IJ} - n^{I}n^J)\partial^y_{J} {\mathfrak{f}}^{(q-3)}_{[0]} =  \frac  {2\sqrt{2}}{(\mu^{6})^2} \, e^{i\beta}  \, y^{[I}n{}^{J]} {\mathfrak{f}}^{(q-1)}_{J[0]}\; .  \qquad
\eea

Notice  that this system of equations contains derivatives with respect to variables $y^I$ only. With this in mind one easily finds  that  \eqref{Psi1=1-0} implies

\be\label{Psi1=1-0n}
\epsilon_{JKL} y^Jn^K(\delta^{[L|P} - n^{[L}n^P)\partial^y_{P} (n^I {\mathfrak{f}}_{I[0]}{}^{(q)})=0 \; , \qquad
\ee

\noindent which indicates that $n^I {\mathfrak{f}}_{I[0]}{}^{(q)}$ depends on $y^J$ only through its square $|y|^2=y^Jy^J$.

Eq. \eqref{Psi0==1-0} with leading order term of \eqref{ffIq-1=}
\be\label{ffIq-1=1}
{\mathfrak{f}}^{(q-1)}_{I[0]}= n^I {\mathfrak{f}}^{(q-1)}_{(n)[0]}+ y^I {\mathfrak{f}}^{(q-1)}_{(y)[0]}
\ee
does not impose any condition of ${\mathfrak{f}}^{(q-1)}_{(n)[0]}$ but requires  ${\mathfrak{f}}^{(q-1)}_{(y)[0]}$ to obey
$(y^I\partial^y_{I} +2){\mathfrak{f}}^{(q-1)}_{(y)[0]}=0$. Nevertheless, as we will see in a moment, we have to choose the trivial solution of this equation.

Indeed, using \eqref{ffIq-1=1} in  \eqref{Psi2==1-0}, we find
\bea
 \label{Psi2==1-01}
 &&  (\delta^{IJ} - n^{I}n^J)\partial^y_{J} {\mathfrak{f}}^{(q-3)}_{[0]} =  \frac  {\sqrt{2}}{(\mu^{6})^2} \, e^{i\beta}  \, \left(y^{I} {\mathfrak{f}}^{(q-1)}_{(n)[0]}-n{}^{I} (y^{J}y^{J}){\mathfrak{f}}^{(q-1)}_{y[0]}\right)\; .  \qquad
\eea

As both l.h.s. and the first term in the r.h.s. of this equation vanish when contracted with $n^I$, the last term should vanish by itself,

\be
{\mathfrak{f}}^{(q-1)}_{y[0]}=0\qquad \Rightarrow \qquad {\mathfrak{f}}^{(q-1)}_{I[0]}= n^I {\mathfrak{f}}^{(q-1)}_{(n)[0]}\; .
\ee
Then   \eqref{Psi2==1-01} reduces to

\bea
 \label{Psi2==1-02}
 &&  (\delta^{IJ} - n^{I}n^J)\partial^y_{J} {\mathfrak{f}}^{(q-3)}_{[0]} =  \frac  {\sqrt{2}}{(\mu^{6})^2} \, e^{i\beta}  \,y^{I} {\mathfrak{f}}^{(q-1)}_{(n)[0]}\; .  \qquad
\eea
This equation implies that both $ {\mathfrak{f}}^{(q-3)}_{[0]}$ and $ {\mathfrak{f}}^{(q-1)}_{(n)[0]}$ depend on $y^I$ vector coordinate only through  its length $|y|=\sqrt{y^Iy^I}$ and that
\be\label{dy2fq-3=} \partial_{|y|^2} {\mathfrak{f}}^{(q-3)}_{[0]}= \frac  {1}{(\mu^{6})^2\sqrt{2}} \, e^{i\beta}  \,{\mathfrak{f}}^{(q-1)}_{(n)[0]}\; . \qquad \ee

On the other hand, with the above conclusion on dependence on   $y^I$ vector only through  its length $|y|=\sqrt{y^Iy^I}$, Eq. \eqref{Psi2=1} reduces to $ \partial_{|y|^2}{\mathfrak{f}}^{(q-1)}_{(n)[0]}= 8\sqrt{2} \, e^{-i\beta}  \, {\mathfrak{f}}^{(q-3)}_{[0]}$. Using this and \eqref{dy2fq-3=}, we find a simple equation

\be\label{dy22fq-1=} \frac {\partial}{\partial |y|^2} \frac {\partial}{\partial |y|^2} {\mathfrak{f}}^{(q-1)}_{(n)[0]}= \frac  {8}{(\mu^{6})^2}   \,{\mathfrak{f}}^{(q-1)}_{(n)[0]}\;  \qquad \ee
which is solved by

\be\label{fq-10=exp}
 {\mathfrak{f}}^{(q-1)}_{(n)[0]}= {\mathfrak{h}}^{(q-1)}(r,\beta, \vec{n})\; {\rm exp}\left(-2\sqrt{2}\frac{|y|^2}{\mu^{6}} \right)\;
\ee
with some function  $ {\mathfrak{h}}^{(q-1)}(r,\beta, \vec{n})$ depending on the remaining coordinates.

Similarly one can study the case of other components of the state vector superfield, the statistics of which (bosonic if the superfield is bosonic) is opposite to the above discussed components (fermionic if the superfield is bosonic) which obey Eqs. \eqref{Psi1=1-0} and  \eqref{Psi1==1-0}. As the study of this case  is a bit more involved  then above, we will simplify it a bit by choosing a special frame in which
$n^I=(0,0,1)=\delta^I_3$ and hence $y^I= (y^1,y^2, 0)= \delta^I_r y^r$.

In this SO$(3)$ frame the solution of \eqref{Psi1=1-0n} can be written in the form ${\mathfrak{f}}_{3[0]}{}^{(q)}(y^J)=  {\mathfrak{f}}_{3[0]}{}^{(q)}(|y|^2)$ and the remaining components of  \eqref{Psi1=1-0} and \eqref{Psi1==1-0} acquire the following simple form

 \bea\label{Psi1=1-00}
&& \partial^y_{r} {\mathfrak{f}}^{(q)}_{[0]}=8i\sqrt{2}
\epsilon_{st} y^t\partial^y_{[s} {\mathfrak{f}}_{r][0]}{}^{(q)} \; , \qquad
\\ \nonumber \\ \label{Psi1==1-00}
&& \partial^y_{s} \, \partial^y_{[r}    {\mathfrak{f}}^{(q)}_{s][0]}=-\frac {i2\sqrt{2} }  {(\mu^{6})^2}\,  \,\epsilon_{rs}y^s{\mathfrak{f}}^{(q)}_{[0]}\; \; .  \qquad
\eea

\noindent Furthermore, as $r,s,t=1,2$, $\partial^y_{[s} {\mathfrak{f}}_{r][0]}{}^{(q)}=\epsilon_{sr}\tilde{{\mathfrak{f}}}^{(q)}_{[0]}$ (with
$\tilde{{\mathfrak{f}}}^{(q)}_{[0]}= \frac 1 2 \epsilon_{sr} \partial^y_{[s} {\mathfrak{f}}_{r][0]}{}^{(q)}$) and the above equations further simplify to

 \be\label{Psi1=1-00=}
 \partial^y_{r} {\mathfrak{f}}^{(q)}_{[0]}=8i\sqrt{2}y^r \tilde{{\mathfrak{f}}}^{(q)}_{[0]}\; , \qquad
 \partial^y_{s} \, \tilde{{\mathfrak{f}}}^{(q)}_{[0]}=-\frac{2\sqrt{2} i}{(\mu^{6})^2}\, y^s  \,{\mathfrak{f}}^{(q)}_{[0]}\; .  \qquad
\ee

\noindent  Then it is easy to conclude that both ${\mathfrak{f}}^{(q)}_{[0]}$ and $\tilde{{\mathfrak{f}}}^{(q)}_{[0]}$ depend on $y^I$ through  its square $|y|^2=y^Iy^I$ and obey the second order linear differential equation in $|y|^2$, in particular

\be
\partial_{|y|^2}\partial_{|y|^2}{\mathfrak{f}}^{(q)}_{[0]}= \frac 8 {(\mu^{6})^2}{\mathfrak{f}}^{(q)}_{[0]}\;
\ee

\noindent which is solved by

\be\label{fq0=exp}
{\mathfrak{f}}^{(q)}_{[0]}= {\mathfrak{h}}^{(q)}_{[0]}(r,\beta,\vec{n}) \exp \left(  -\frac {2\sqrt{2}} {\mu^{6}}\, |y|^2\right)\; .
\ee

The sign of the expressions in the exponents of \eqref{fq-10=exp} and \eqref{fq0=exp} is chosen to be such that the wavefunction is convergent in $y^I$. We will not consider higher orders of the Born-Oppenheimer-like approximation, as this goes beyond the scope of this paper, but refer to
\cite{Frohlich:1999zf} for such a study. Notice that there, besides the case of
Matrix model obtained by dimensional reduction of $\text{D}=3$ SYM model to $\text{d}=1$ also the Matrix models proceeding from reduction of higher $\text{D}$  SYM were studied and the exceptional properties of $\text{D}=10$ case were noticed. This will appear in a setup similar to \eqref{nuXi=0}-\eqref{cHXi=0} from the field theory  of our 10D mD$0$ model which we plan to quantize in the near future.

\end{document}